%% file: main.tex
\documentclass[preprint]{elsarticle}

\usepackage[utf8]{inputenc}
\usepackage{graphicx,amsmath,amsfonts,subfiles,caption,subcaption,geometry,amssymb,tikz}
\usetikzlibrary{patterns,decorations.markings,arrows.meta,arrows}
\graphicspath{{./images}{../images}}
\newif\ifinlinefigures
\inlinefigurestrue
\usepackage{lineno,hyperref}
\modulolinenumbers[5]

\journal{Combustion and Flame}
\biboptions{square,sort&compress}
\usepackage{ulem}
\newcommand{\Htwo}{\textsc{h$_{2}$}}
\newcommand{\OH}{\textsc{oh}}
\newcommand{\HtwoO}{\textsc{h$_{2}$o}}
\newcommand{\Otwo}{\textsc{o$_{2}$}}
\newcommand{\Ntwo}{\textsc{n$_{2}$}}
\newcommand{\HOtwo}{\textsc{ho$_{2}$}}
\newcommand{\Ka}{\textrm{Ka}}

\renewcommand{\Re}{\textrm{Re}}
\newcommand{\ddz}{\frac{\textrm{d}}{\textrm{d}z}}

\begin{document}
\subfile{title}
\section{Introduction}
    \subfile{intro}

\section{Background}\label{sec:background}

\subfile{background}

\section{Direct numerical simulation configuration}\label{sec:DNS}
    \subfile{configuration/preamble}
    \subsection{Numerical solver: PeleLMeX}
        \subfile{configuration/Pele}

    \subsection{Jet-in-crossflow configuration}
        \subfile{configuration/JIC}

    \subsection{Combustor configuration}
        \subfile{configuration/DNS}

\section{Flame structure}\label{sec:structure}
    \subfile{structure/preamble}
    \subsection{Temporally-averaged jet and flame structures}\label{sec:structure:steady}
        \subfile{structure/global}

    \subsection{Instantaneous flame structure}\label{sec:structure:instant}
        \subfile{structure/instantaneous}

\section{Flame stabilisation mechanism}\label{sec:stabilisation}
    \subfile{stabilisation/preamble}
    \subsection{Flame propagation}\label{sec:stabilisation:propagation}
        \subfile{stabilisation/propagation}

    \subsection{Ignition events}\label{sec:stabilisation:autoignition}

\subfile{stabilisation/autoignition}

    \subsection{Entrainment and mixing}\label{sec:stabilisation:entrainment}
        \subfile{stabilisation/entrainment}
\section{Discussion and conclusions}\label{sec:conclusions}
    \subfile{conclusion}

\section{Acknowledgements}
    \subfile{acknowledgements}
\bibliography{references}
\newpage
\appendix
\section{Resolution independence}\label{appendix:resolution}
    \subfile{appendices/resolution}

\end{document}


\begin{frontmatter}
\title{Direct numerical simulation of a high-pressure hydrogen micromix combustor:\\
  flame structure and stabilisation mechanism -\\Supplementary material}

\author[ncl]{T.~L.~Howarth\corref{cor1}}
\ead{T.Howarth2@newcastle.ac.uk}
\author[rel,sbs]{M.~A.~Picciani}
\author[soton]{E.~S.~Richardson}
\author[nrel]{M.~S.~Day}
\author[ncl]{A.~J.~Aspden}
\cortext[cor1]{
  Corresponding author
}
\address[ncl]{School of Engineering, Newcastle University, Newcastle-Upon-Tyne, NE1 7RU, UK}
\address[rel]{Reaction Engines Ltd., Culham Science Centre, Abingdon, OX14 3DB, UK}
\address[sbs]{Sunborne Systems Ltd., The Lambourn, Wyndyke Furlong, Abingdon, OX14 1UJ, UK}
\address[soton]{Faculty of Engineering and Physical Sciences, University of Southampton, Southampton, SO17 1BJ, UK}
\address[nrel]{Computational Science Center, National Renewable Energy Laboratory, Golden, CO 80401‑3305, USA}

\begin{abstract}
In the interest of visualising the structure and stabilisation mechanism of the flame, various animations of the flame have been attached as supplementary material. Here the contents of each animation are described.
\end{abstract}
\end{frontmatter}
\section{Mid-plane animations}
Slices through the $x=0$ and $y=0$ are collated into various animations. In all cases the top of the animation are the $x$-slices, and the bottom half are the $y$-slices. A list is provided below:
\begin{itemize}
    \item \textbf{z-velocity.mp4}: $u_{z}$, with colours given by the colour bar from figure 5. 
    \item \textbf{Y(H2).mp4}: $Y_{\textsc{h}_{2}}$, with colours given by the colour bar from figure 4.
    \item \textbf{temp.mp4}: $T$, with colours given by the colour bar from figure 3.
    \item \textbf{mag\_vort.mp4}: Magnitude of vorticity $|\mathbf{\omega}| = |\nabla \times \mathbf{u}|$. A reduction in the intensity of the turbulence can be seen within a short distance downstream of the stream-wise vortex tubes entering the domain at the inlet.
    \item \textbf{Y(HO2).mp4}: $Y_{\textsc{ho}_{2}}$. Stabilisation through ignition can be seen through the rapid increasing in concentration of HO2$_{2}$ between the core and peripheral flames. 
    \item \textbf{combo.mp4}: A colour palette is derived from a combination of fuel and temperature plots. Blue corresponds to the presence of fuel, and red scales with temperature. Black regions indicate regions of low temperature and no fuel, i.e.~the non-reacting air buffer.
    \end{itemize}
\section{Diagonal animations}
Identical fields are reproduced as the mid-plane animations, however, for these images slices are taken through the diagonals of the domain, with $x=y$ slices on the top half and $x=-y$ slices on the bottom half.
\begin{itemize}
    \item \textbf{z-velocity-diag.mp4}: diagonal version of \textbf{z-velocity.mp4}.
    \item \textbf{Y(H2)-diag.mp4}: diagonal version of \textbf{Y(H2).mp4}.
    \item \textbf{temp-diag.mp4}: diagonal version of \textbf{temp.mp4}.
    \item \textbf{mag\_vort-diag.mp4}: diagonal version of \textbf{mag\_vort.mp4}.
    \item \textbf{Y(HO2)-diag.mp4}: diagonal version of \textbf{Y(HO2).mp4}.
    \item \textbf{combo-diag.mp4}: diagonal version of  \textbf{combo.mp4}.
\end{itemize}
\section{Line-of-sight animation}
Following equation (17), a line-of-sight has been constructed to examine the formation of ignition spots without being misled by out-of-plane effects. An animation (cropped down to the region of interest) is provided under the filename \textbf{los.mp4}. The left side shows the integration through the $x$-direction and the right side shows the integration through the $y$-direction. The sporadic kernels below the flame base can be observed here, as well as the formation of the igniting flame sheets, discussed in section 5.

%% file: title.tex
\begin{frontmatter}
\title{Direct numerical simulation of a high-pressure hydrogen micromix combustor:\\
  flame structure and stabilisation mechanism}

\author[ncl]{T.~L.~Howarth\corref{cor1}}
\ead{T.Howarth2@newcastle.ac.uk}
\author[rel,sbs]{M.~A.~Picciani}
\author[soton]{E.~S.~Richardson}
\author[nrel]{M.~S.~Day}
\author[ncl]{A.~J.~Aspden}
\cortext[cor1]{
  Corresponding author
}
\address[ncl]{School of Engineering, Newcastle University, Newcastle-Upon-Tyne, NE1 7RU, UK}
\address[rel]{Reaction Engines Ltd., Culham Science Centre, Abingdon, OX14 3DB, UK}
\address[sbs]{Sunborne Systems Ltd., The Lambourn, Wyndyke Furlong, Abingdon, OX14 1UJ, UK}
\address[soton]{Faculty of Engineering and Physical Sciences, University of Southampton, Southampton, SO17 1BJ, UK}
\address[nrel]{Computational Science Center, National Renewable Energy Laboratory, Golden, CO 80401‑3305, USA}

\begin{abstract}
A high-pressure hydrogen micromix combustor has been investigated using direct numerical simulation with detailed chemistry to examine the flame structure and stabilisation mechanism. The configuration of the combustor was based on the design by Schefer \cite{Schefer2003EvaluationBurner}, using numerical periodicity to mimic a large square array. A precursor simulation of an opposed jet-in-crossflow was first conducted to generate appropriate partially-premixed inflow boundary conditions for the subsequent reacting simulation.  The resulting flame can be described as an predominantly-lean inhomogeneously-premixed lifted jet flame.  Five main zones were identified: a jet mixing region, a core flame, a peripheral flame, a recirculation zone, and combustion products.  The core flame, situated over the jet mixing region, was found to burn as a thin reaction front, responsible for over 85\% of the total fuel consumption.  The peripheral flame shrouded the core flame, had low mean flow with high turbulence, and burned at very lean conditions (in the distributed burning regime).  It was shown that turbulent premixed flame propagation was an order-of-magnitude too slow to stabilise the flame at these conditions.  Stabilisation was identified to be due to ignition events resulting from turbulent mixing of fuel from the jet into mean recirculation of very lean hot products.  Ignition events were found to correlate with shear-driven Kelvin-Helmholtz vortices, and increased in likelihood with streamwise distance.  At the flame base, isolated events were observed, which developed into rapidly burning flame kernels that were blown downstream.  Further downstream, near-simultaneous spatially-distributed ignition events were observed, which appeared more like ignition sheets.  The paper concludes with a broader discussion that considers generalising from the conditions considered here.
\end{abstract}

\begin{keyword}
Direct numerical simulation, hydrogen, flame structure, flame stabilisation, micromix combustor
\end{keyword}

\end{frontmatter}

%% file: intro.tex
Micromix combustor design seeks to meet the competing objectives of low flashback susceptibility and low NO$_{x}$ emission by replacing larger burners with a large number of smaller non-premixed fuel injectors, such that NO$_{x}$ formation is limited by the short residence time within the small flames. In particular, micromix has received attention as a means to manage the elevated flashback risks associated with burning hydrogen in aircraft, spacecraft, industrial and domestic combustors \cite{Funke2019AnApplications}.  Over the years, several generations of designs have been proposed; a comprehensive overview can be found in \cite{Funke202130Activities}. Micromix combustors are designed based on the concept of a large number of small non-premixed injectors, and a sudden expansion to establish recirculating flow.  Common designs use a jet-in-crossflow configuration, injecting fuel into an oxidiser crossflow. Flame behaviour in these combustors is influenced by several factors, including global equivalence ratio, injection velocities, and combustor geometry. 


While there have been extensive experimental and low-fidelity numerical studies (e.g.~\cite{HajAyed2015ExperimentalTurbine,Funke2019NumericalApplications,Lopez-Ruiz2021StudyApproach,Boerner2013DevelopmentTurbine}) of the factors affecting flame stability and NO$_{x}$ emissions, a fundamental investigation into the structure of the flame, and detailed stabilisation mechanism in particular, has not been reported experimentally or numerically for hydrogen micromix burners. Understanding both flame structure and stabilisation is crucial for selecting appropriate models in lower-fidelity simulations, as well as for the conceptualisation and design of new micromix combustors. 

The particular configuration examined in this paper follows the NASA design used by Schefer \cite{Schefer2003EvaluationBurner}, whereby the fuel injector plate is comprised of an array of circular air nozzles, with two diametrically opposed fuel jets within the nozzle. To enable direct numerical simulation (DNS) with detailed chemistry, periodicity is exploited to mimic the behaviour of an individual injector within a large square array of ports.  A range of inlet temperatures and pressures are relevant for hydrogen micromix combustion systems, from ambient temperatures and pressures that might be found in a domestic boiler, to elevated temperatures and pressures resulting from compression within aerospace propulsion systems. The inlet conditions affect flame stabilisation, due to their effect on flame speed and ignition behaviour, and strongly affect NO$_{x}$ production. The present study employs high-pressure high-temperature operating conditions that might be encountered in a high power-density aerospace application, such as the preburner in Reaction Engines' SABRE cycle (see \cite{Zhang2017ThermodynamicMode} for an outline of the thermodynamic cycle). The aim of the study is to identify the flame structure and the stabilisation mechanism. A review of technical background is provided in section \ref{sec:background}, followed by a description of the numerical solver and DNS configuration in section \ref{sec:DNS}. The simulations are analysed in parts.  The first part considers both the steady (section \ref{sec:structure:steady}) and instantaneous (section \ref{sec:structure:instant}) flame structure, the combustion regimes and turbulence-flame interactions. The second part considers the stabilisation mechanism, assessing the role of flame propagation (section \ref{sec:stabilisation:propagation}), ignition events (section \ref{sec:stabilisation:autoignition}) and entrainment and mixing (section \ref{sec:stabilisation:entrainment}). The paper is concluded by a summary of the structure and stabilisation (section \ref{sec:conclusion:structure}), and a broader discussion of the factors affecting micromix combustor design (section \ref{sec:conclusion:conjecture}).

%% file: background.tex
\subfile{tikz/micromix}

A schematic of the micromix combustor considered is shown in figure \ref{fig:micromix_schematic}, where opposing sub-millimeter injectors blow hydrogen at high speed into an air crossflow inside the region is referred to as the `air port'. Each fuel jet forms a counter-rotating vortex pair, which may or may not interact depending on the level of penetration. Some distance downstream from the fuel injection, there is a sudden expansion into the combustion chamber which forms a jet. The corner between the air port and combustion chamber is referred to as the `lip'. 

A key contributing factor to the flame structure and stabilisation is the bulk flow and turbulent mixing in the air port ahead of the combustion chamber. The hydrogen injection into the air stream forms a jet-in-crossflow configuration; to avoid confusion with the main jet in the combustion chamber, the hydrogen inflow is referred to as an injector, and the configuration abbreviated to JICF.  The JICF is described non-dimensionally by the momentum flux ratio, given by
\begin{equation}
    J = \frac{\rho_{\textsc{i}}u_{\textsc{i}}^{2}}{\rho_{\textsc{c}}u_{\textsc{c}}^{2}},
\end{equation}
where subscripts $\textsc{i}$ and $\textsc{c}$ denote injector and crossflow quantities, respectively, and $\rho$ and $u$ denote density and velocity, respectively. For a given global equivalence ratio $\phi$, assuming ideal gas behaviour and no change in pressure across the injector, a relation can be found
\begin{equation}
    J = \frac{\phi^{2}\overline{W}_{\textsc{c}}T_{\textsc{i}}Y_{\textsc{o},\textsc{c}}^{2}A_{\textsc{c}}^{2}}{N^{2}\overline{W}_{\textsc{i}}T_{\textsc{c}}s^{2}Y_{\textsc{f},\textsc{i}}^{2}A_{\textsc{i}}^{2}},
\end{equation}
where $\overline{W}$ is the mean molecular weight, $T$ is the temperature, $N$ is the number of fuel injectors per air port, $A$ is the area, $Y_{\textsc{o,c}}$ is the oxygen mass fraction in the air crossflow, $Y_{\textsc{f,i}}$ is the fuel mass fraction in the fuel injector, and $s$ is the mass stoichiometric ratio. This equation shows the direct relationship between global equivalence ratio and momentum flux ratio. At a lower momentum flux ratio, the fuel does not penetrate as far into the crossflow, similarly at a higher momentum flux ratio the fuel penetrates further, and may interact with an opposed fuel injector, depending on geometry. A review of the effect of momentum flux ratio (and other factors) on JICF trajectory and mixing can be found in \cite{Mahesh2013TheCrossflow}.

For a fixed geometry, the penetration $y$ typically scales with $J^{\alpha}$, where $1/3 < \alpha < 2/3$. Fluid regions in a canonical JICF configuration can be said to be on the windward- or leeward-side which distinguish between fluid flowing above or below the counter-rotating vortex pair respectively. Furthermore, regions within a small distance downstream of the injector (typically $x<0.3Jd_{\textsc{i}}$, for $d_{\textsc{i}}$ the injector diameter, \cite{Smith1998MixingCrossflow}) are classed as the near-field, and further downstream as far-field. In the case of far-field mixing, the scalar field fluctuations are directly attributed to the Reynolds number of the injector ($\Re_{\textsc{i}} = u_{\textsc{i}}d_{\textsc{i}}/\nu_{\textsc{i}}$), with improving mixedness up to $\Re_{\textsc{i}} \sim 20000$ \cite{Shan2006Reynolds-numberMixing}. Naturally, increasing $J$ (or $\phi$) will increase the Reynolds number through increased velocities $\Re_{\textsc{i}} \sim \sqrt{J} \, (\sim \phi)$. In the near field, mixing is driven by the formation of the counter-rotating vortex pair rather than turbulent mixing in the far-field, with decay of scalar concentration along the JICF centerline scaling with arc length of the centreline to the power of -1.3 \cite{Smith1998MixingCrossflow}. However, the study of JICF trajectories and mixing in variable-density JICF is still an active area of research (e.g.~\cite{Gevorkyan2016TransverseCharacteristics,Gevorkyan2018InfluenceJets,Getsinger2014StructuralCrossflow}), and more studies are required.

Flame stabilisation mechanisms depend simultaneously on the fluid dynamics, flame regime and interaction with external factors, such as walls or acoustics. Assuming no direct interaction with external sources, four loosely categorised lifted flame stabilisation theories are 
(\cite{Lyons2007TowardExperiments,Lawn2009LiftedAir,Peters1983LiftoffFlames,Peters2000TurbulentCombustion,Broadwell1985BlowoutFlames,Buckmaster1996Edge-flame-holding,Karami2015MechanismsFlame}):
\begin{enumerate}
    \item Premixed flame propagation: The base of the lifted flame is premixed, and hence stabilised by premixed flame propagation, where the propagation can either be driven by the laminar or turbulent flame speed.
    \item Critical scalar dissipation rate: The extinction of diffusion flamelets controls the position of flame stabilisation, which can be determined by the scalar dissipation rate dropping below some critical value \cite{Peters1983LiftoffFlames}. This theory has become less popular in recent years \cite{Peters2000TurbulentCombustion}.
    \item Large eddy theory: Large scale structures in the fluid are able to repeatedly transport either the flame, or its products upstream to stabilise the reaction zone \cite{Broadwell1985BlowoutFlames}.
    \item Edge flame propagation: This theory assumes that the flame is partially premixed, and can propagate upstream in balance with the local flow-field, with a triple-flame structure observed \cite{Buckmaster1996Edge-flame-holding}.
\end{enumerate}
\cite{Karami2015MechanismsFlame} presents a visualisation of each theory (see figure 1 therein), and it is noted that autoignition may also be a factor that can contribute to any of these mechanisms. The stabilisation mechanism of lifted flames has been investigated using three-dimensional DNS in several studies, but almost exclusively focussing on non-premixed configurations stabilised through autoignition in hot coflows (e.g. \cite{Pantano2004DirectChemistry,YOO2009Three-dimensionalStructure,Kerkemeier2013DirectAir,Karami2016EdgeStudy, Yoo2011ACoflow}).

It is noted by Schefer \cite{Schefer2003EvaluationBurner} that various flame behaviours are observed in the prescribed micromix combustor including both stable and unstable lifted flames, flames attached to the lip, blow-off and flashback. Flashback only occurred in low speed, higher equivalence ratio environments which are rarely used in micromix combustors. Blow-off occurs at very lean equivalence ratios, with this limit increasing with air crossflow velocity. Mixedness will have a strong effect on flame structure, stability and whether the flame becomes attached, all of which is poorly understood. Flame stabilisation in micromix combustors has been attributed to `flame stabilising vortices' (e.g.\ \cite{Funke202130Activities}), although the precise mechanism by which this enables stabilisation has not yet been investigated, nor has the combustion regime that lifted micromix flames burn in been explored.

%% file: tikz/micromix.tex
\begin{figure}[ht!]
    \centering
    \resizebox{60mm}{!}{
    \begin{tikzpicture}[scale=4]
\draw[dotted] (-2,1) -- (-2,5);
\draw (-1,0) -- (-1,0.4);
\fill[pattern=north east lines] (-1,0) rectangle (-1.05,0.35);
\draw (-1,0.6) -- (-1,1.5) -- (-2,1.5);
\fill[pattern=north east lines] (-1,0.65) rectangle (-1.05,1.45);
\fill[pattern=north east lines] (-1,1.45) rectangle (-2,1.5);

\draw[dotted](-2,1.5) -- (-2,5);

\draw (-1,0.6) -- (-2,0.6);
\fill[pattern=north east lines] (-1,0.6) rectangle (-2,0.65);
\draw (-1,0.4) -- (-2,0.4);
\fill[pattern=north east lines] (-1,0.4) rectangle (-2,0.35);
\draw[dashed] (-1,0.6) to[out=0, in=270] (-0.4,1.5);
\draw[dashed] (-1,0.4) to[out=0, in=270] (0,1.5);
\draw[dashed] (-1,1.5) -- (-2,4.4);

\draw (1,0) -- (1,0.4);
\fill[pattern=north east lines] (1,0) rectangle (1.05,0.35);
\draw (1,0.6) -- (1,1.5) -- (2,1.5);
\fill[pattern=north east lines] (1,0.65) rectangle (1.05,1.45);
\fill[pattern=north east lines] (1,1.45) rectangle (2,1.5);

\draw[dotted](2,1.5) -- (2,5);

\draw (1,0.6) -- (2,0.6);
\fill[pattern=north east lines] (1,0.6) rectangle (2,0.65);
\draw (1,0.4) -- (2,0.4);
\fill[pattern=north east lines] (1,0.4) rectangle (2,0.35);
\draw[dashed] (1,0.6) to[out=180, in=270] (0.4,1.5);
\draw[dashed] (1,0.4) to[out=180, in=270] (0,1.5);
\draw[dashed] (1,1.5) -- (2,4.4);

\node at (1.5,0.5) {\huge fuel injector}; 
\node[text width=6cm,align=center] at (0,1.8) {\huge interacting counter-rotating vortex pair}; 
\node at (0,0.1) {\huge air crossflow in air port}; 
\node at (0.8, 1.4) {\huge lip};
\node at (0,2.8) {\huge main jet};
\node at (0,4.5) {\huge combustion chamber};
\end{tikzpicture}
}
\caption{Schematic of a micromix combustor in 2D, including nomenclature used throughout.}
\label{fig:micromix_schematic}
\end{figure}
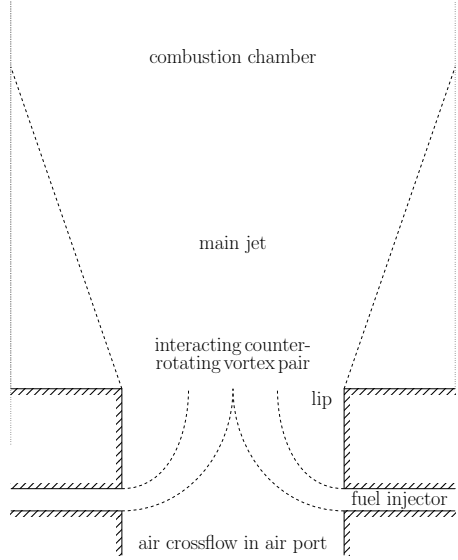

%% file: configuration/preamble.tex
This section provides a description of the numerical solver, the generation of micromix combustor inflow conditions using a precursor simulation of a non-reacting JICF, and the configuration for the reacting combustor simulation.

%% file: configuration/pele.tex
The simulations presented here were conducted with the PeleLMeX solver \cite{Day2022PeleLMeX}, developed as part of the US DoE Exascale Computing project. PeleLMeX is the non-subcycling version of the well-established code PeleLM, which has been extensively used to study hydrogen flames in the canonical flame-in-a-box configuration in 2D (e.g.\ \cite{Bell2007NumericalFlames,Howarth2022AnFlames}), 3D freely-propagating (e.g.\ \cite{Aspden2011CharacterizationFlames,Howarth2023Thermodiffusively-unstablePoints}) and turbulent (e.g.\ \cite{Aspden2015Turbulence-chemistryCombustion,Aspden2019TowardsFlames,Howarth2023Thermodiffusively-unstablePoints}) flames, as well as in experimental configurations (e.g.\ \cite{Bell2013SimulationBurner,Day2015AFlames}). PeleLMeX is capable of running on massively parallel CPU- and GPU-based computing systems, both of which were used for the present study.

The equations of motion are based on a low-Mach-number formulation of the reacting Navier-Stokes equations, where the fluid is treated as a mixture of ideal gases \cite{Day2000NumericalChemistry}. A mixture-averaged model for differential species diffusion is used, where each species has its own temperature- and composition-dependent diffusivity; the Soret effect is neglected. The discretisation couples a multi-implicit spectral deferred correction approach for the integration of mass, species and energy equations with a density-weighted approximate projection method, which has incorporated the equation of state through a velocity divergence constraint \cite{Nonaka2012AChemistry}. The resulting discretisation is integrated with timesteps determined by the advective transport through an advective CFL number, with the faster diffusion and chemical processes treated implicitly. This scheme is embedded in a parallel adaptive mesh refinement (AMR) algorithm framework based on a hierarchical system of rectangular grid patches. The complete integration algorithm is second-order accurate in space and time; the reader is referred to \cite{Day2000NumericalChemistry,Nonaka2012AChemistry,Nonaka2017AIntegration} for further details. The transport coefficients, thermodynamic relationships and chemical kinetics are obtained from a comprehensive H$_{2}$/O$_{2}$ kinetic model \cite{Burke2012ComprehensiveCombustion}, suitable for high-pressure hydrogen combustion. 

%% file: configuration/JIC.tex
To establish partially-premixed inflow boundary conditions suitable for the micromix combustor, a precursor simulation was conducted using opposed hydrogen injectors in an air crossflow. A schematic for the simulation can be seen at the top of figure \ref{fig:dns_schematic}. Two round injectors of diameter 0.5\,mm were set up 2\,mm downstream from the crossflow inflow face in a cubic domain with 7\,mm sides. The mass flow rate of hydrogen was determined from the global equivalence ratio specified for the combustor, using a crossflow velocity of 100\,m/s. A base grid of 256$^3$ was used, with one level of AMR based on adjacent differences in density to cover the injected fuel, and an additional level of AMR based on hydrogen mass fractions to cover the potential core of each JICF. A further precursor simulation of maintained homogeneous isotropic turbulence (following \cite{Aspden2008AnalysisMethods}) was first run to generate the crossflow and injector inflow boundary conditions with a turbulence intensity of approximately 3\%. Hyperbolic tangent functions were employed to smooth the inflow velocities to zero at the crossflow and injector walls. The parameters associated with the non-reacting simulation can be found in table \ref{table:jic}. Note that the injector Reynolds number is high enough that improved mixedness would not be anticipated in the far-field with a higher equivalence ratio.

\begin{table*}
\centering
\begin{tabular}{ c c }
 \hline
 Domain size & 7\,mm $\times$ 7\,mm $\times$ 7\,mm \\ 
 Resolution (base) & 256 $\times$ 256 $\times$ 256 \\
$\Delta x$ (base) &  27.3\,$\mu$m \\
Resolution (effective) & 1024 $\times$ 1024 $\times$ 1024 \\
$\Delta x$ (finest) & 6.84\,$\mu$m \\
Ambient pressure & 24\,atm \\
Crossflow velocity $\bar{u}_{\textsc{c}}$ & 100\,m/s \\
Injector velocity $\bar{u}_{\textsc{i}}$ & 1360\,m/s \\
Crossflow temperature $T_{\textsc{c}}$ & 750\,K \\
Injector temperature $T_{\textsc{i}}$ & 550\,K \\
Crossflow density $\rho_{\textsc{c}}$ & 11.27\,kg/m$^{3}$\\
Injector density $\rho_{\textsc{i}}$ & 1.04\,kg/m$^{3}$\\
Crossflow rms velocity fluctuation $u'_{\textsc{c}}$ & 3\,m/s \\
Injector rms velocity fluctuation $u'_{\textsc{i}}$ & 40\,m/s \\
Momentum flux ratio $J = \rho_{\textsc{i}}u_{\textsc{i}}^{2} / \rho_{\textsc{c}} u_{\textsc{c}}^{2}$ & 17.2\\
Injector Reynolds number $\Re_{\textsc{i}} = u_{\textsc{i}}d_{\textsc{i}}/\nu_{\textsc{i}}$ & 48021\\
\hline
\end{tabular}
\caption{\label{table:jic}Parameters for the non-reacting JICF simulation}
\end{table*}

To construct the partially-premixed turbulent inflow for the combustor simulation, planes of velocities, species and temperatures were sampled from the JICF simulation 3\,mm downstream from hydrogen injectors; the sampling frequency was based on assuming an inflow velocity of approximately 100\,m/s and computational cell size of the reacting combustor simulation. The distance of the sampling plane is just beyond the near-field criteria devised by \cite{Smith1998MixingCrossflow} (in this case is $x = 2.58$\,mm). A total of 784 planes were sampled and collated (covering approximately 1.5 jet times $t_{\textsc{j}} = d_{\textsc{j}}/u_{\textsc{j}}$), with the combustor inflow condition cycling periodically through the planes, using quadratic interpolation between planes.

\subfile{../tikz/dns}

%% file: tikz/dns.tex
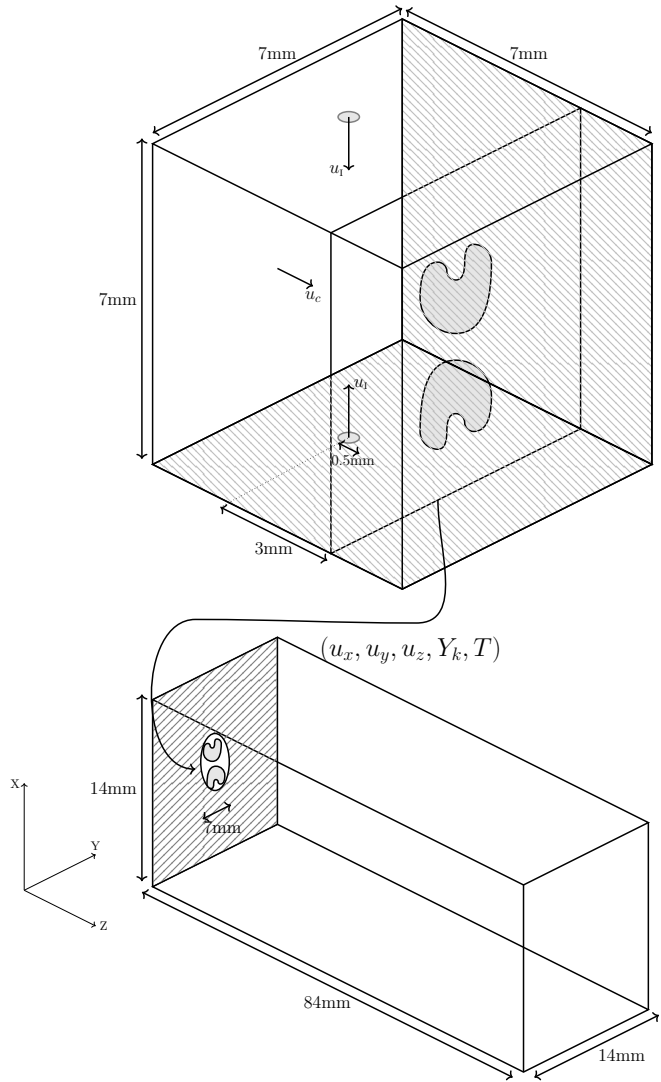
\begin{figure}[ht!]
    \centering
    \resizebox{88mm}{!}{
\begin{tikzpicture}[x={(1cm,-0.5cm)},y={(1cm,0.5cm)},z={(0cm,1.5cm)}]

\draw[very thick,fill=white] (-0.2, -1.2, -0.7) -- (-0.2, -1.2, 5.3) -- (-0.2, 5.8, 5.3) -- (-0.2, 5.8, -0.7) -- cycle;
\draw[very thick, fill=gray!20] (-0.2, 2.3, 1.75) to[out=180, in=90] (-0.2, 1.3, 0.75) to[out=270, in=180] (-0.2, 1.8, 0.25) to[out=0, in=270] (-0.2, 2.05, 0.5) to[out=90, in=180] (-0.2, 2.3, 0.75) to[out=0, in=90] (-0.2, 2.55, 0.5) to[out=270, in=180] (-0.2, 2.8, 0.25) to[out=0, in=270] (-0.2, 3.3, 0.75) to[out=90, in=0] cycle;

\draw[very thick, fill=gray!20] (-0.2, 2.3, 2.25) to[out=180, in=270] (-0.2, 1.3, 3.25) to[out=90, in=180] (-0.2, 1.8, 3.75) to[out=0, in=90] (-0.2, 2.05, 3.5) to[out=270, in=180] (-0.2, 2.3, 3.25) to[out=0, in=270] (-0.2, 2.55, 3.5) to[out=90, in=180] (-0.2, 2.8, 3.75) to[out=0, in=90] (-0.2, 3.3, 3.25) to[out=270, in=0] cycle;

\filldraw[pattern=north west lines, pattern color=gray!50, line width=0.4pt, very thick] (-5.2, -1.2, -0.7) -- (1.8, -1.2, -0.7) -- (1.8, 5.8, -0.7) -- (-5.2, 5.8, -0.7) -- cycle;
\filldraw[pattern=north west lines, pattern color=gray!50, line width=0.4pt, very thick] (-5.2, 5.8, -0.7) -- (1.8, 5.8, -0.7) -- (1.8, 5.8, 5.3) -- (-5.2, 5.8, 5.3) -- cycle;

\draw[very thick] (-5.2, -1.2, -0.7) -- (1.8, -1.2, -0.7) -- (1.8, 5.8, -0.7) -- (-5.2, 5.8, -0.7) -- cycle;
\draw[very thick] (-5.2, -1.2, -0.7) -- (-5.2, -1.2, 5.3);
\draw[very thick] (1.8, -1.2, -0.7) -- (1.8, -1.2, 5.3);
\draw[very thick] (1.8, 5.8, -0.7) -- (1.8, 5.8, 5.3);
\draw[very thick] (-5.2, 5.8, -0.7) -- (-5.2, 5.8, 5.3);
\draw[very thick] (-5.2, -1.2, 5.3) -- (1.8, -1.2, 5.3) -- (1.8, 5.8, 5.3) -- (-5.2, 5.8, 5.3) -- cycle;

\draw[<->,very thick] (-5.5, -1.2, -0.7) -- (-5.5, -1.2, 5.3) node[midway, left] {\Large 7mm};
\draw[<->,very thick] (-5.2, -1.2, 5.5) -- (-5.2, 5.8, 5.5) node[midway, above=2mm] {\Large 7mm};
\draw[<->,very thick] (-5.1, 5.8, 5.5) -- (1.8, 5.8, 5.5) node[midway,above=2mm] {\Large 7mm};

\draw[very thick, gray, fill=gray!20] (-3.2,2.3,5.3) ellipse (0.3cm and 0.15cm);
\draw[->, very thick] (-3.2,2.3,5.3) -- (-3.2, 2.3, 4.3) node[left] {\Large $u_{\textsc{i}}$};
\draw[very thick, gray, fill=gray!20] (-3.2,2.3,-0.7) ellipse (0.3cm and 0.15cm);
\draw[->, very thick] (-3.2,2.3,-0.7) -- (-3.2,2.3,0.3) node[right] {\Large $u_{\textsc{i}}$};
\draw[<->,very thick] (-3.5,2.3,-0.9) -- (-2.9,2.3,-0.9) node[pos=0.7, below] {\large 0.5mm};

\draw[->, very thick] (-5.2, 2.3, 1.8) -- (-4.2, 2.3, 1.8) node[below] {\Large $u_{c}$};
\draw[<->,very thick] (-3.2, -1.3, -0.8) -- (-0.2, -1.3, -0.8) node[midway, below=1mm] {\Large 3mm};
\draw[dotted] (-3.2,-1.3,-0.8) -- (-3.2,2.3,-0.7);

\draw[very thick] (-5.2, -1.2, -8.6) -- (5.2, -1.2, -8.6) -- (5.2, 2.3, -8.6) -- (-5.2, 2.3, -8.6) -- cycle;
\draw[very thick] (-5.2, -1.2, -8.6) -- (-5.2, -1.2, -5.1);
\draw[very thick] (5.2, -1.2, -8.6) -- (5.2, -1.2, -5.1);
\draw[very thick] (5.2, 2.3, -8.6) -- (5.2, 2.3, -5.1);
\draw[very thick] (-5.2, 2.3, -8.6) -- (-5.2, 2.3, -5.1);
\draw[very thick] (-5.2, -1.2, -5.1) -- (5.2, -1.2, -5.1) -- (5.2, 2.3, -5.1) -- (-5.2, 2.3, -5.1) -- cycle;

\draw[<->,very thick] (-5.5, -1.2, -8.6) -- (-5.5, -1.2, -5.1) node[midway, left] {\Large 14mm};
\draw[<->,very thick] (5.5, -1.2, -8.6) -- (5.5, 2.3, -8.6) node[pos=0.7, below=3mm] {\Large 14mm};
\draw[<->,very thick] (-5.2, -1.5, -8.6) -- (5.2, -1.5, -8.6) node[midway, below=2mm] {\Large 84mm};

\draw[pattern=north east lines, pattern color=gray] (-5.2, -1.2, -8.6) -- (-5.2, 2.3, -8.6) -- (-5.2, 2.3, -5.1) -- (-5.2, -1.2, -5.1) -- cycle;

\draw[very thick,fill=white] (-5.2, 0.55, -6.85) ellipse (0.4cm and 0.8cm);

\node[below left] at (1, 2.5, -3) {\huge $(u_{x},u_{y},u_{z},Y_{k},T)$};

\draw[<->,very thick] (-5.2, 0.2, -7.8) -- (-5.2, 1, -7.8) node[pos=0.7, below=2mm] {\Large 7mm};
\begin{scope}[scale=0.25,shift={(-23.8, 0.8, -29.5)}]
    \draw[very thick, fill=gray!20] (3, 1.5, 1.75) to[out=180, in=90] (3, 0.5, 0.75) to[out=270, in=180] (3, 1, 0.25) to[out=0, in=270] (3, 1.25, 0.5) to[out=90, in=180] (3, 1.5, 0.75) to[out=0, in=90] (3, 1.75, 0.5) to[out=270, in=180] (3, 2, 0.25) to[out=0, in=270] (3, 2.5, 0.75) to[out=90, in=0] cycle;
\end{scope}

\begin{scope}[scale=0.25,shift={(-23.8, 0.4, -29.5)}]
    \draw[very thick, fill=gray!20] (3, 1.5, 2.25) to[out=180, in=270] (3, 0.5, 3.25) to[out=90, in=180] (3, 1, 3.75) to[out=0, in=90] (3, 1.25, 3.5) to[out=270, in=180] (3, 1.5, 3.25) to[out=0, in=270] (3, 1.75, 3.5) to[out=90, in=180] (3, 2, 3.75) to[out=0, in=90] (3, 2.5, 3.25) to[out=270, in=0] cycle;
\end{scope}

\draw[->,very thick] (-0.2, 1.8, -0.7) to[out=270, in=0] (-0.5, 1.5, -3) to[out=180, in=0] (-5.2, 0, -4) to[out=180, in=180] (-5.2, 0, -6.8);
\draw[->] (-6,-4,-8) -- (-4,-4,-8) node[right] {Z};
\draw[->] (-6,-4,-8) -- (-6,-2,-8) node[above] {Y};
\draw[->] (-6,-4,-8) -- (-6,-4,-6) node[left] {X};

\end{tikzpicture}
}
\caption{Schematic of the JICF precursor simulation (top) and the reacting combustor simulation (bottom).}
\label{fig:dns_schematic}

\end{figure}

%% file: configuration/DNS.tex
The combustor domain was a 14\,mm $\times$ 14\,mm $\times$ 84\,mm cuboid, with a 7\,mm diameter circular inflow imposed on the bottom ($z$ low) face using the pre-generated inflow described above. Outside of the circular inflow, a no-slip wall condition was imposed on the remainder of the bottom face. The domain is centred with (0,0,0)\,mm at the centre of the circular inflow. Periodic boundary conditions were used laterally to mimic a Cartesian grid of inlet ports. An outflow condition was employed at the top boundary. A schematic is shown at the bottom of figure \ref{fig:dns_schematic}. A first level of AMR was defined to cover the entire jet, as defined by large values of $u_{z}$, $Y_{\Htwo}$ and magnitude of vorticity $|\boldsymbol{\omega}|$, leading to a resolution of 27.3\,$\mu$m inside the jet. Two further levels of AMR were added based on the intermediate species $Y_{\HOtwo}$ to cover the flame, leading to a resolution in the most reactive parts of the flame of 6.4\,$\mu$m. The base grid is referred to as level 0, increasing up to ``maximum level 3'' (ML3), giving an effective resolution in excess of 50 billion computational cells. The ML3 simulations were run on ARCHER2, the UK national facility at EPCC. Given the varied equivalence ratio of the inflow, there is no single thermal thickness that could be chosen before simulation, however 1D unstretched values of thermal thickness obtained from Cantera \cite{Goodwin2018Cantera:Processes} at the mean conditions of the inflow suggest an approximate thickness of 17.4\,$\mu$m. 
A further simulation was conducted with an additional level of AMR (i.e.\ ML4) at the flame to ensure statistical invariance. This resulted in a simulation with an effective resolution of over 400 billion cells, which was run on Polaris at the Argonne Leadership Computing Facility (ALCF). An assessment of resolution is provided in \ref{appendix:resolution} for these particular conditions. 

The following strategy was used to establish the statistically-steady flame. The simulation was first run at the base grid (i.e.\ without AMR) for approximately 20 jet times.  Ignition was induced synthetically by depositing heat in four locations at $(x,y,z)=(\pm3,\pm3,30)$\,mm for 3\,$\mu$s over a radius of 400\,$\mu$m, which was sufficient to establish sustained reactions.  The simulation was then run for a further 95 jet times at the base grid, over which time the flame propagated upstream to a statistically-stationary location, and the temperature of the recirculation zone settled to a statistically-stationary value.  The AMR was then enabled, and the simulation was run for another 8 jet times before any statistics were taken; the subsequent statistical averaging period was 4 jet times, which corresponds to approximately 25 flame times (based on the mean reactant conditions).  Animations of the statistically-steady period used for analysis are included as supplementary material, which provide valuable context and insight into the flame structure and stabilisation discussed in the following results sections.

A summary of the parameters for the combustor is given in table \ref{table:combustor}, where the integral length scale has been calculated as the mean integral of the auto-correlation function for the velocities on each 2D inflow plane. Effective inflow viscosities and Kolmogorov length scales were calculated following \cite{Aspden2008AnalysisMethods}. 

\begin{table*}[ht]
\centering
\begin{tabular}{ c c }
 \hline
 Domain size & 14\,mm $\times$ 14\,mm $\times$ 84\,mm \\ 
 Resolution (base) & 256 $\times$ 256 $\times$ 1536 \\
$\Delta x$ (base) &  54.7\,$\mu$m \\
Resolution (effective) & 2048 $\times$ 2048 $\times$ 12288 \\
$\Delta x$ (finest, at flame) & 6.84\,$\mu$m \\
$\Delta x$ (inflow) & 27.3\,$\mu$m \\
Jet diameter $d_{\textsc{j}}$ & 7\,mm \\
Mean inflow velocity $\bar{u}_{\textsc{j}}$ & 111\,m/s \\
Inflow r.m.s. velocity fluctuation $u'$ & 26.1\,m/s \\
Inflow integral length scale $\ell$ & 1.0\,mm\\
Inflow turbulent energy dissipation rate $\epsilon = u'^{3}/\ell$ & $1.79 \times 10^{7}$\,m$^{2}$/s$^{3}$ \\
Mean inflow viscosity $\bar{\nu}$ & $3.33 \times 10^{-5}$ m$^{2}$/s \\
Inflow Kolmogorov length scale $\eta = (\bar{\nu}^{3}/\epsilon)^{1/4}$ & 6.77\,$\mu$m\\
Effective mean inflow viscosity $\bar{\nu}_{e}$ &  $5.90 \times 10^{-5}$\,m$^{2}$/s\\
Effective inflow Kolmogorov length scale $\eta_{e} =(\bar{\nu}_{e}^{3} / \epsilon)^{1/4}$ &  10.4\,$\mu$m\\
Jet Reynolds number Re$_{\textsc{j}} = \bar{u}_{\textsc{j}}d_{\textsc{j}}/\bar{\nu}$ & 23200\\
Inflow turbulent Reynolds number Re$_{\textsc{t}} = u'\ell/\bar{\nu}$ & 781\\
Mean inflow equivalence ratio $\phi_{\textsc{j}}$ & 0.46\\
Mean inflow temperature $T_{\textsc{j}}$ & 732\,K\\
Mean thermal thickness $\ell_{\textsc{l}}(\phi_{\textsc{j}},T_{\textsc{j}},p)$ & 17.4\,$\mu$m \\
Flame resolution $\ell_{\textsc{l}}/\Delta x$ & 2.54 \\
 \hline
\end{tabular}
\caption{\label{table:combustor}Parameters for the combusting simulation.}
\end{table*}

%% file: structure/preamble.tex
This section considers the general structure of the flow, first in a temporally-averaged sense
in \ref{sec:structure:steady}, and then through the evolution of instantaneous snapshots
in \ref{sec:structure:instant}.

%% file: structure/global.tex
Temporal averaging was performed according to 
\begin{equation}
    \left<q\right>(x,y,z) = \frac{1}{N_{t}}\sum_{t=t_{s}}^{t_{f}}q(x,y,z,t),
\end{equation}
where $N_{t}$ is the number of time points averaged over the period $[t_s, t_f]$, and the corresponding Favre average is defined as $\left<q\right>_{\textsc{f}}(x,y,z) = \left< \rho q \right>/\left<\rho\right>$.

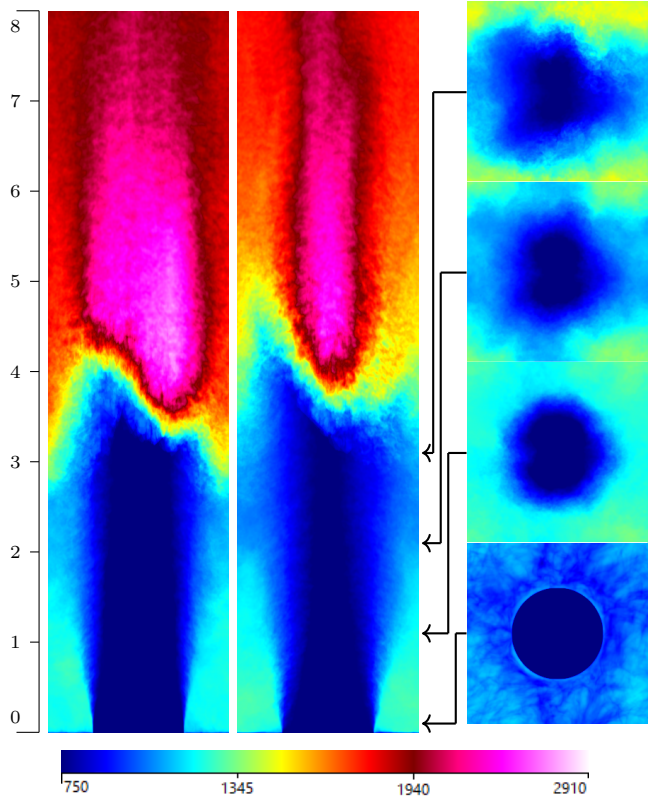
\begin{figure}[ht]
\centering
    \centering
    \begin{tikzpicture}
    \draw (0, 0) -- (0, 96mm);
    
    \draw (0mm,0mm) -- (-3.0mm, 0mm);
    \node [anchor = south east] at (-1.2mm,0mm) {\scriptsize 0};
    \draw (0mm,96mm) -- (-3.0mm, 96mm);
    \node [anchor = north east] at (-1.2mm,96mm) {\scriptsize 8};
    \foreach \y in {1,...,7} {
        \draw (-1.2mm, \y*12 mm) -- (0mm, \y*12 mm) ;
        \node[anchor=east] at (-1.2mm, \y*12 mm) {\scriptsize \y};
    }
    \end{tikzpicture}
    \includegraphics[width=24mm]{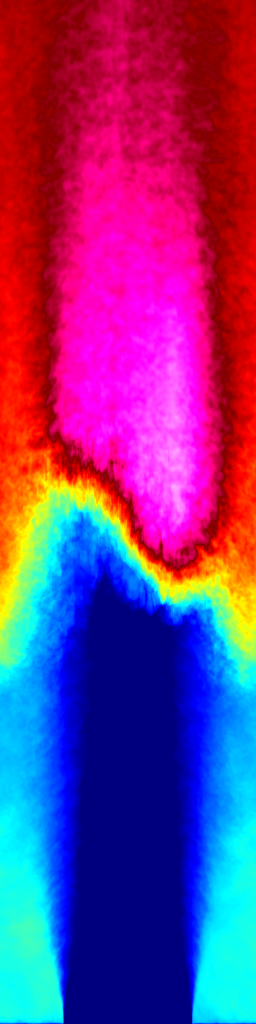}
    \includegraphics[width=24mm]{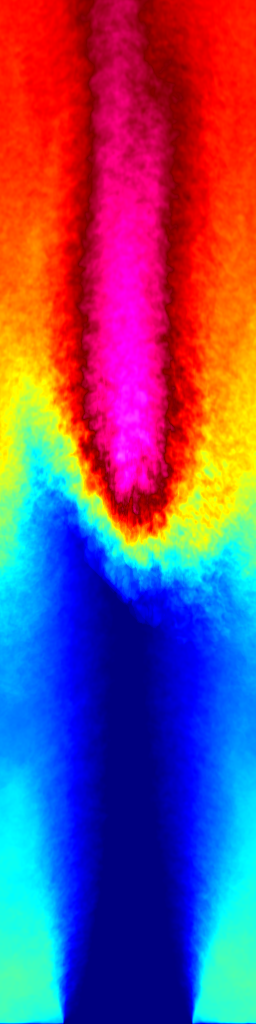}
    \hspace{-2mm}
     \begin{tikzpicture}    
    \node[inner sep=0pt] (image) at (0,0) {
    \includegraphics[width=24mm]{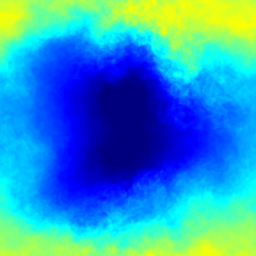}};
    \draw[thick] (image.west) -- (-16.5mm,0mm);
    \draw[thick] (-16.5mm,0mm) -- (-16.5mm,-48mm);
    \draw[thick,->] (-16.5mm,-48mm) -- (-18mm,-48mm);
    \draw[thick,white] (-12mm,-12mm) -- (12mm,-12mm);
    
    \node[inner sep=0pt] (image) at (0,-24mm) {
    \includegraphics[width=24mm]{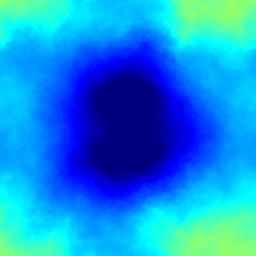}};
    \draw[thick] (image.west) -- (-15.5mm,-24mm);
    \draw[thick] (-15.5mm,-24mm) -- (-15.5mm,-60mm);
    \draw[thick,->] (-15.5mm,-60mm) -- (-18mm,-60mm);
    \draw[thick,white] (-12mm,-36mm) -- (12mm,-36mm);
    
    \node[inner sep=0pt] (image) at (0,-48mm) {
    \includegraphics[width=24mm]{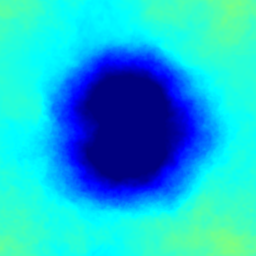}};
    \draw[thick] (image.west) -- (-14.5mm,-48mm);
    \draw[thick] (-14.5mm,-48mm) -- (-14.5mm,-72mm);
    \draw[thick,->] (-14.5mm,-72mm) -- (-18mm,-72mm);
    \draw[thick,white] (-12mm,-60mm) -- (12mm,-60mm);
    
    \node[inner sep=0pt] (image) at (0,-72mm) {
    \includegraphics[width=24mm]{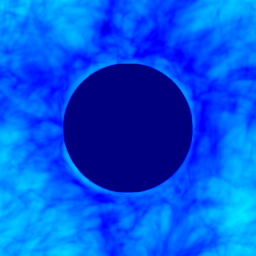}};   
    \draw[thick] (image.west) -- (-13.5mm,-72mm);
    \draw[thick] (-13.5mm,-72mm) -- (-13.5mm,-84mm);
    \draw[thick,->] (-13.5mm,-84mm) -- (-18mm,-84mm);
    \end{tikzpicture}\\
    \vspace{2mm}
    \includegraphics[width=70mm]{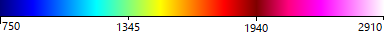}
\caption{$\left<T\right>$ slices through L: $x/d_{\textsc{j}}=0$ and $y/d_{\textsc{j}}=0$ and R: $z/d_{\textsc{j}}=0,1,2,3$. The scale on the left is streamwise distance in jet diameters. At heights below the flame stabilisation height, outside the jet, the temperature of the fluid is consistently around 1200\,K, with pockets of even hotter fluid near the wall closer to 1400\,K.}
\label{fig:meanSlicesTemp}
\end{figure}

\begin{figure}[ht]
\centering
    \begin{tikzpicture}
    \draw (0, 0) -- (0, 96mm);
    
    \draw (0mm,0mm) -- (-3.0mm, 0mm);
    \node [anchor = south east] at (-1.2mm,0mm) {\scriptsize 0};
    \draw (0mm,96mm) -- (-3.0mm, 96mm);
    \node [anchor = north east] at (-1.2mm,96mm) {\scriptsize 8};
    \foreach \y in {1,...,7} {
        \draw (-1.2mm, \y*12 mm) -- (0mm, \y*12 mm) ;
        \node[anchor=east] at (-1.2mm, \y*12 mm) {\scriptsize \y};
    }
    \end{tikzpicture}
    \includegraphics[width=24mm]{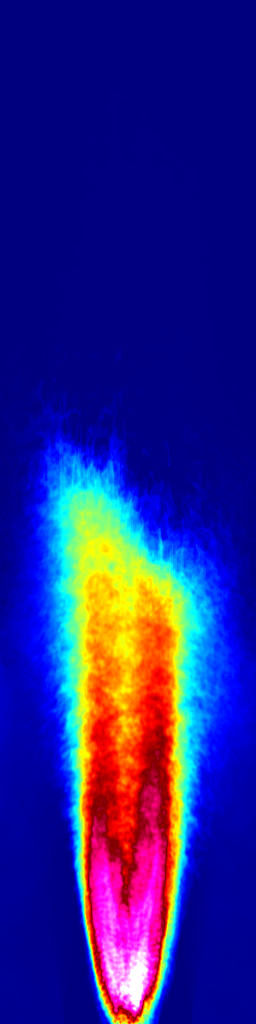}
    \includegraphics[width=24mm]{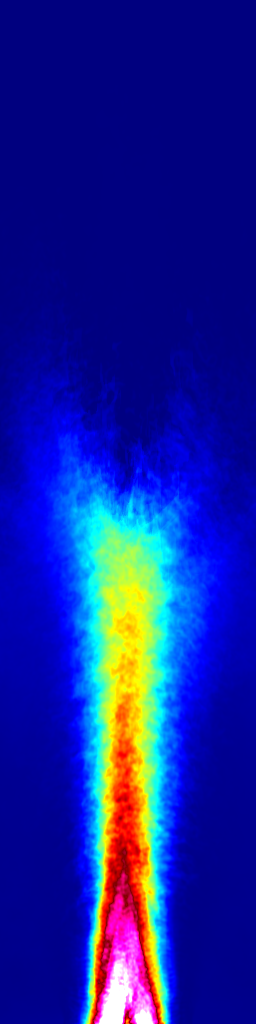}
    \hspace{-2mm}
    \begin{tikzpicture}
    \node[inner sep=0pt] (image) at (0,0) {
    \includegraphics[width=24mm]{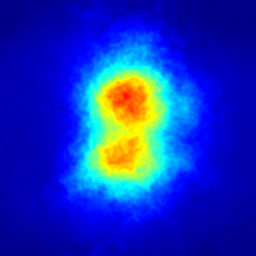}};
    \draw[thick] (image.west) -- (-16.5mm,0mm);
    \draw[thick] (-16.5mm,0mm) -- (-16.5mm,-48mm);
    \draw[thick,->] (-16.5mm,-48mm) -- (-18mm,-48mm);
    \draw[thick,white] (-12mm,-12mm) -- (12mm,-12mm);
    
    \node[inner sep=0pt] (image) at (0,-24mm) {
    \includegraphics[width=24mm]{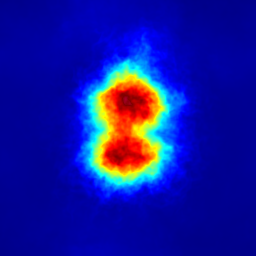}};
    \draw[thick] (image.west) -- (-15.5mm,-24mm);
    \draw[thick] (-15.5mm,-24mm) -- (-15.5mm,-60mm);
    \draw[thick,->] (-15.5mm,-60mm) -- (-18mm,-60mm);
    \draw[thick,white] (-12mm,-36mm) -- (12mm,-36mm);
    
    \node[inner sep=0pt] (image) at (0,-48mm) {
    \includegraphics[width=24mm]{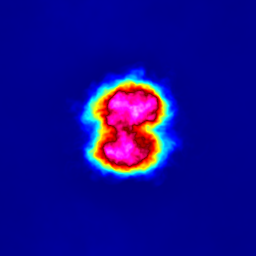}};
    \draw[thick] (image.west) -- (-14.5mm,-48mm);
    \draw[thick] (-14.5mm,-48mm) -- (-14.5mm,-72mm);
    \draw[thick,->] (-14.5mm,-72mm) -- (-18mm,-72mm);
    \draw[thick,white] (-12mm,-60mm) -- (12mm,-60mm);
    
    \node[inner sep=0pt] (image) at (0,-72mm) {
    \includegraphics[width=24mm]{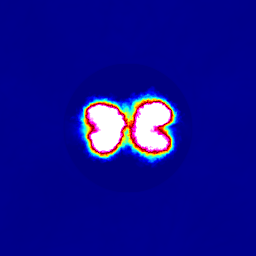}};
    \draw[thick] (image.west) -- (-13.5mm,-72mm);
    \draw[thick] (-13.5mm,-72mm) -- (-13.5mm,-84mm);
    \draw[thick,->] (-13.5mm,-84mm) -- (-18mm,-84mm);   
    \draw[draw=black] (image.south) ++(0mm,12mm) circle (6.5mm);
    \draw[draw=white, dashed] (image.south) ++(0mm,12mm) circle (6.5mm);
    \end{tikzpicture}\\
    \vspace{2mm}
    \includegraphics[width=70mm]{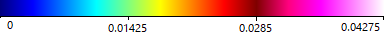}
\caption{$\left<Y_{\Htwo}\right>_{\textsc{f}}$ slices through L: $x/d_{\textsc{j}}=0$ and $y/d_{\textsc{j}}=0$ and R: $z/d_{\textsc{j}}=0,1,2,3$. The black and white circle on the $z/d_{\textsc{j}}=0$ image illustrates the airport. Notice the approximate kidney shape taken by the hydrogen entering the domain due to the upstream jet in crossflow, which interact and rotate into circular profiles in the opposite axis, leading to the asymmetry in the flame which can be seen in the temperature profile in figure \ref{fig:meanSlicesTemp}.}
\label{fig:meanSlicesSpecies}
\end{figure}
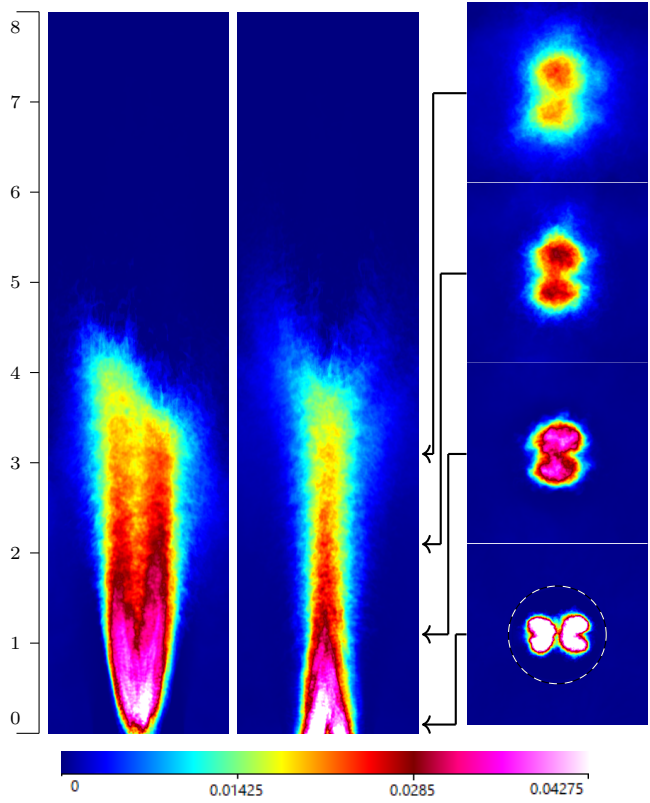

\begin{figure}[ht]
\centering
    \begin{tikzpicture}
    \draw (0, 0) -- (0, 96mm);
    
    \draw (0mm,0mm) -- (-3.0mm, 0mm);
    \node [anchor = south east] at (-1.2mm,0mm) {\scriptsize 0};
    \draw (0mm,96mm) -- (-3.0mm, 96mm);
    \node [anchor = north east] at (-1.2mm,96mm) {\scriptsize 8};
    \foreach \y in {1,...,7} {
        \draw (-1.2mm, \y*12 mm) -- (0mm, \y*12 mm) ;
        \node[anchor=east] at (-1.2mm, \y*12 mm) {\scriptsize \y};
    }
    \end{tikzpicture}
    \includegraphics[width=24mm]{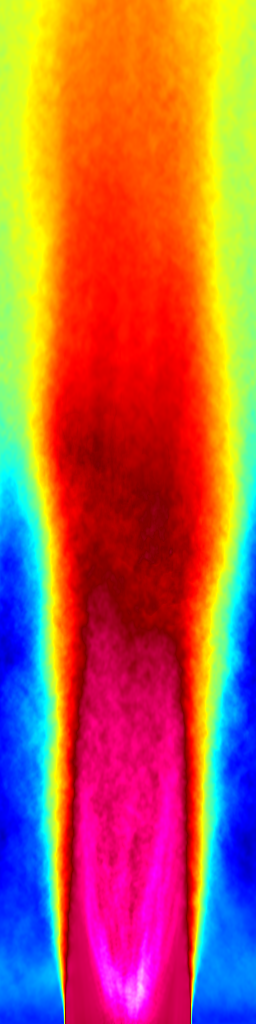}
    \includegraphics[width=24mm]{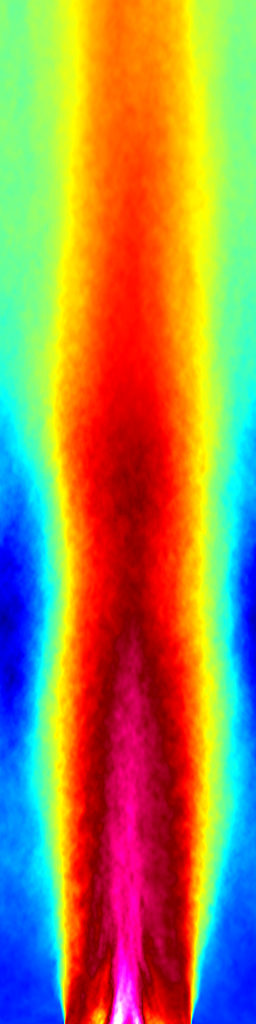}
    \hspace{-2mm}
    \begin{tikzpicture}
    \node[inner sep=0pt] (image) at (0,0) {
    \includegraphics[width=24mm]{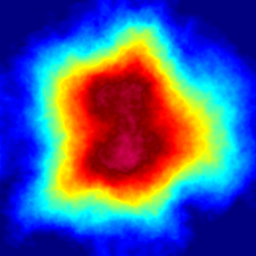}};
    \draw[thick] (image.west) -- (-16.5mm,0mm);
    \draw[thick] (-16.5mm,0mm) -- (-16.5mm,-48mm);
    \draw[thick,->] (-16.5mm,-48mm) -- (-18mm,-48mm);
    \draw[thick,white] (-12mm,-12mm) -- (12mm,-12mm);
    
    \node[inner sep=0pt] (image) at (0,-24mm) {
    \includegraphics[width=24mm]{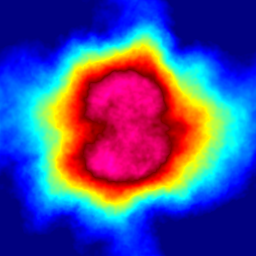}};
    \draw[thick] (image.west) -- (-15.5mm,-24mm);
    \draw[thick] (-15.5mm,-24mm) -- (-15.5mm,-60mm);
    \draw[thick,->] (-15.5mm,-60mm) -- (-18mm,-60mm);
    \draw[thick,white] (-12mm,-36mm) -- (12mm,-36mm);
    
    \node[inner sep=0pt] (image) at (0,-48mm) {
    \includegraphics[width=24mm]{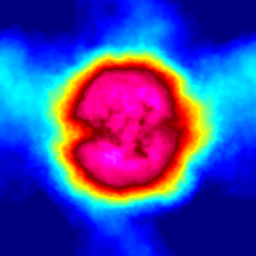}};
    \draw[thick] (image.west) -- (-14.5mm,-48mm);
    \draw[thick] (-14.5mm,-48mm) -- (-14.5mm,-72mm);
    \draw[thick,->] (-14.5mm,-72mm) -- (-18mm,-72mm);
    \draw[thick,white] (-12mm,-60mm) -- (12mm,-60mm);
    
    \node[inner sep=0pt] (image) at (0,-72mm) {
    \includegraphics[width=24mm]{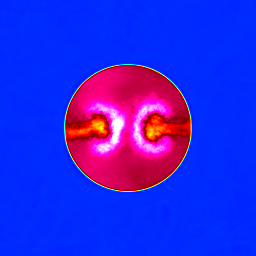}};   
    \draw[thick] (image.west) -- (-13.5mm,-72mm);
    \draw[thick] (-13.5mm,-72mm) -- (-13.5mm,-84mm);
    \draw[thick,->] (-13.5mm,-84mm) -- (-18mm,-84mm);
    
    \end{tikzpicture}\\
    \vspace{2mm}   
    \includegraphics[width=70mm]{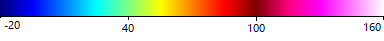}
\caption{$\left<u_{z}\right>_{\textsc{f}}$ slices through L: $x/d_{\textsc{j}}=0$ and $y/d_{\textsc{j}}=0$ and R: $z/d_{\textsc{j}}=0,1,2,3$. Upon entry into the domain, the wake vortices are still clear through the mushroom-shaped profile leading from the wall of the air port.}
\label{fig:meanSlicesZvel}

\end{figure}

To paint a picture of the flame, profiles of average temperature $\left<T\right>$, fuel mass fraction $\left<Y_{\Htwo}\right>_{\textsc{f}}$ and streamwise velocity $\left<u_{z}\right>_{\textsc{f}}$ are shown as two-dimensional slices through the centre of the jet in both the $x$ and $y$ directions (the top third of the domain has been cropped) in figures \ref{fig:meanSlicesTemp}, \ref{fig:meanSlicesSpecies} and \ref{fig:meanSlicesZvel}, respectively, along with horizontal slices at four streamwise locations. Colours are normalised by the adiabatic flame temperature of the mean inflow condition ($\phi=0.46, T_{u}=732$\,K, $p=$24\,atm), the stoichiometric mass fraction of hydrogen in air, and the velocity of the jet, allowing for negative velocities.
The flame base can be seen to be lifted at a height approximately $3.5\,d_{\textsc{j}}$ ($\approx 25\,$mm) from the bottom, with a higher temperature observed in the core of the flame. 
The flame is visibly asymmetric; there is a degree of asymmetry in each panel, which is likely due to the short temporal averaging window, but more importantly, the slices through $x=0$ and $y=0$ are different from each other.
This can also be seen in the slices taken through the fuel mass fraction in figure \ref{fig:meanSlicesSpecies}. This asymmetry is inherently due to the inflow condition; the counter-rotating vortex pair originating from the upstream JICF results in kidney-shaped structures in temperature, species and velocity (see $z=0$ slices in the bottom right of figures \ref{fig:meanSlicesTemp}, \ref{fig:meanSlicesSpecies} and \ref{fig:meanSlicesZvel}). After entering the domain, the two pairs of vortices propagate towards the $x=0$ plane, interact with each other, and then propagate away from the $y=0$ plane. This results in larger mass fractions of fuel and velocities visible in the slices through the $x$ planes, and therefore the asymmetry observed. The interaction of the crossflow with the jet can be seen in the $z=0$ slices in figure \ref{fig:meanSlicesZvel}, where the fluid on the windward side has been accelerated (the white crescent), and surrounds slower-moving fluid on the leeward side of the JICF (the yellow/orange mushroom-like structure).

Further planar averaging results in one-dimensional profiles as a function of streamwise distance, i.e.
\begin{equation}
    \hat{q}(z) = \frac{1}{L_{x}L_{y}}\int_{0}^{L_{y}}\int_{0}^{L_{x}}\left<{q}\right>(x,y,z)\,\textrm{d}x\,\textrm{d}y,
\end{equation}
with the natural Favre average $\check{q}(z) = \widehat{\rho q}/\hat{\rho}$.

\begin{figure}[ht]
    \centering
    \includegraphics[width=88mm]{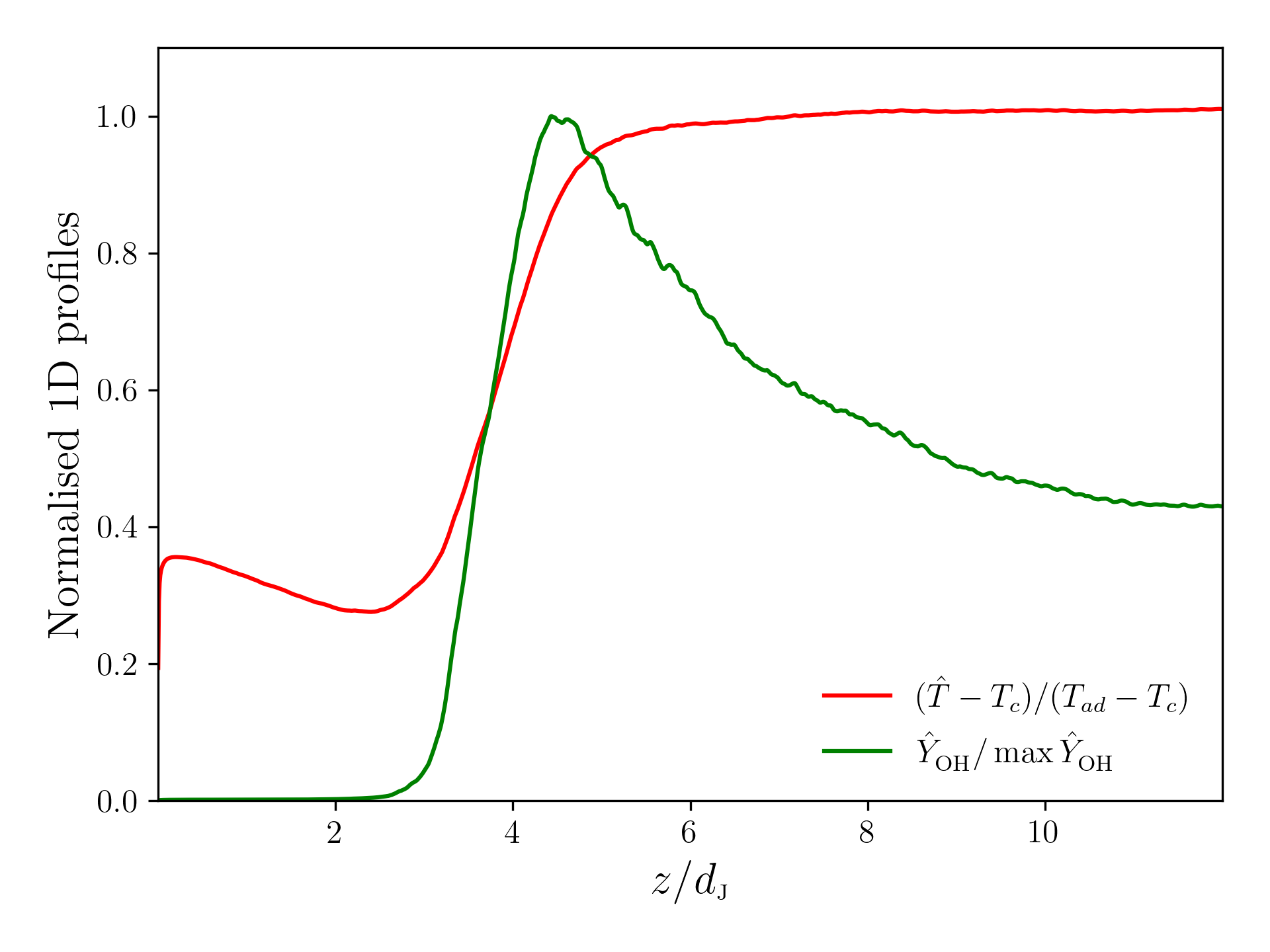}
    \caption{Integrated average profiles of normalised temperature and $Y_{\OH}$. These profiles suggest the flame sits at a height of between 20 and 36\,mm.}
    \label{fig:1dprofiles}
\end{figure}

Normalised profiles of $\hat{T}$ and $\check{Y}_{\OH}$ are given in figure \ref{fig:1dprofiles}. Between the temperature and OH profile, it can be concluded that the majority of the flame sits $3$--$4.5\,d_{\textsc{j}}$ (21--30.5\,mm) above the jet inflow. Defining a flame brush thickness as
\begin{equation}
    L_{\textsc{fb}} = \frac{T_{b}-T_{\textsc{j}}}{\max\{\textrm{d} \hat{T}/\mathrm{d} z\}},
\end{equation}
gives an approximate thickness of $L_{\textsc{fb}} = 1.2$\,cm (1.7\,$d_{\textsc{j}})$. 

\begin{figure}[ht!]
    \centering
    \includegraphics[width=75mm]{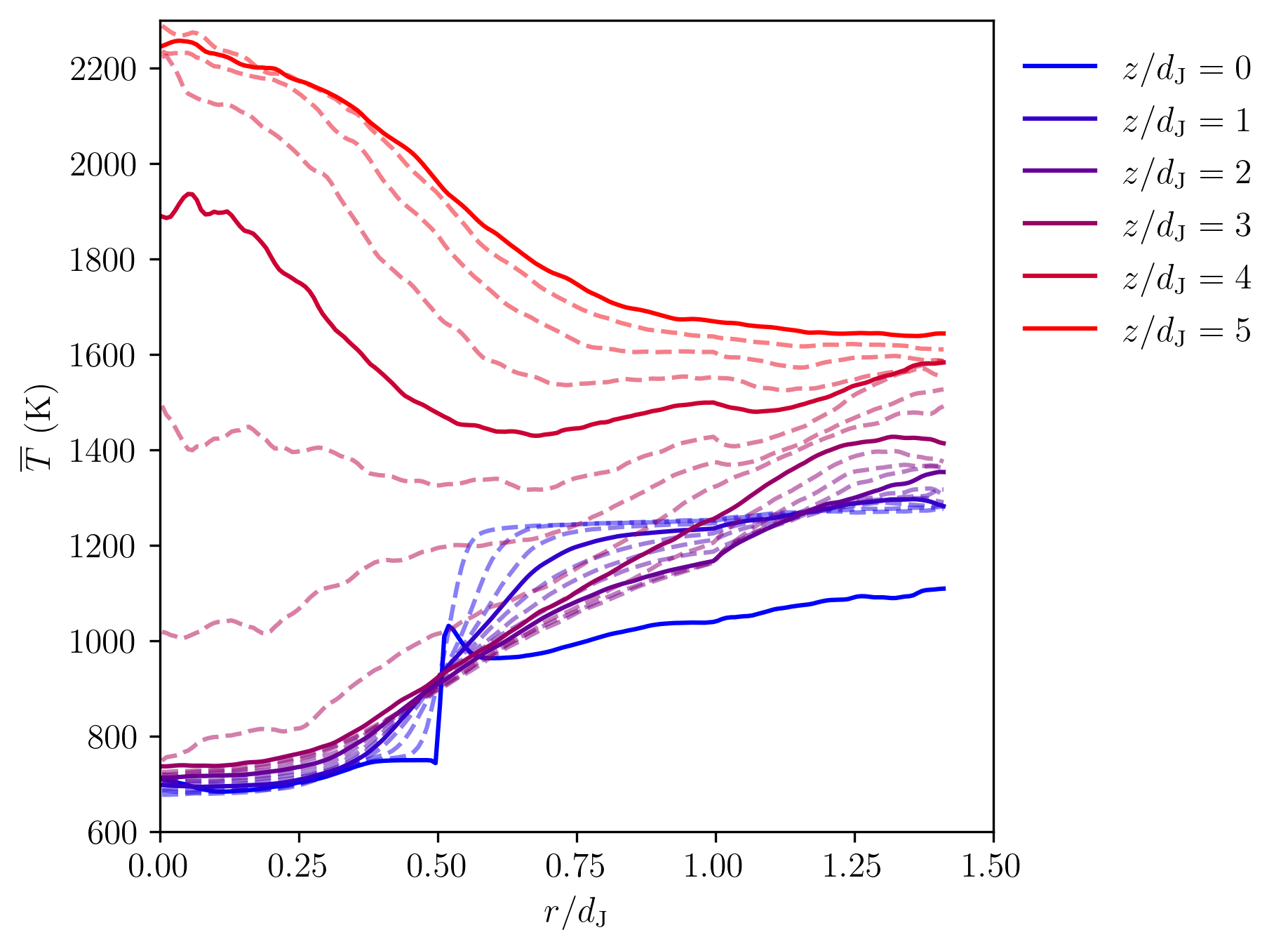}
    
    \includegraphics[width=75mm]{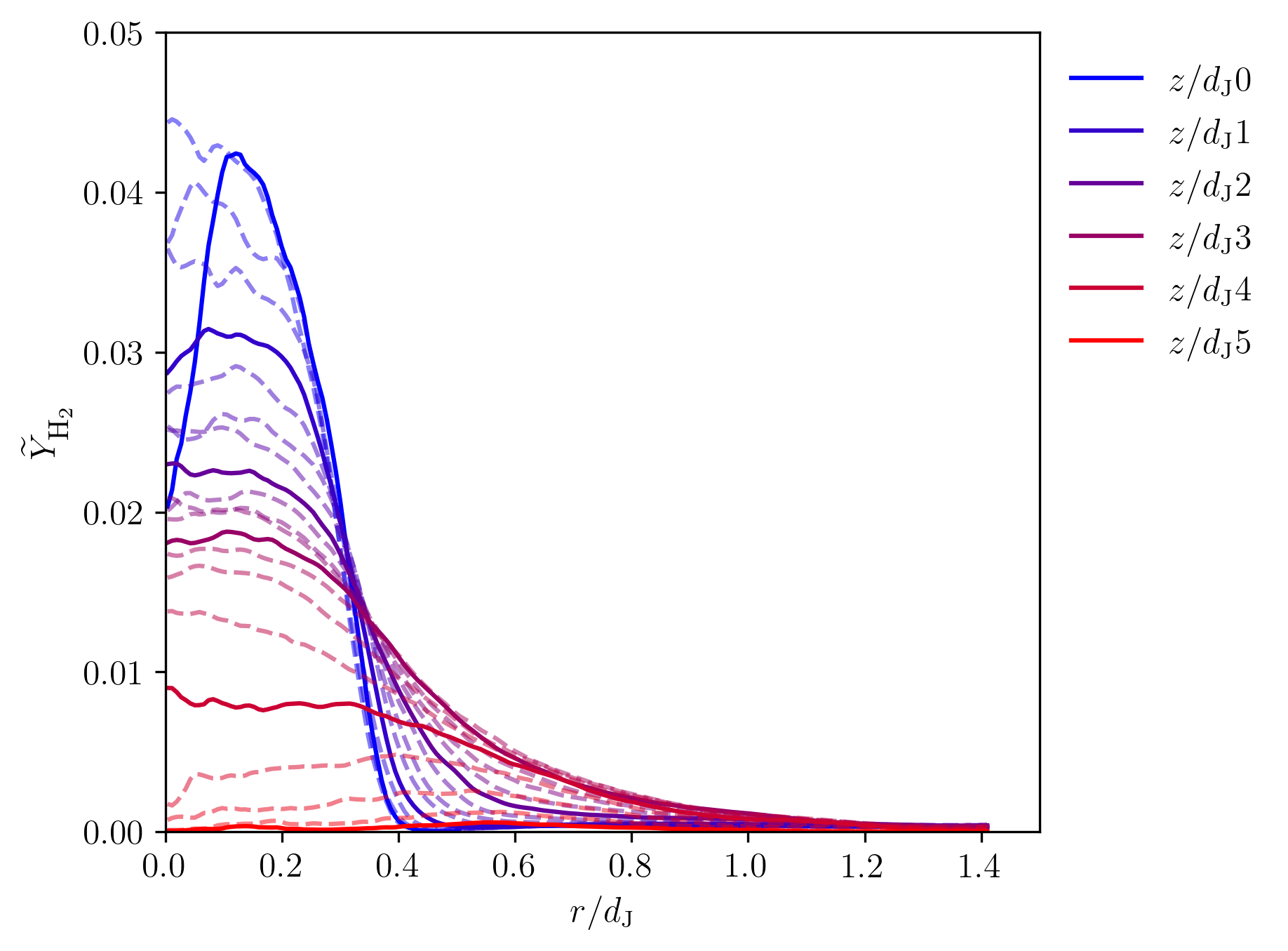}
    
    \includegraphics[width=75mm]{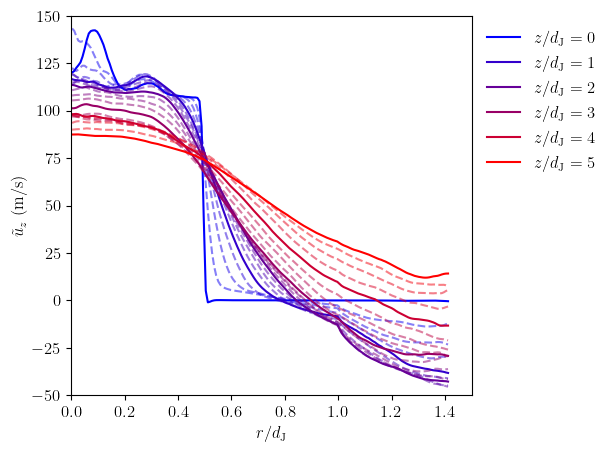}
    \caption{Profiles of $\overline{T}$, $\tilde{Y}_{\Htwo}$ and $\tilde{u}_{z}$ at $z/d_{\textsc{j}} = 0,1,2,3,4,5$. Note the jump in temperature and drop in H$_{2}$ between $z/d_{\textsc{j}}=3$ and $4$ as expected from the flame height, while the combination of jet spreading and thermal expansion cause almost no change in streamwise velocity at these heights within the jet.}
    \label{fig:rz_profiles}
\end{figure}

Despite the jet not being perfectly axisymmetric, for modelling and averaging purposes, azimuthal averaging can also be performed in terms of the radius $r$ and streamwise distance $z$
\begin{equation}
    \overline{q}(r,z) = \begin{cases} \displaystyle\frac{1}{2\pi}\int_{0}^{2\pi} \left<q\right>(r,\theta,z)\,\textrm{d}\theta &\mbox{ for } r \leq L_{x}/2,\\
     \displaystyle \sum_{n=0}^{3}\frac{1}{\beta_{n}-\alpha_{n}}\int_{\alpha_{n}}^{\beta_{n}}\left<q\right>(r,\theta,z)\,\textrm{d}\theta &\mbox{ otherwise.} 
    \end{cases}
\end{equation}
where
\begin{equation}
    \alpha_{n} = \frac{n\pi}{2}+\cos^{-1}\bigg(\frac{L_{x}}{2r}\bigg), \quad \beta_{n} = \frac{(n+1)\pi}{2}-\cos^{-1}\bigg(\frac{L_{x}}{2r}\bigg)
\end{equation}
after converting from Cartesian to cylindrical coordinates, with corresponding Favre average $\tilde{q}(r,z) = \overline{\rho q}/\overline{\rho}$. 
Figure \ref{fig:rz_profiles} shows the profiles of $\overline{T}$, $\tilde{Y}_{\Htwo}$ and $\tilde{u}_{z}$ at streamwise locations from $z=0$ to 5\,$d_{\textsc{j}}$; the black vertical line represents the point beyond which the radius exceeds half the domain width, meaning profiles extend into the corners of the domain.

\begin{figure}[ht!]
    \centering
    \includegraphics[width=70mm]{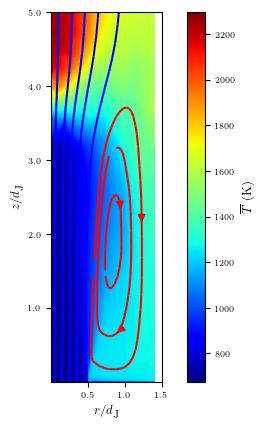}
    \caption{Streamlines of velocity in $r-z$ space, with blue lines originating in the jet inflow, and red lines originating in the recirculation region.}
    \label{fig:streamlines}
\end{figure}

There is a clear recirculation region ($\tilde{u}_{z} < 0$), especially in the corners, which extends down from flame stabilisation height. Crucially, relatively-hot (partially) reacted fluid is transported upstream all the way to the inlet plane. The upstream transport of heat is further illustrated by streamlines in figure \ref{fig:streamlines}, where recirculation can be seen outside of the jet up to approximately $3.6\,d_{\textsc{j}}$ ($\approx 25\,$mm). The background is coloured by temperature, indicating fluid upwards of 1500\,K is recirculated.

%% file: structure/instantaneous.tex
\begin{figure}[ht]
    \centering
    \begin{tikzpicture}
    \draw (0, 0) -- (0, 96mm);
    
    \draw (0mm,0mm) -- (-3.0mm, 0mm);
    \node [anchor = south east] at (-1.2mm,0mm) {\scriptsize 0};
    \draw (0mm,96mm) -- (-3.0mm, 96mm);
    \node [anchor = north east] at (-1.2mm,96mm) {\scriptsize 8};
    \foreach \y in {1,...,7} {
        \draw (-1.2mm, \y*12 mm) -- (0mm, \y*12 mm) ;
        \node[anchor=east] at (-1.2mm, \y*12 mm) {\scriptsize \y};
    }
    \end{tikzpicture}
    \includegraphics[width=24mm]{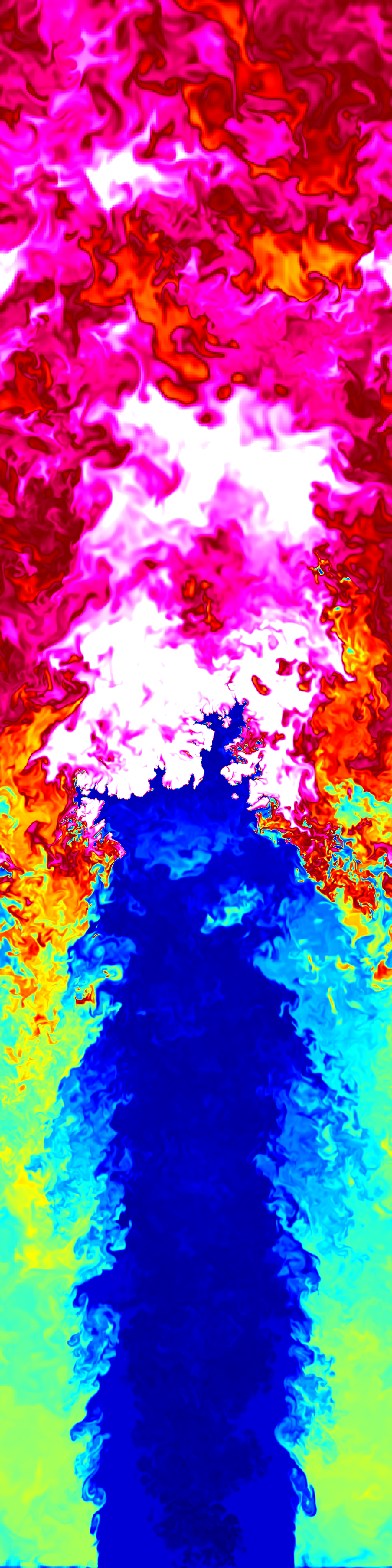}
    \includegraphics[width=24mm]{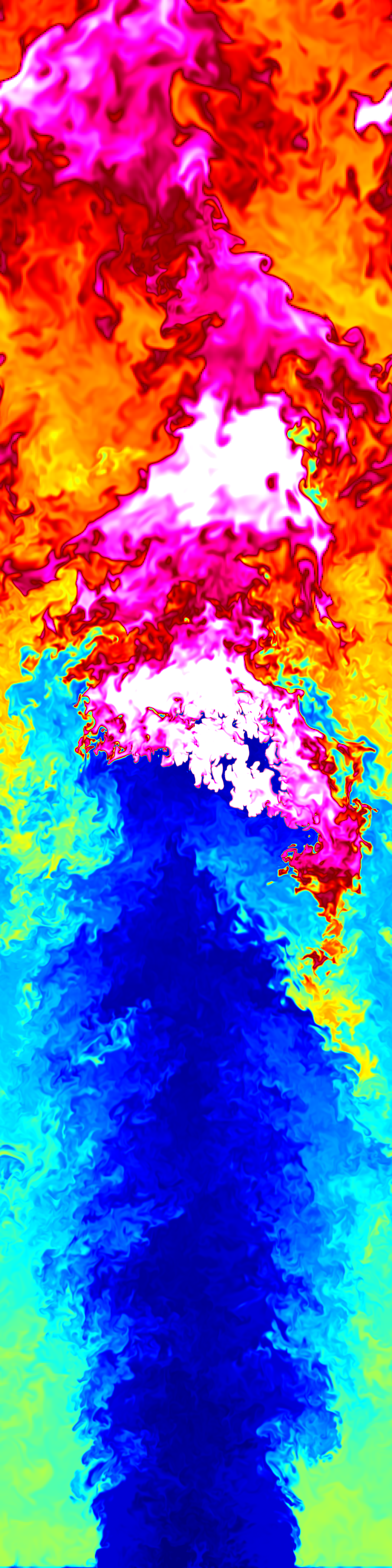}
    \hspace{-2mm}
    \begin{tikzpicture}
    \node[inner sep=0pt] (image) at (0,0) {
    \includegraphics[width=24mm]{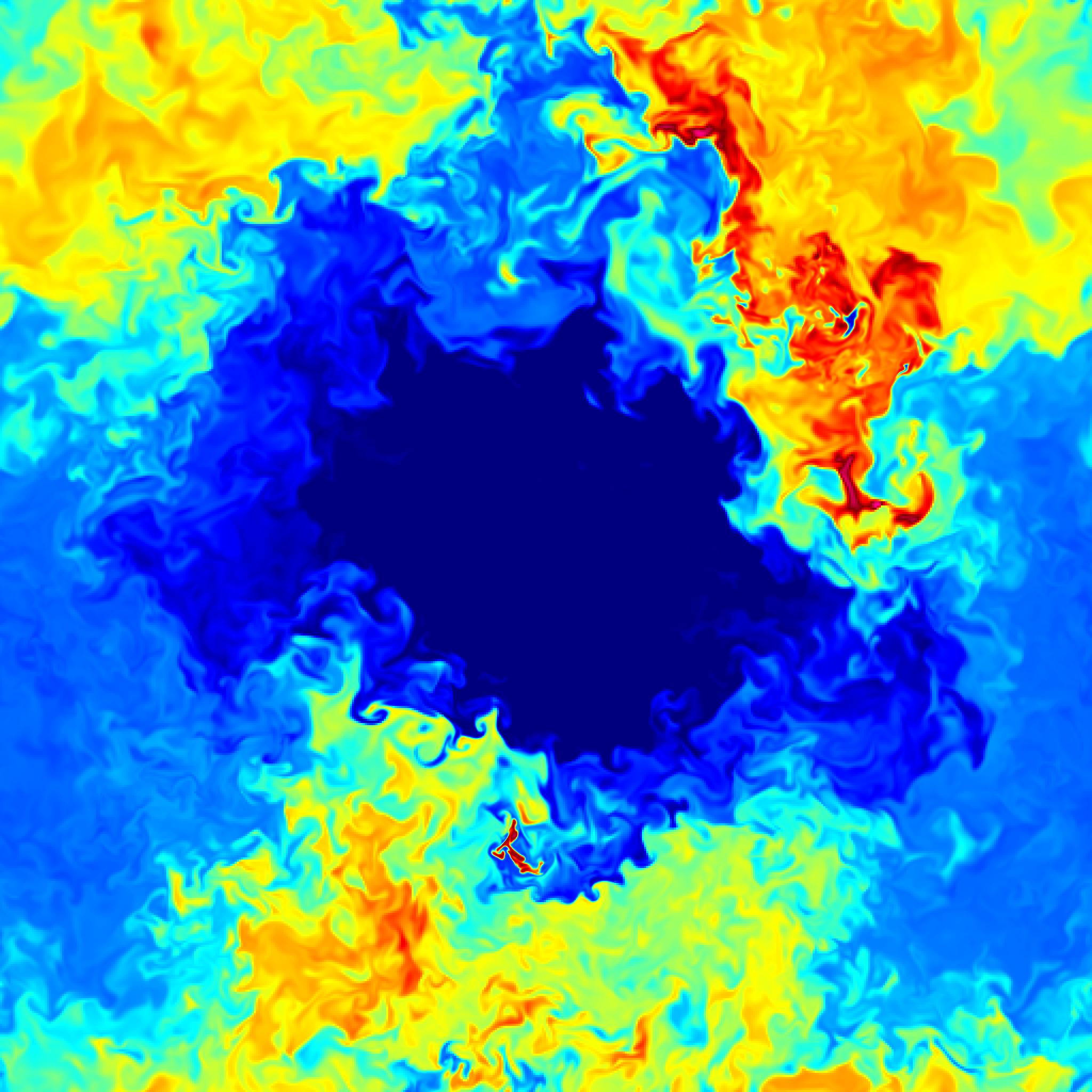}};
    \draw[thick] (image.west) -- (-16.5mm,0mm);
    \draw[thick] (-16.5mm,0mm) -- (-16.5mm,-48mm);
    \draw[thick,->] (-16.5mm,-48mm) -- (-18mm,-48mm);
    \draw[thick,white] (-12mm,-12mm) -- (12mm,-12mm);
    
    \node[inner sep=0pt] (image) at (0,-24mm) {
    \includegraphics[width=24mm]{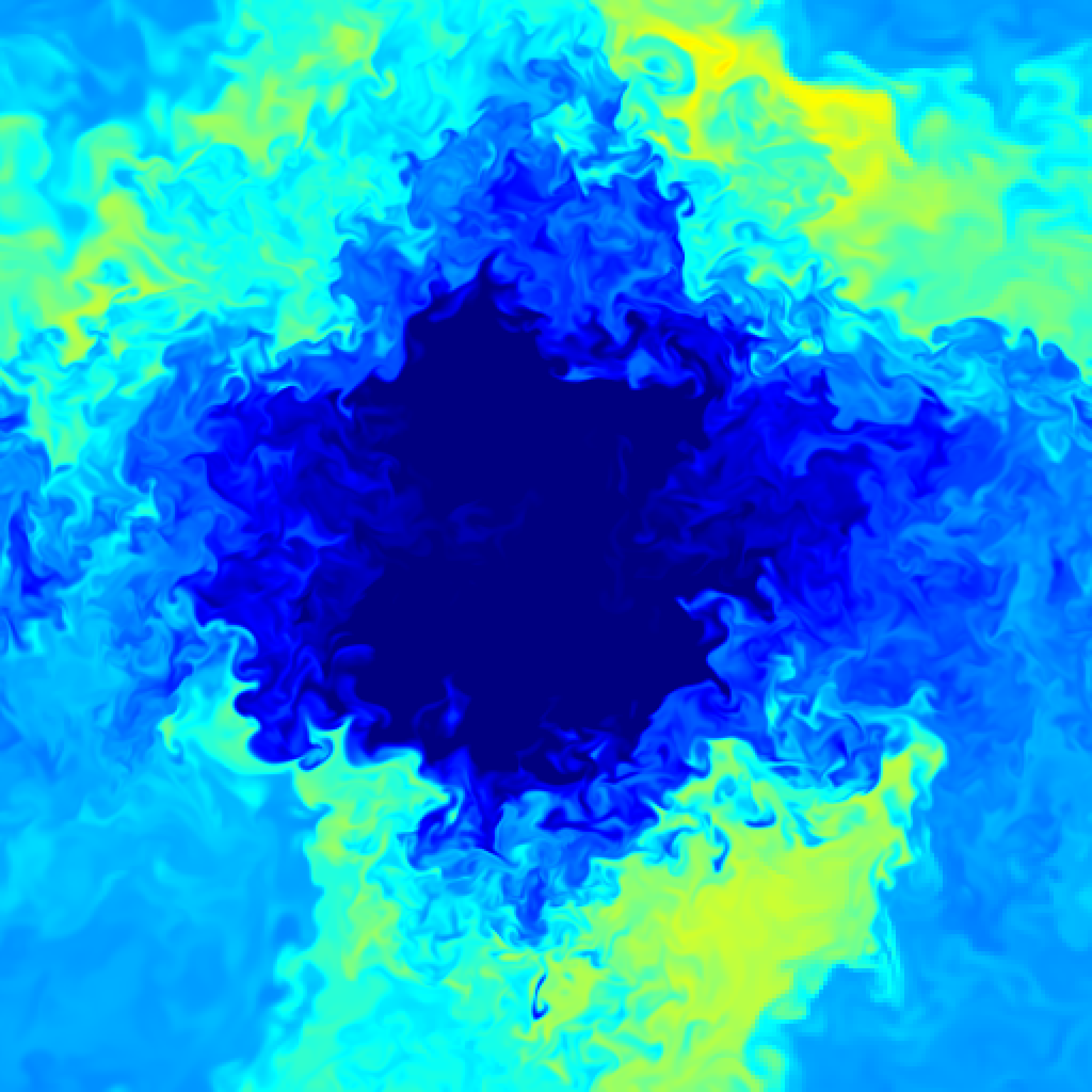}};
    \draw[thick] (image.west) -- (-15.5mm,-24mm);
    \draw[thick] (-15.5mm,-24mm) -- (-15.5mm,-60mm);
    \draw[thick,->] (-15.5mm,-60mm) -- (-18mm,-60mm);
    \draw[thick,white] (-12mm,-36mm) -- (12mm,-36mm);
    
    \node[inner sep=0pt] (image) at (0,-48mm) {
    \includegraphics[width=24mm]{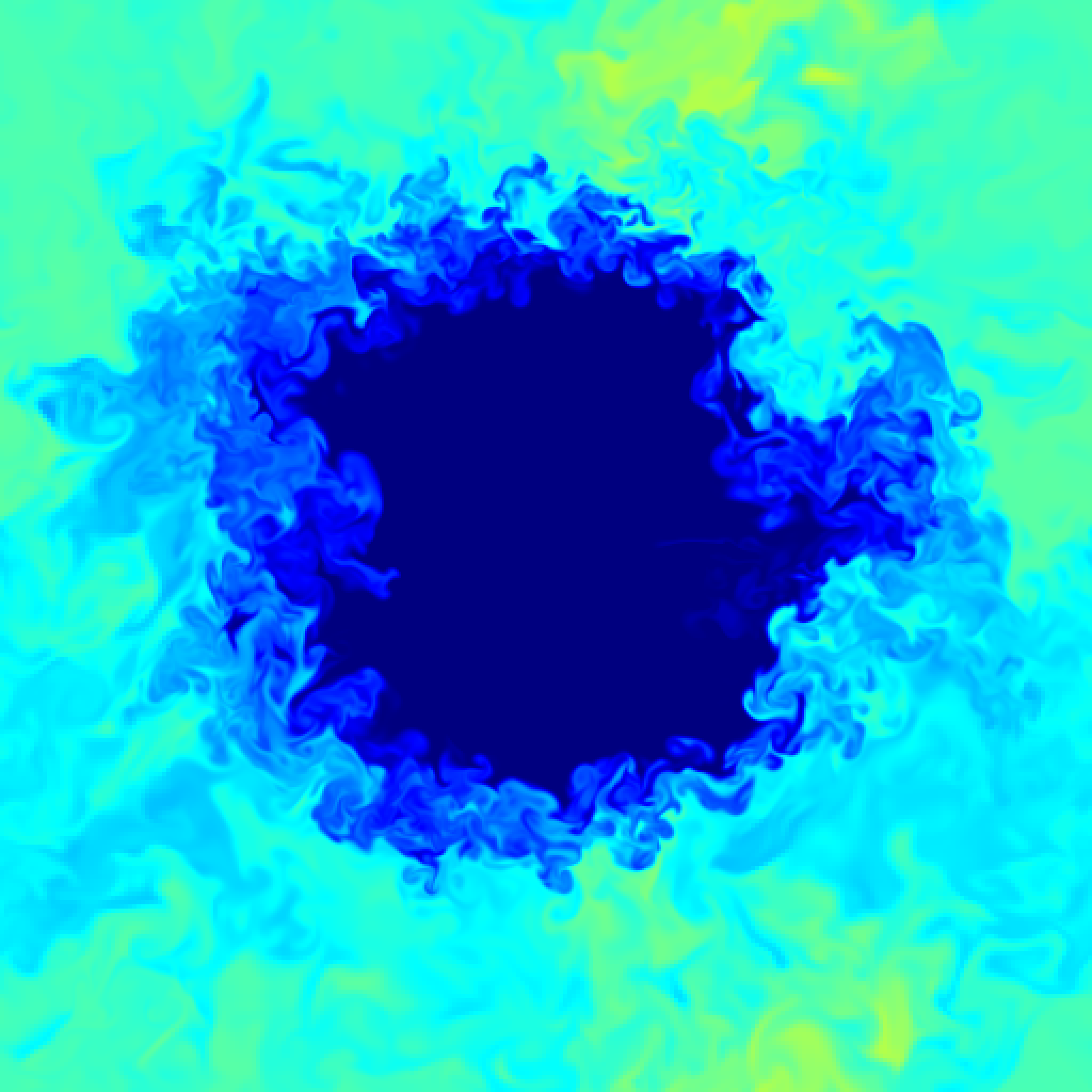}};
    \draw[thick] (image.west) -- (-14.5mm,-48mm);
    \draw[thick] (-14.5mm,-48mm) -- (-14.5mm,-72mm);
    \draw[thick,->] (-14.5mm,-72mm) -- (-18mm,-72mm);
    \draw[thick,white] (-12mm,-60mm) -- (12mm,-60mm);
    
    \node[inner sep=0pt] (image) at (0,-72mm) {
    \includegraphics[width=24mm]{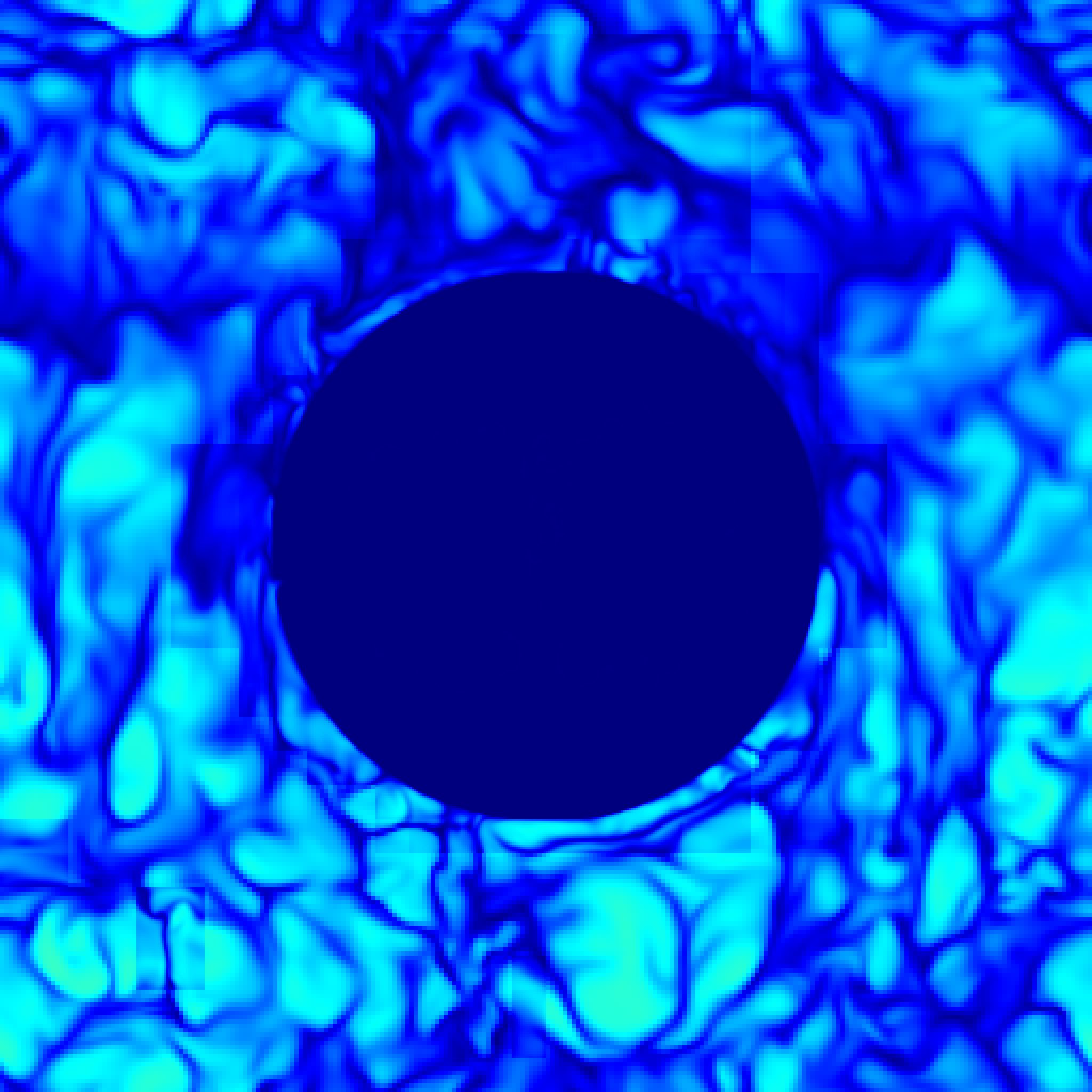}};
    \draw[thick] (image.west) -- (-13.5mm,-72mm);
    \draw[thick] (-13.5mm,-72mm) -- (-13.5mm,-84mm);
    \draw[thick,->] (-13.5mm,-84mm) -- (-18mm,-84mm);   
    \end{tikzpicture}\\
    \vspace{2mm}
    \includegraphics[width=70mm]{colourBar_temp.png}
    
    \caption{Snapshots of $T$ in the L: $x/d_{\textsc{j}}=0$ and $y/d_{\textsc{j}}=0$, and R: $z/d_{\textsc{j}}=0,1,2,3$ planes. Again, the scale is streamwise distance in jet diameters.}
    \label{fig:tempInstant}
\end{figure}
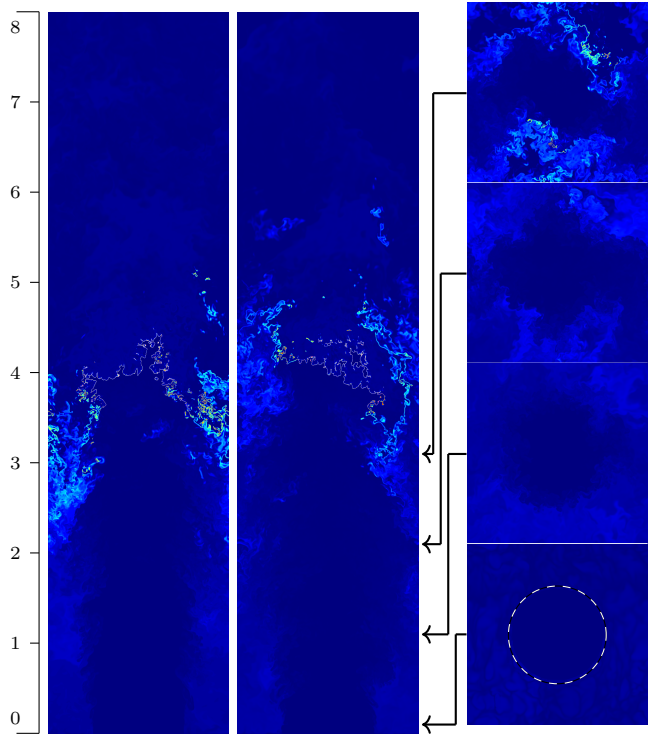
\begin{figure}[ht]
    \centering
    \begin{tikzpicture}
    \draw (0, 0) -- (0, 96mm);
    
    \draw (0mm,0mm) -- (-3.0mm, 0mm);
    \node [anchor = south east] at (-1.2mm,0mm) {\scriptsize 0};
    \draw (0mm,96mm) -- (-3.0mm, 96mm);
    \node [anchor = north east] at (-1.2mm,96mm) {\scriptsize 8};
    \foreach \y in {1,...,7} {
        \draw (-1.2mm, \y*12 mm) -- (0mm, \y*12 mm) ;
        \node[anchor=east] at (-1.2mm, \y*12 mm) {\scriptsize \y};
    }
    \end{tikzpicture}
    \includegraphics[width=24mm]{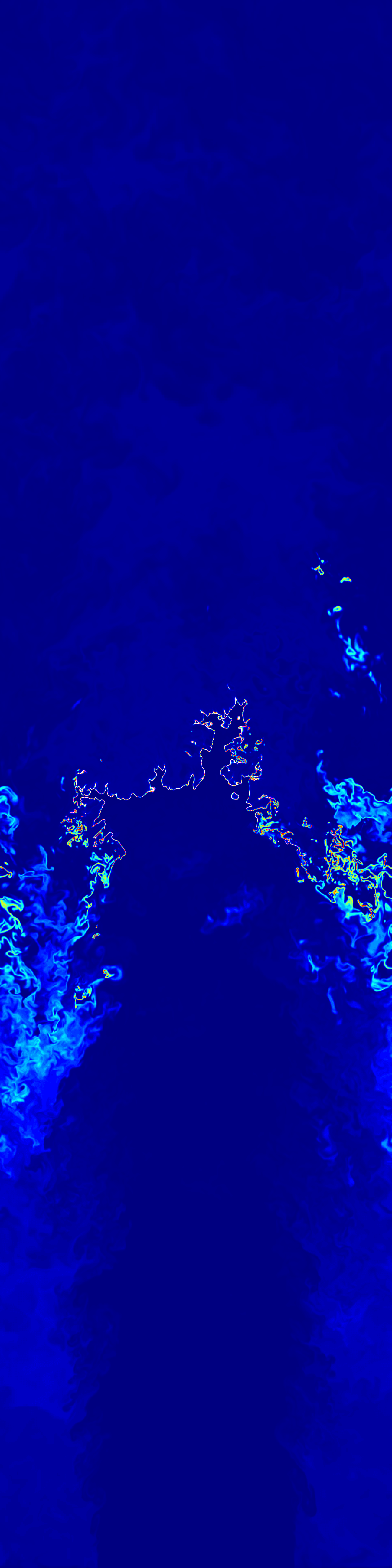}
    \includegraphics[width=24mm]{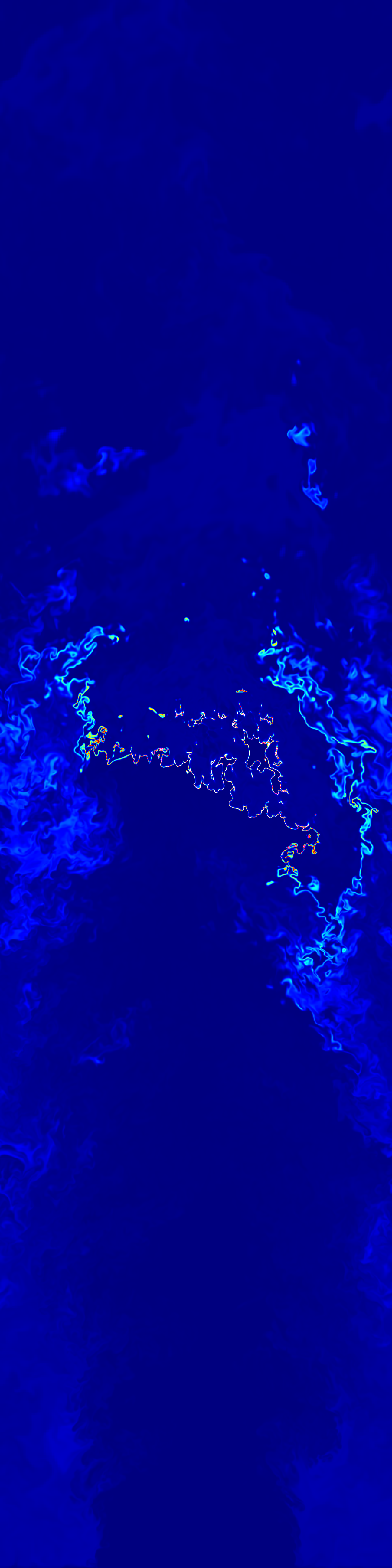}    
    \hspace{-2mm}
    \begin{tikzpicture}
    \node[inner sep=0pt] (image) at (0,0) {
    \includegraphics[width=24mm]{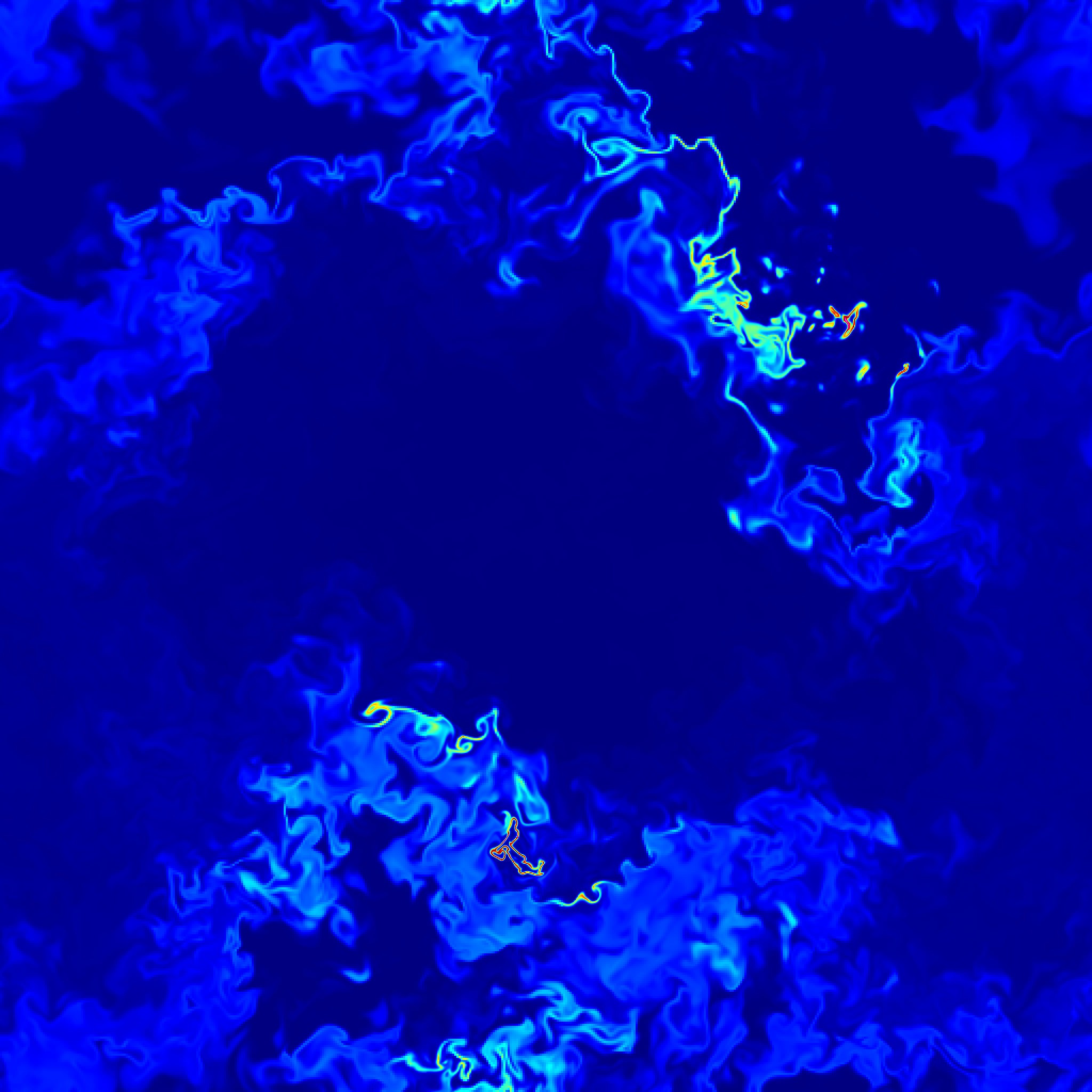}};
    \draw[thick] (image.west) -- (-16.5mm,0mm);
    \draw[thick] (-16.5mm,0mm) -- (-16.5mm,-48mm);
    \draw[thick,->] (-16.5mm,-48mm) -- (-18mm,-48mm);
    \draw[thick,white] (-12mm,-12mm) -- (12mm,-12mm);
    
    \node[inner sep=0pt] (image) at (0,-24mm) {
    \includegraphics[width=24mm]{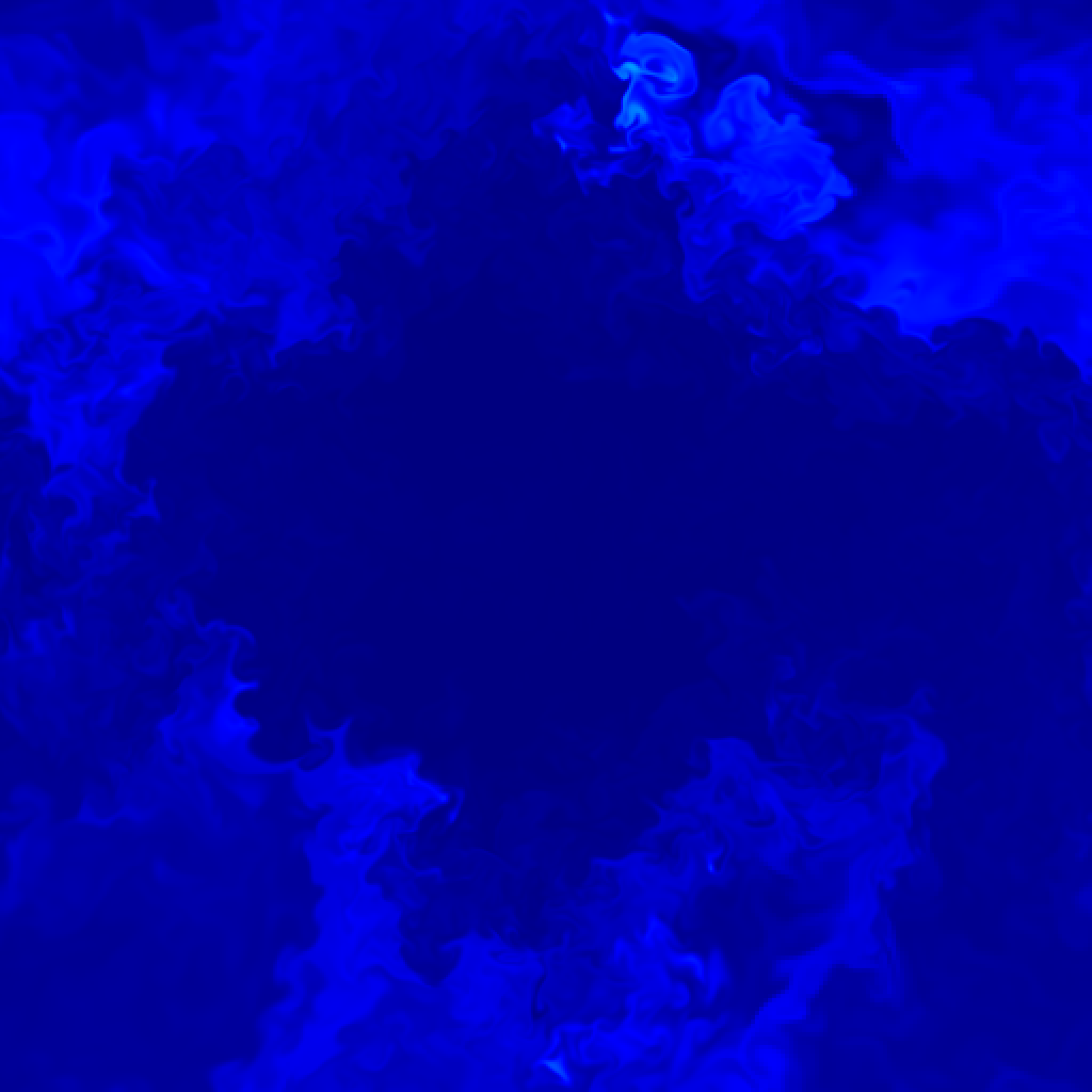}};
    \draw[thick] (image.west) -- (-15.5mm,-24mm);
    \draw[thick] (-15.5mm,-24mm) -- (-15.5mm,-60mm);
    \draw[thick,->] (-15.5mm,-60mm) -- (-18mm,-60mm);
    \draw[thick,white] (-12mm,-36mm) -- (12mm,-36mm);
    
    \node[inner sep=0pt] (image) at (0,-48mm) {
    \includegraphics[width=24mm]{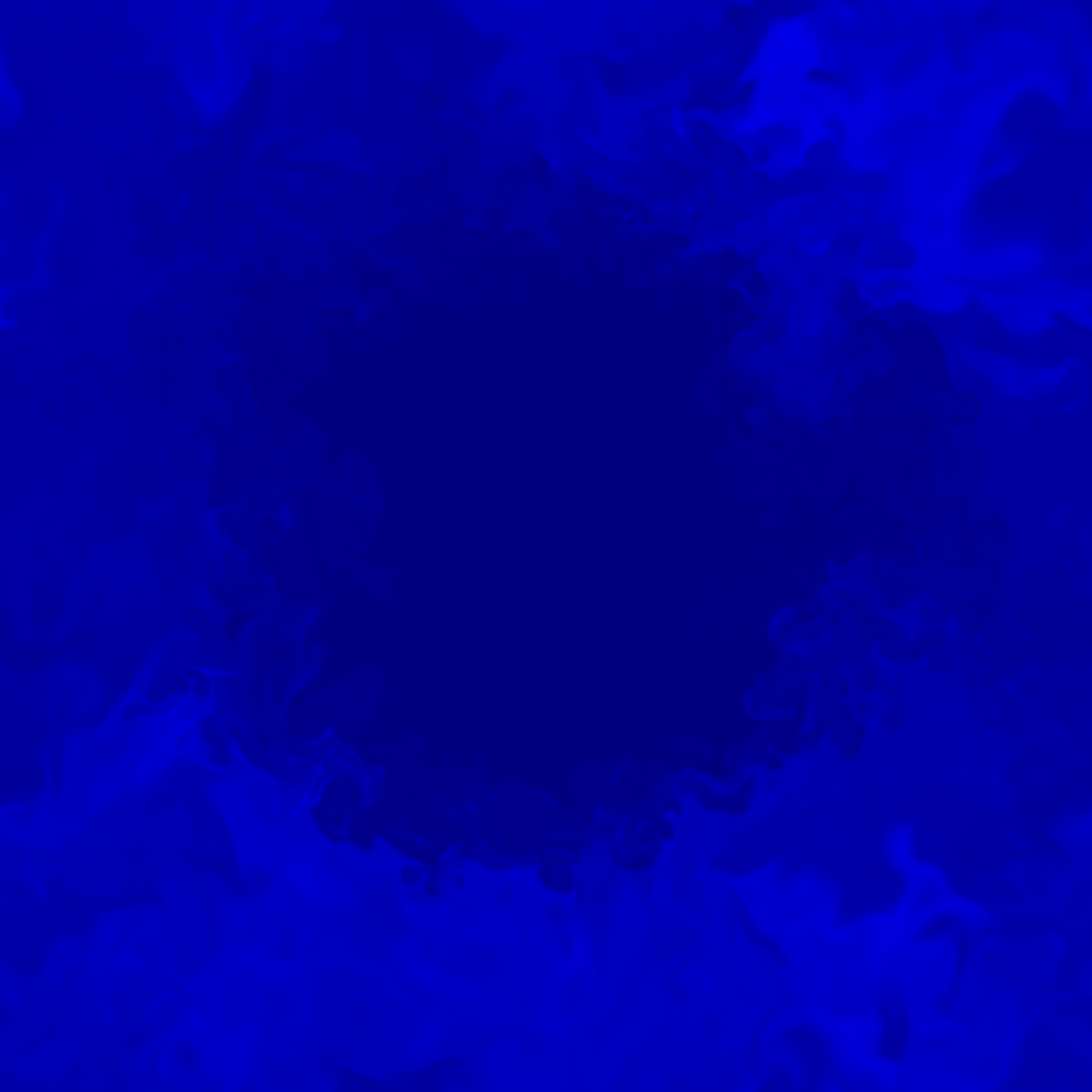}};
    \draw[thick] (image.west) -- (-14.5mm,-48mm);
    \draw[thick] (-14.5mm,-48mm) -- (-14.5mm,-72mm);
    \draw[thick,->] (-14.5mm,-72mm) -- (-18mm,-72mm);
    \draw[thick,white] (-12mm,-60mm) -- (12mm,-60mm);
    
    \node[inner sep=0pt] (image) at (0,-72mm) {
    \includegraphics[width=24mm]{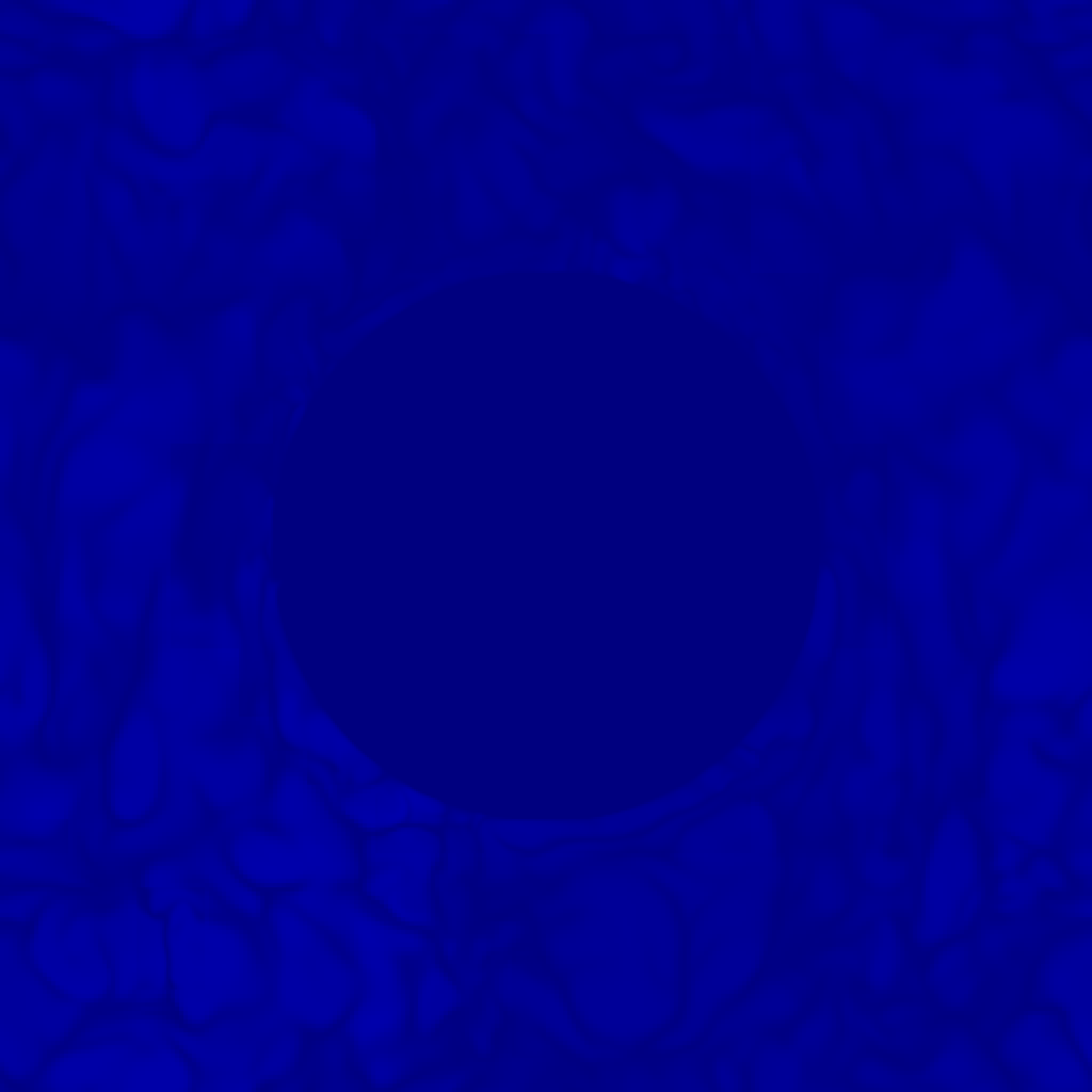}};
    \draw[thick] (image.west) -- (-13.5mm,-72mm);
    \draw[thick] (-13.5mm,-72mm) -- (-13.5mm,-84mm);
    \draw[thick,->] (-13.5mm,-84mm) -- (-18mm,-84mm);   
    \draw[draw=black] (image.south) ++(0mm,12mm) circle (6.5mm);
    \draw[draw=white, dashed] (image.south) ++(0mm,12mm) circle (6.5mm);
    \end{tikzpicture}
    
    \caption{Snapshots of $Y_{\HOtwo}$ in the L: $x/d_{\textsc{j}}=0$ and $y/d_{\textsc{j}}=0$ planes, and R: $z$ slices at $z/d_{\textsc{j}}=0,1,2,3$. The black and white circle on the $z/d_{\textsc{j}}=0$ image illustrates the airport.}
    \label{fig:HO2instant}
\end{figure}

While the analysis above provides a temporally-averaged description of flame height and jet structure, an in-depth analysis of instantaneous snapshots is required to understand the flame structure more completely. An example temperature field in $x$ and $y$ planes is provided in figure \ref{fig:tempInstant}. Figure \ref{fig:HO2instant} contains $x$ and $y$ slices of mass fractions of HO$_{2}$. In addition to highlighting the thin flame front in the middle of the domain, HO$_{2}$ is present in the recirculation zone in low (but non-negligible) concentrations. It is also found  in higher concentrations in the region separating the regions of recirculated and downstream products; this suggests the existence of a region of distributed reactions. The presence of this burning region can be highlighted by considering measures of flame progress. Typically in the literature, a progress variable for partially premixed combustion (e.g.~\cite{Bray2005RoleCombustion}) is defined as a function of fuel mass fraction and mixture fraction:
\begin{equation}
    c_{\textsc{f}} = \begin{cases}
    \displaystyle1-\frac{Y_{\textsc{f}}}{Z} &\mbox{ where } Z \leq Z_{\textrm{st}},\vspace{2mm}\\
    \displaystyle\frac{1-Z_{\mathrm{st}}}{Z_{\mathrm{st}}}\frac{Z-Y_{\textsc{f}}}{1-Z} &\mbox{ otherwise}.
    \end{cases}
\end{equation}
A progress variable based on product mass fraction can also be derived,
\begin{equation}
    c_{\textsc{p}} = \begin{cases}
    \displaystyle\frac{Y_{\textsc{p}}Z_{\textrm{st}}}{Z(Y_{\textsc{o},\textsc{cf}}[1-Z_{\textrm{st}}]+Z_{\textrm{st}})} &\mbox{where } Z \leq Z_{\textrm{st}},\vspace{2mm}\\
    \displaystyle\frac{Y_{\textsc{p}}(1-Z_{\textrm{st}})}{(1-Z)(Y_{\textsc{o},\textsc{cf}}[1-Z_{\textrm{st}}]+Z_{\textrm{st}})} & \mbox{otherwise},
    \end{cases}
\end{equation}
where the mixture fraction $Z$ is given by
\begin{equation}
    Z = \frac{sY_{\textsc{f}} - Y_{\textsc{o}} + Y_{\textsc{o},\textsc{cf}}}{s + Y_{\textsc{o},\textsc{cf}}},
\end{equation}
with $s=8$ the stoichiometric ratio, and $Y_{\textsc{o},\textsc{cf}}=0.232$ the oxygen mass fraction in the crossflow. 

\begin{figure}[ht]
    \begin{tikzpicture}
    \draw (0, 0) -- (0, 72mm);
    
    \draw (0mm,0mm) -- (-3.0mm, 0mm);
    \node [anchor = south east] at (-1.2mm,0mm) {\scriptsize 0};
    \draw (0mm,72mm) -- (-3.0mm, 72mm);
    \node [anchor = north east] at (-1.2mm,72mm) {\scriptsize 8};
    \foreach \y in {1,...,7} {
        \draw (-1.2mm, \y*9 mm) -- (0mm, \y*9 mm) ;
        \node[anchor=east] at (-1.2mm, \y*9 mm) {\scriptsize \y};
    }
    \end{tikzpicture}
    \centering
    \centering
    \includegraphics[width=18mm]{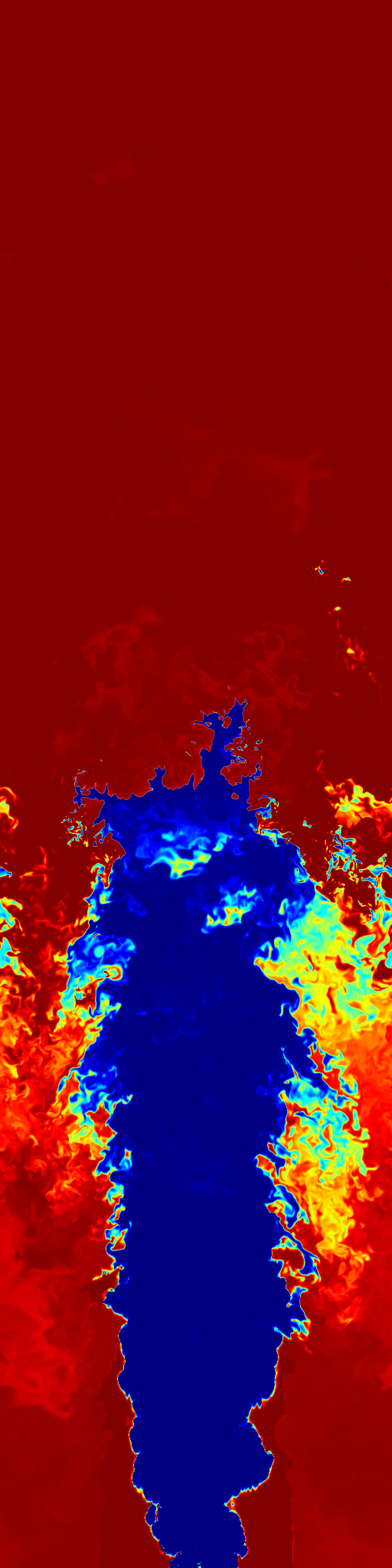}
    \includegraphics[width=18mm]{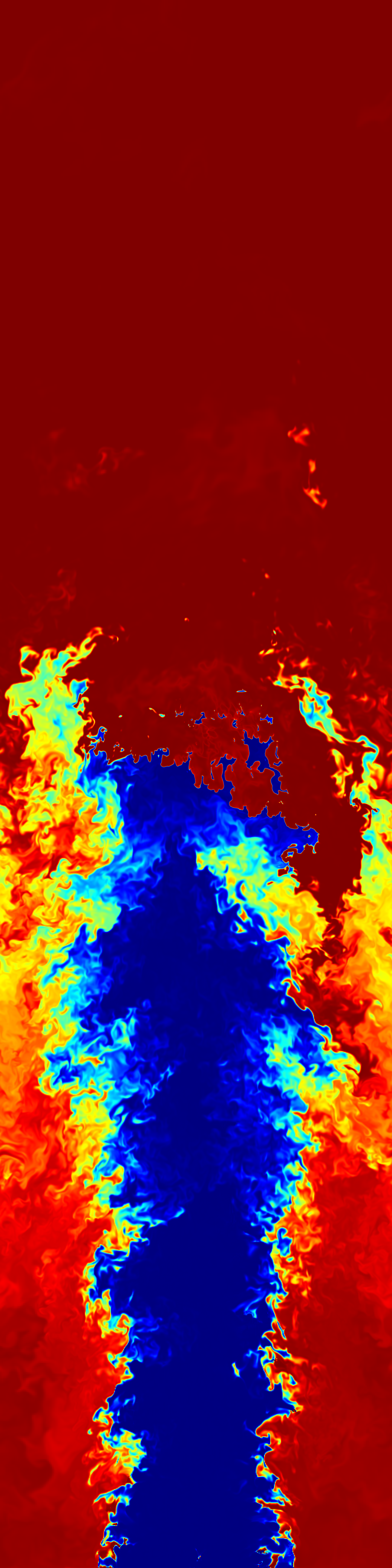}
    \hspace{2mm}
    \includegraphics[width=18mm]{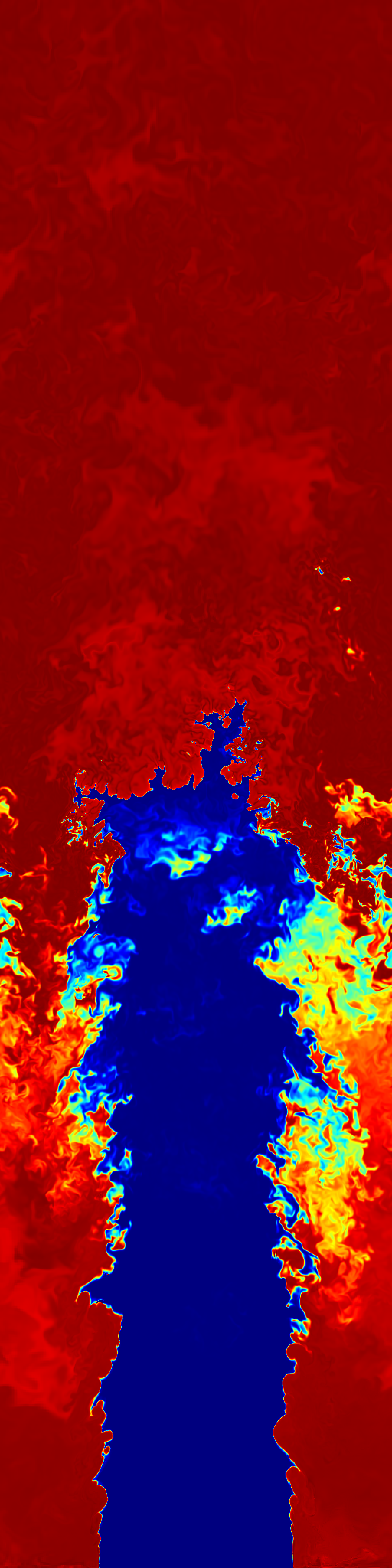}
    \includegraphics[width=18mm]{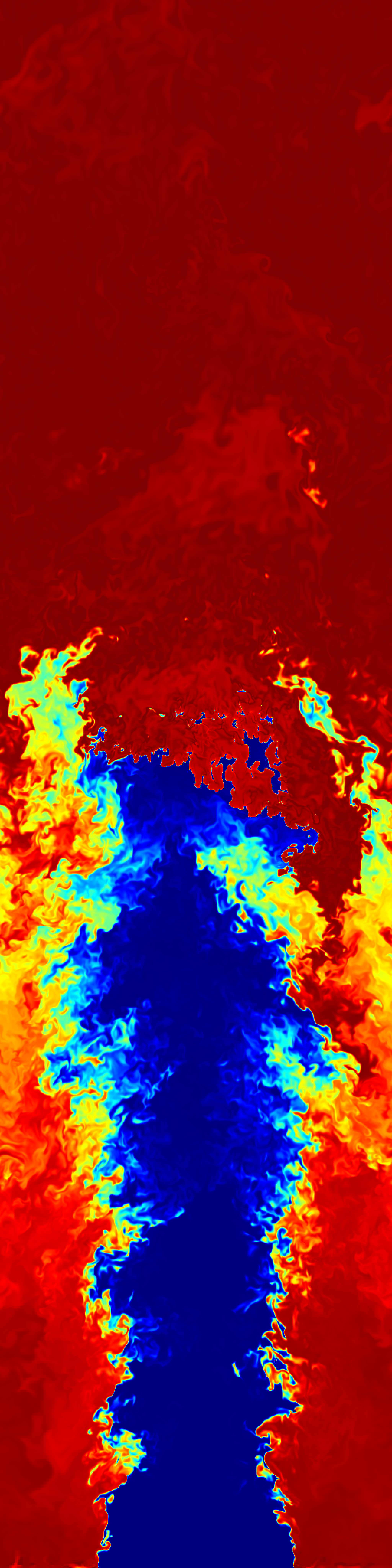}
    \caption{A comparison between the fuel-based and products-based progress variables (L: $c_{\textsc{f}}$, R: $c_{\textsc{p}})$. There are differences between the two where pure air is present, and identifies both a very thin core flame and significant turbulent mixing on the periphery.}
    \label{fig:cInstant}
\end{figure}

An $x$ and $y$ slice of the fuel- and product-based progress variable are given in figure \ref{fig:cInstant}. The two progress variables are naturally similar, and both identify a thin flame in the middle of the domain, as well as what appears to be a distributed flame in the periphery (there is broad spatial region with intermediate values of progress).  Both variables also identify the recirculation region as products (hot/wet and absent of fuel).  It is worth noting that near the jet inlet, there is a region of near-pure air surrounding the fuel, which is identified differently by the two progress variables; the fuel-based progress variable identifies this region as products (no fuel) and the products-based progress variable identifies it as reactants (cold/dry).

\subsubsection{Flame zones}
A zonal classification can be defined to segregate the flow into five regions as described by the following criteria, ordered by priority:
\begin{enumerate}
    \item\label{zone:core} Core flame: $Y_{\HOtwo} > 4\times 10^{-4}$;
    \item Products: $T > 1600$K (or $z > 3.8$\,cm);
    \item Jet: $T < 800$\,K;
    \item\label{zone:peripheral} Peripheral flame: $Y_{\HOtwo} > 1\times 10^{-4}$;
    \item Recirculation otherwise.
\end{enumerate}
Furthermore, a buffer region is added to each of the core and peripheral flame zones to ensure comprehensive coverage of the reacting regions, while avoiding spurious classification away from the flames. Specifically, any point within 24 (fine-level) computational cells of a point satisfying criterion \ref{zone:core} is also classified in the core flame zone; similarly 8 cells within a point satisfying criterion \ref{zone:peripheral} is classified in the peripheral flame zone.

\begin{figure}[ht]
    \centering
    \begin{tikzpicture}
    \draw (0, 0) -- (0, 120mm);
    
    \draw (0mm,0mm) -- (-3.0mm, 0mm);
    \node [anchor = south east] at (-1.2mm,0mm) {\scriptsize 0};
    \draw (0mm,120mm) -- (-3.0mm, 120mm);
    \node [anchor = north east] at (-1.2mm,120mm) {\scriptsize 8};
    \foreach \y in {1,...,7} {
        \draw (-1.2mm, \y*15 mm) -- (0mm, \y*15 mm) ;
        \node[anchor=east] at (-1.2mm, \y*15 mm) {\scriptsize \y};
    }
    \end{tikzpicture}
    \includegraphics[width=30mm]{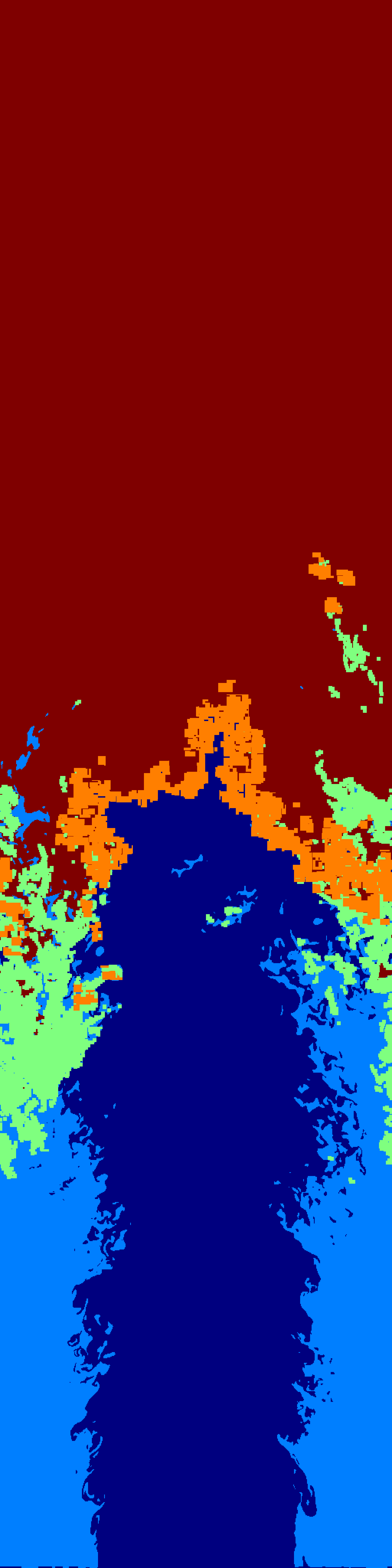}
    \includegraphics[width=30mm]{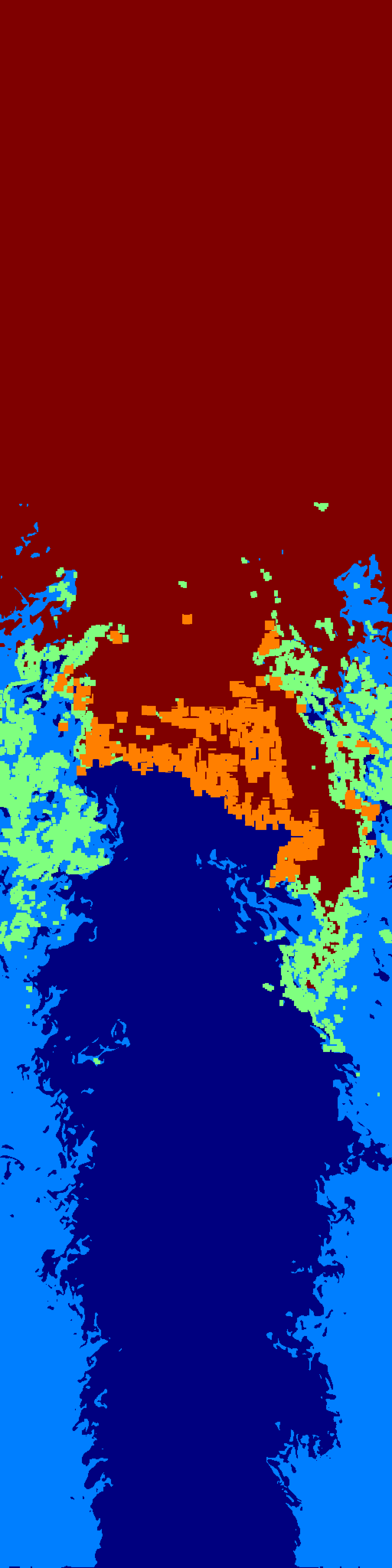}
    \caption{Zone classification L: $x$ slice, R: $y$ slice. Orange corresponds to the core flame, products are dictated by red, peripheral flame by green, recirculating fluid is light blue and the jet is dark blue.}
    \label{fig:zones}
\end{figure}
\begin{figure}[ht]
    \centering
    \includegraphics[width=88mm]{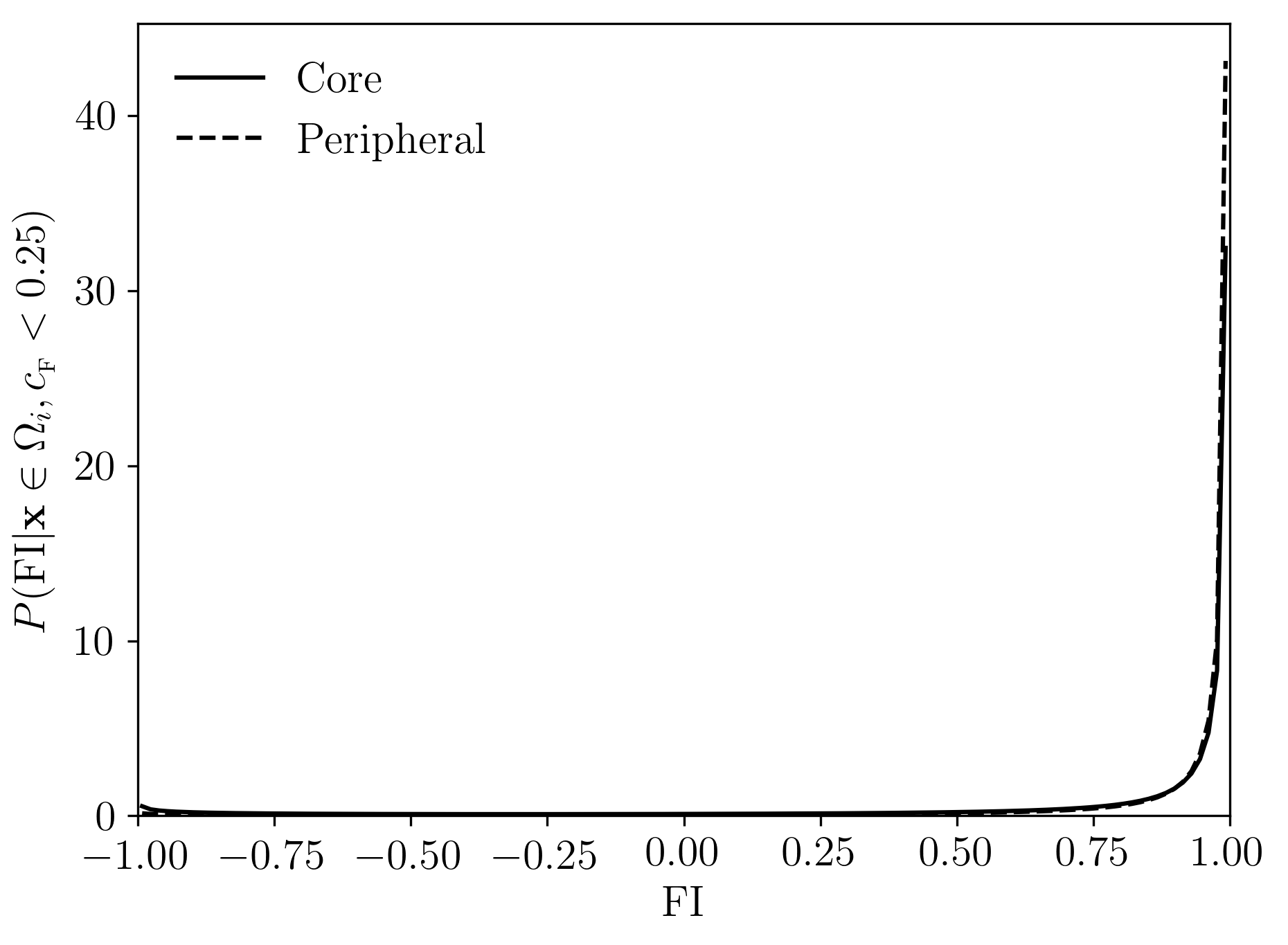}
    \caption{Flame index in preheat region of the core and peripheral flame.}
    \label{fig:flameIndex}
\end{figure}

To determine the combustion regime in each zone, a flame index is formulated following \cite{Yamashita1996AFlames} and \cite{Domingo2002PartiallyCombustion} as
\begin{equation}
    \textsc{fi} = \frac{\nabla Y_{\textsc{f}} \cdot \nabla Y_{\textsc{o}}}{|\nabla Y_{\textsc{f}}||\nabla Y_{\textsc{o}}|},
\end{equation}
conditioned on flame location and is shown in figure \ref{fig:flameIndex}. In this case, the flame index is evaluated conditional on $c_{\textsc{f}}<0.25$ and conditional on location within either the core or peripheral zone. This value was chosen to isolate gradients in the preheat region and avoid division by small numbers close to the products. The flame index is essentially unity everywhere, meaning the flame burns in a premixed regime, but may burn at varying equivalence ratio across the flame surface, as discussed below; the regime can be described as `inhomogeneously premixed'. 

\begin{figure}[ht!]
    \centering
    \includegraphics[width=88mm]{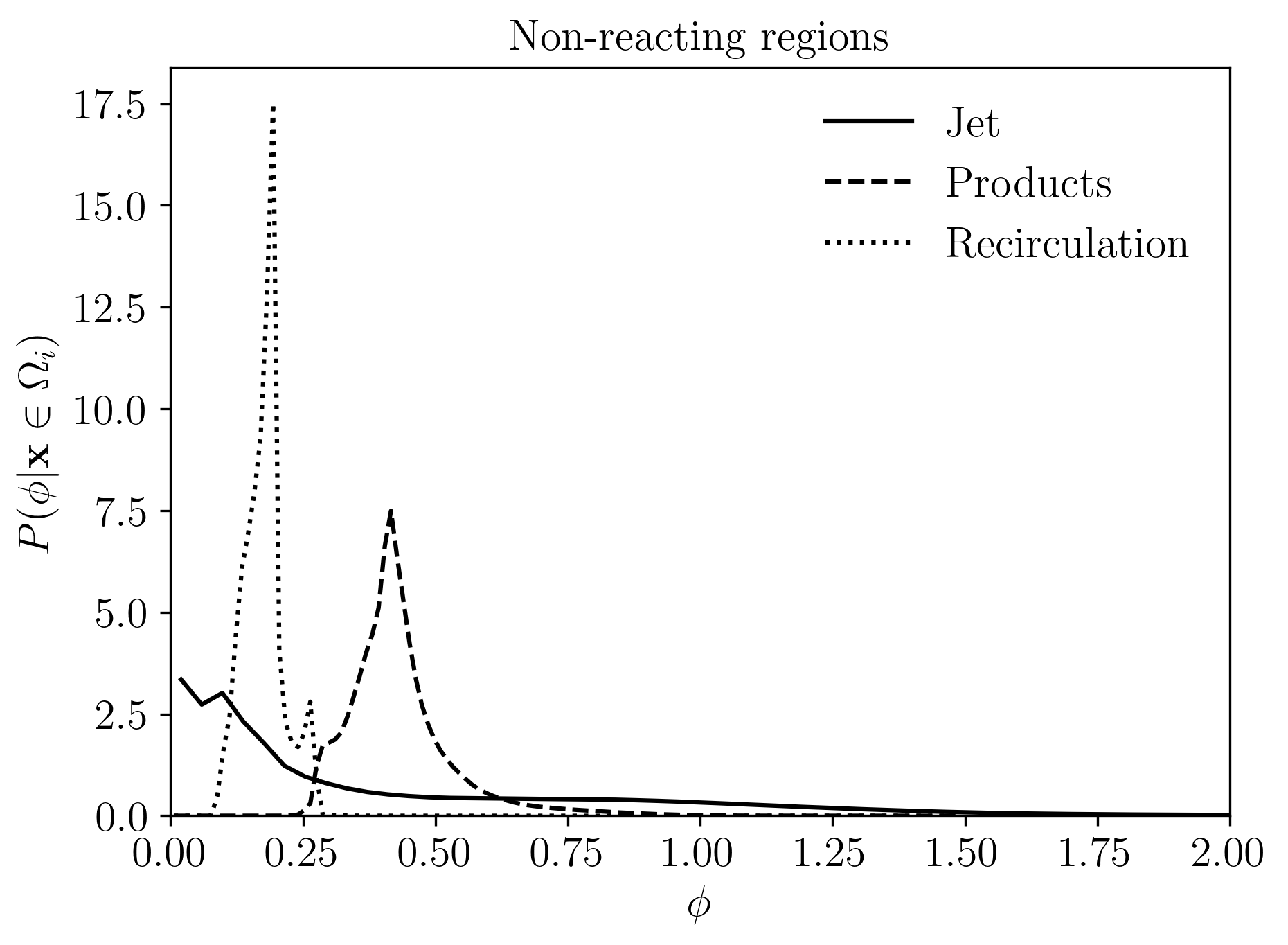}
    \includegraphics[width=88mm]{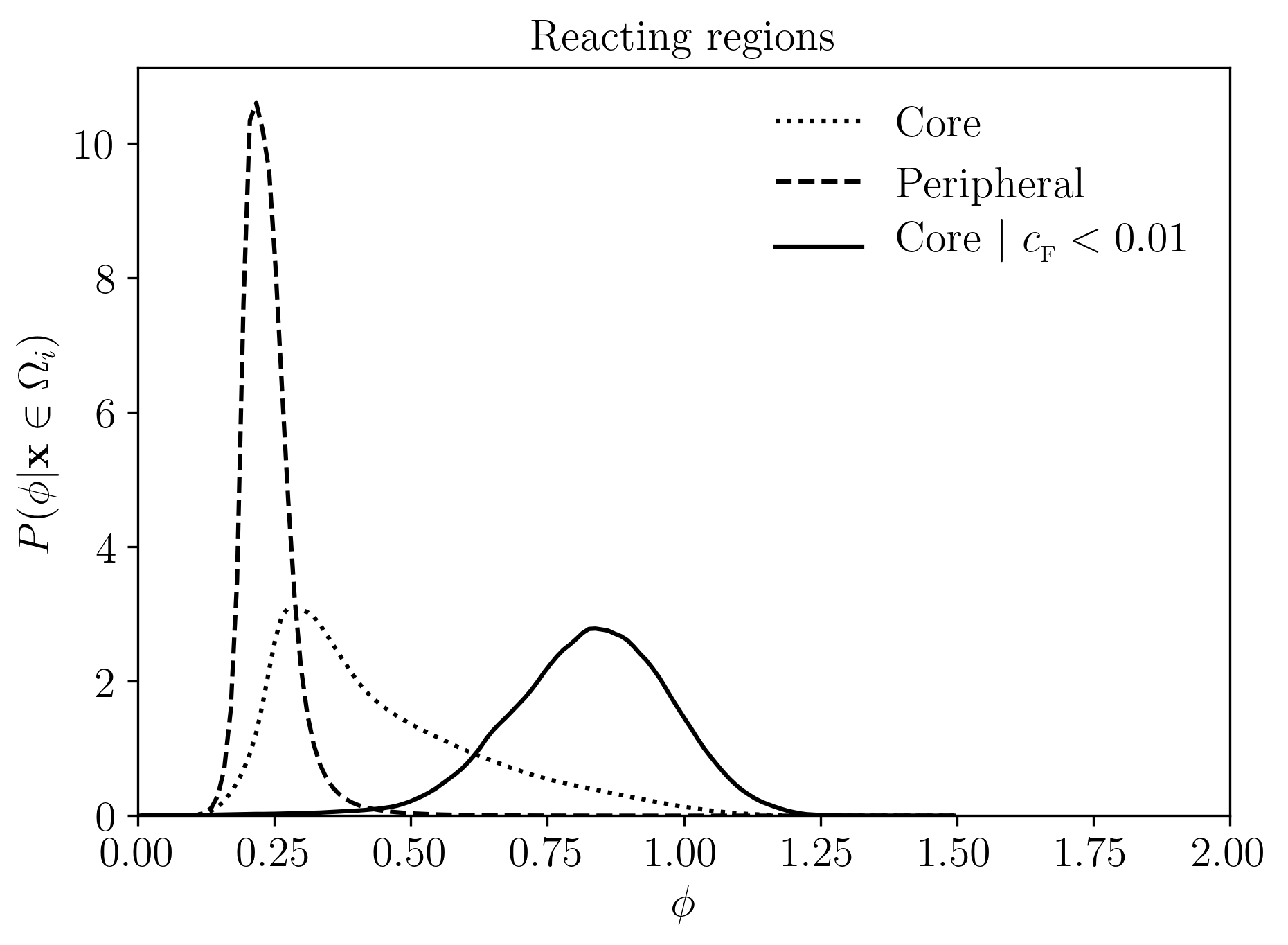}
    \caption{PDFs of local equivalence ratio $\phi$ conditioned on each zone in the non-reacting (top) and reacting (bottom) regions}
    \label{fig:phiConditioned}
\end{figure}

To explore the burning regimes further, distributions of local equivalence ratio is considered in the different zones. The local equivalence ratio is defined as $\phi = 2\xi_{\textsc{h}}/\xi_{\textsc{o}}$, where $\xi_{\textsc{e}}$ denotes the molar concentration of element \textsc{e}.
Probability density functions (PDFs) of $\phi$ conditioned on each zone in the non-reacting and reacting regions are shown separately in figure \ref{fig:phiConditioned}.  The jet region has an equivalence ratio below $\phi=0.3$ almost everywhere, but low probabilities extend up to about 1.3; this suggests that it may be possible to observe diffusion flames (in the context of an edge flame) but they should not be anticipated to be prevalent.  
In the core flame, preferential diffusion of H$_{2}$ alters the local equivalence ratio distribution and so may not be representative of the unburned/fully-burned equivalence ratio; consequently, the $\phi$ distribution doubly-conditioned on zone and $c_{\textsc{f}} < 0.01$ is also shown in figure \ref{fig:phiConditioned}. This suggests that the actual range of unburned equivalence ratios in the core flame is around 0.4--1.2, in an approximately Gaussian distribution centred at $\phi \approx 0.8$.  On the other hand, the equivalence ratio distribution in the peripheral flame is far narrower ranging from approximately 0.18--0.3, with the distribution centred around $\phi \approx 0.22$. 

\begin{figure}[ht!]
    \centering
    \includegraphics[width=88mm]{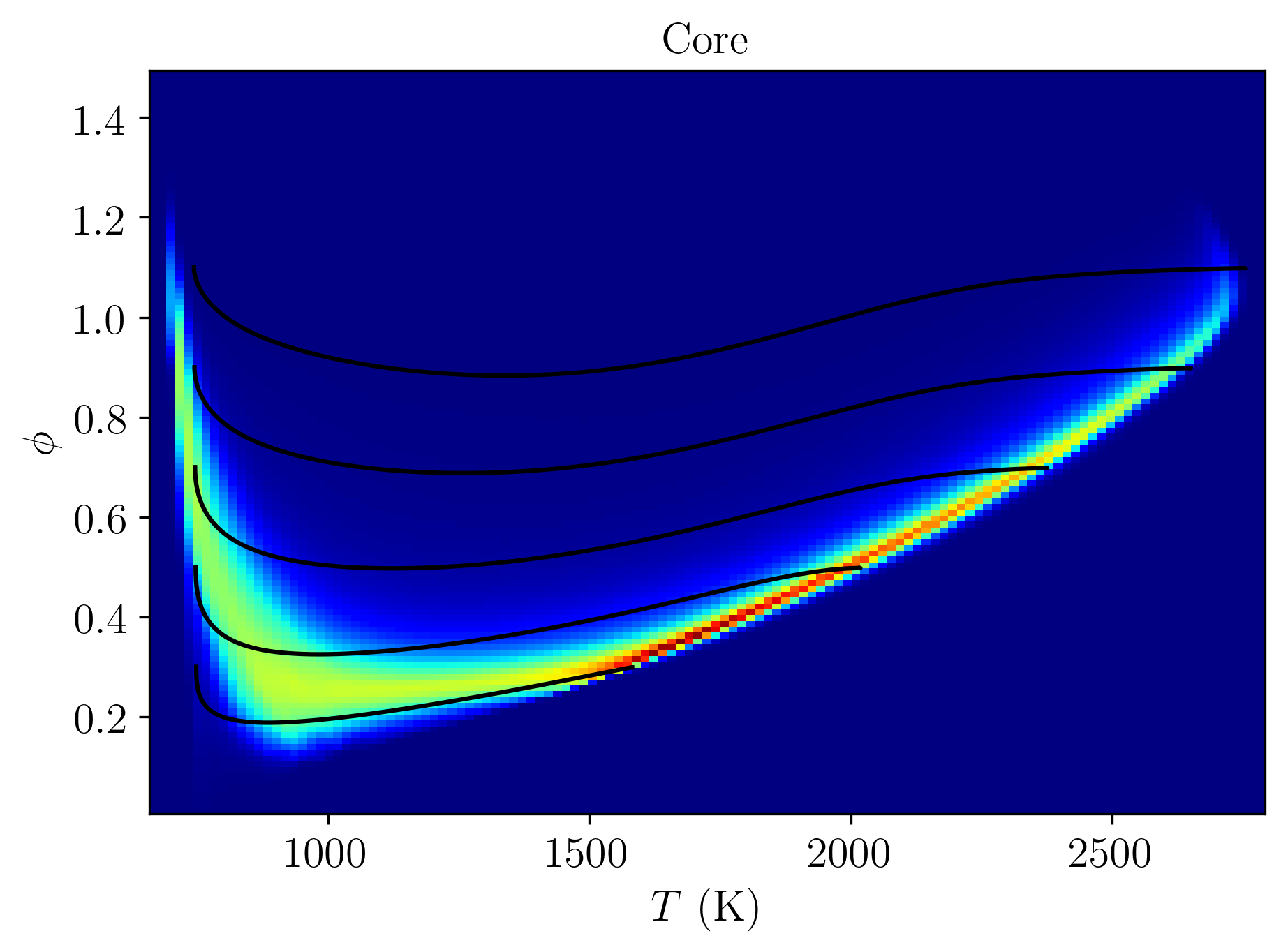}
    \includegraphics[width=88mm]{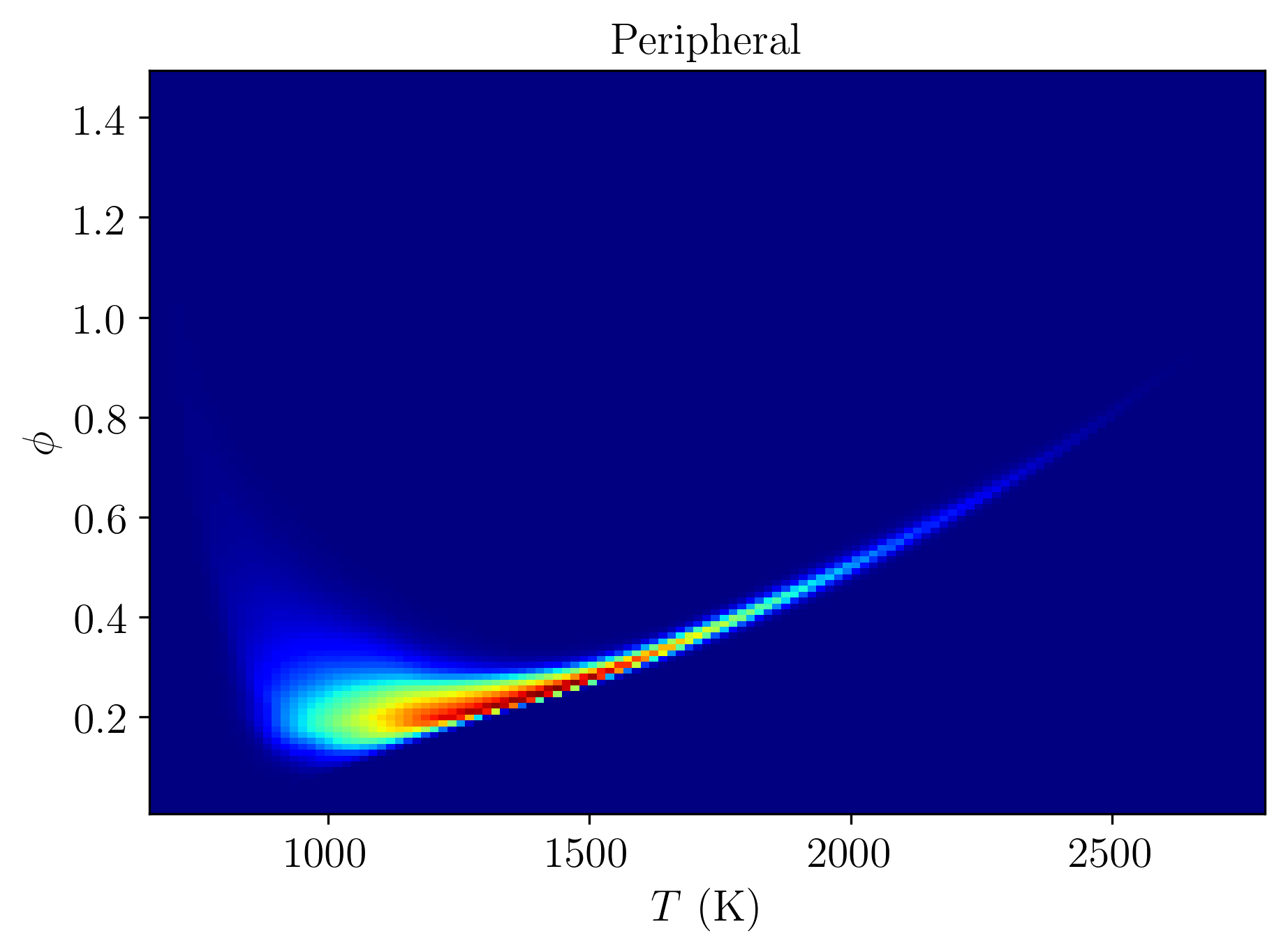}
    \caption{JPDFs of temperature and local equivalence ratio in the core (top) and peripheral (bottom) flame. Black lines are profiles from 1D Cantera simulations with equivalence ratios from 0.3 - 1.2 in the case of the core flame, and 0.2-0.36 in the peripheral flame.}
    \label{fig:phiTJPDF}
\end{figure}

The variation in equivalence ratio through the flame can be examined in more detail in a joint probability distribution function (JPDF) of temperature and local equivalence ratio, shown in figure \ref{fig:phiTJPDF}. One-dimensional premixed flame profiles (from Cantera) are superimposed on the JPDF at equivalence ratios 0.3 - 1.1 (in $\phi$ increments of 0.2) and unburned temperature of 750\,K, which show the decrease in equivalence ratio due to preferential diffusion effects. The decrease in equivalence ratio in the JPDF is more pronounced than the one-dimensional profiles, which may be due to differential/preferential diffusion laterally due to the inhomogeneously-premixed reactants. The corresponding JPDF of $T$ and $\phi$ in the peripheral flame is given in figure \ref{fig:phiTJPDF}, with black lines from 1D Cantera simulations at equivalence ratios 0.2 - 0.36 in $\phi$ increments of 0.04 at an unburned temperature of 750\,K. 

In summary, there is a broad range of equivalence ratios (i.e.\ not fully-premixed) but those equivalence ratios are almost-everywhere below unity (i.e.\ not non-premixed); therefore, this flame can be classified as burning in an `essentially-lean inhomogenously-premixed' mode.  At the operating conditions considered, there is a low probability of finding diffusion flamelets; this may differ at other conditions, such as lower injector Reynolds number of momentum flux ratios, where insufficient mixing may occur.

\subsubsection{Turbulent burning regimes}

Turbulence-flame interactions in the core and peripheral flames are very different. The core flame appears as a distinct, if corrugated, flame surface, with burning cells characteristic of low Lewis number effects. On the other hand, the peripheral flame is spatially distributed over a large region with no clear flame surface. 

\begin{figure}[ht]
    \centering
    \includegraphics[width=88mm]{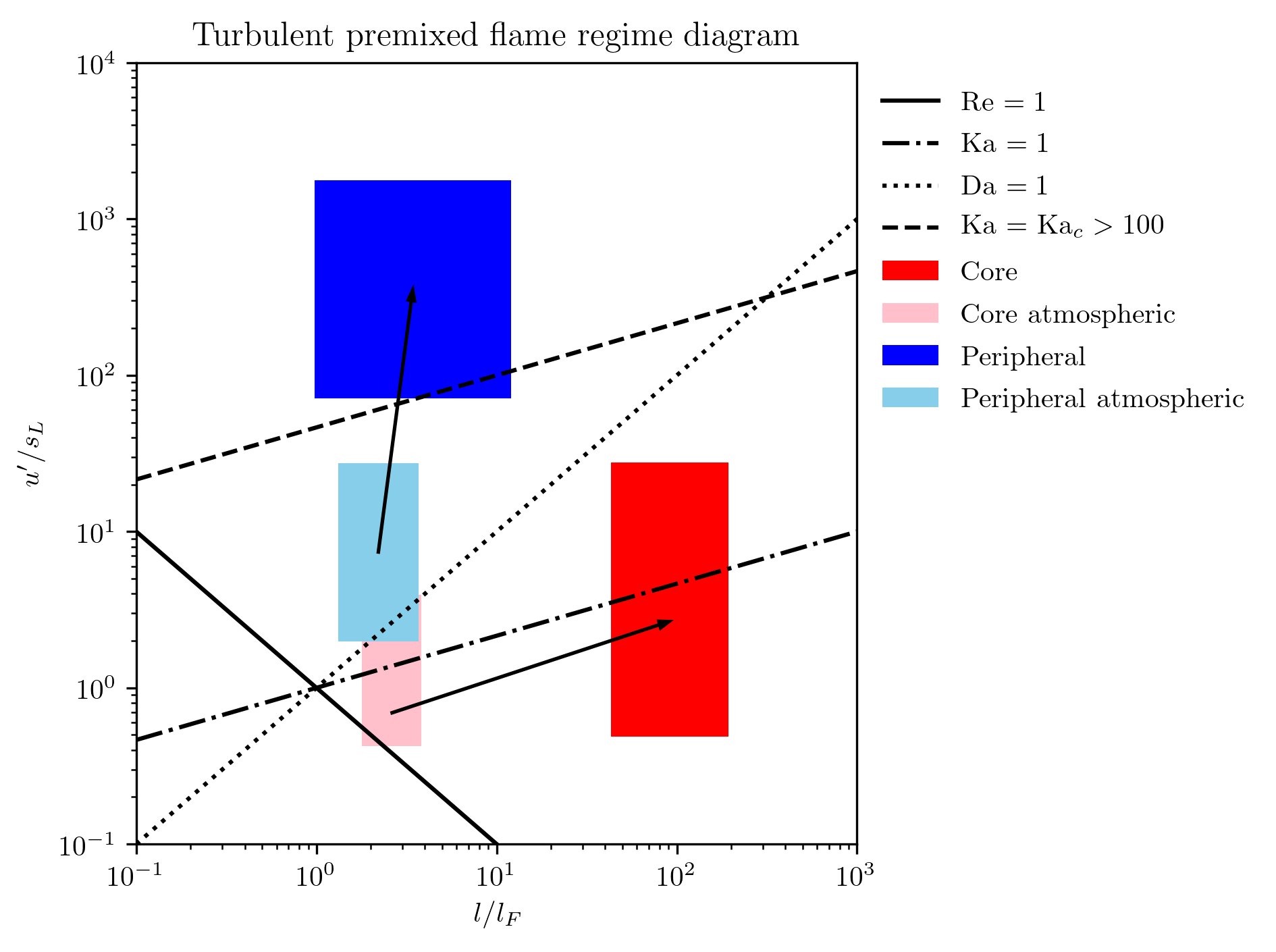}
    \caption{Turbulent premixed flame regime diagram, with the core and peripheral flame shown by the red and blue blocks respectively. The black arrow indicates the effect of increasing pressure.}
    \label{fig:borghi}
\end{figure}

Based on an equivalence ratio distribution, and unburned temperature of 750\,K (which appears to be the case for almost all the flamelets observed), and a pressure of 24\,atm, maximum and minimum values of characteristic flame speed and thickness can be obtained from Cantera. Assuming a range of turbulent characteristics of $u^\prime = 5-20$\,m/s and $\ell = 1-2$\,mm, regions of the premixed turbulent flame regime diagram can be highlighted using these values. From the PDFs, a range of $0.4 < \phi < 1.2$ is observed in the core flame, and $0.18 < \phi < 0.3$ in the peripheral flame; corresponding indicative regions are shown in figure \ref{fig:borghi}. These two regions imply Karlovitz number ranges of $0.05 < \Ka < 22$ for the core flame and $350 < \Ka < 66,000$ for the peripheral flame. This would place the core flame either in the thin reaction zone or corrugated/wrinkled flame regime, whereas the peripheral flame would be in the distributed burning regime. The large range of burning regimes is only present due to the high pressure of the system; if an identical assessment were performed at atmospheric pressure, there would be less of a distinction between the regimes, as indicated by the pale regions in figure \ref{fig:borghi}.  In the case of the core flame, the Karlovitz number is mostly unchanged by an increase in pressure because the flame speed increases and thickness decreases along an approximately-constant Karlovitz line; Damk\"ohler number, on the other hand, increases significantly with increasing pressure. The lower value of $\ell/\ell_{\textsc{f}}$ in the core flame also indicates significantly less resolution would be required to resolve the flame. This is not the case in the peripheral flame, where the flame speed is substantially reduced at increased pressure at lean conditions, which moves the flame from being moderately-turbulent to the distributed regime at high pressure. 

%% file: stabilisation/preamble.tex
There are several mechanisms by which a flame can stabilise, depending on the thermochemical composition and mixedness of the flow. The present flame is inhomogeneously premixed with elevated reactant temperature, situated in a recirculatory flow. Therefore, it may stabilise through purely deflagrative flame propagation, autoignition, or in some mixed mode whereby deflagrative or ignitive burning is accelerated by both pre-ignition reactions and pre-mixing with combustion products. Stabilisation through deflagrative flame propagation is reliant on diffusive and turbulent transport from the burned to unburned side of the flame. On the other hand, autoignition allows isolated flame kernels to develop ahead of the flame base.

%% file: stabilisation/propagation.tex
The feasibility of flame stabilisation by purely deflagrative turbulent flame propagation is assessed by comparing relevant flow and flame speeds. First, a comparison is made between the turbulent flame speed and jet velocity to establish the feasibility of flame stabilisation by flame propagation; specifically, by estimating the upper bound on the jet inlet velocity that can be stabilised by turbulent flame propagation without blowing off. Following \cite{PETERS1999TheTurbulence}, in the large-scale turbulence limit (i.e.\ giving the highest feasible enhancement of surface area), an estimate for turbulent flame speed can be written as
\begin{equation}\label{eq:peters}
    s_{T} = s + 2u',
\end{equation}
where $s$ is some reference local flame speed (i.e.\ laminar or freely-propagating \cite{Howarth2022AnFlames}). By taking turbulent intensity $I = u' / \bar{u}$, turbulent flame propagation can only stabilise the flame if
\begin{equation}\label{eq:speedcondition}
    \bar{u} < \frac{s}{1-2I}.
\end{equation}
Given an area of the jet $A_{\textsc{j}}$ and area of the total bottom plate representative of jet separation $A_{\textsc{p}}$, the velocity could decay from the inlet velocity of $u_{\textsc{j}}$ to $A_{\textsc{j}}u_{\textsc{j}} / A_{\textsc{p}}$. This would be the minimum possible flow velocity.  Therefore, taking a maximum laminar flame speed of $s_{\textrm{max}}$, the flame can be stabilised by flame propagation where
\begin{equation}
    u_{\textsc{j}} < \frac{A_{\textsc{p}}s_{\textrm{max}}}{A_{\textsc{j}}(1-2I)}.
\end{equation}
In the present setup, the area ratio $A_{\textsc{p}}/A_{\textsc{j}} = 5.09$, the flame speed of a stoichiometric hydrogen flame with unburned temperature 750\,K and pressure 24\,atm is 8.5\,m/s.  While Lewis number effects are present (as seen in the variation of equivalence ratio through the flame), thermodiffusive effects at these conditions are not expected to accelerate the flame, primarily due to the the stabilising effect of the high unburned temperature of the reactants (see \cite{Howarth2022AnFlames,Howarth2023Thermodiffusively-unstablePoints,Berger2022IntrinsicEnhancement}). Evaluation of the turbulent intensity suggests a maximum value of 22\% reached near the inflow before decaying to approximately 10\% further downstream. Combining all of these conservative estimates, the upper bound on the jet velocity is approximately 77\,m/s.  This would require the bulk flow have reached the theoretical minimum, with the highest feasible turbulent intensity, and the flame burning everywhere at the highest feasible speed, but still results in a mean inflow velocity that is 30\% smaller than the actual value. 

For completeness, the actual mean local flame speed in the core flame region can be calculated as follows. Given a flame surface $\Gamma$, with area $A_{\Gamma}$, a mean local flame speed can be determined by
\begin{equation}\label{eq:consumptionSpeed}
    s = \frac{-1}{(\rho Y_{\Htwo})_{\textsc{j}}A_{\Gamma}}\int_{\Omega}\rho\dot{\omega}_{\Htwo} \textrm{d}\Omega,
\end{equation}
where $(\rho Y_{\Htwo})_{\textsc{j}}$ is the mean density-weighted fuel mass fraction at the inflow. Furthermore, the mean local flame speed in the core flame of interest can be found as
\begin{equation}\label{eq:coreConsumptionSpeed}
    s_{c} = \frac{-1}{(\rho Y_{\Htwo})_{\textsc{j}}A_{\Gamma,c}}\int_{\Omega_{c}}\rho\dot{\omega}_{\Htwo} \textrm{d}\Omega_{c},
\end{equation}
where $\Omega_{c}$ denotes the core zone defined in section \ref{sec:structure:instant} and $A_{\Gamma,c}$ denotes the area of the instantaneous isosurface inside the core zone.

An isosurface can be defined using the progress variable constructed in section \ref{sec:structure:instant}, with $c_{\textsc{f}} = 0.9$; the fuel-based progress variable and particular value of 0.9 was chosen based on obtaining accurate mean local flame speeds in lean hydrogen flames (see \cite{Howarth2022AnFlames,Berger2019CharacteristicFlames}). Calculating a mean local flame speed based using this surface and then ensemble averaging over time gives a value of $s_{c} = 1.75$\,m/s (compare this with the value of 8.5\,m/s used above for $s_{\textrm{max}}$). Using the relation above, and assuming a turbulent intensity of 10\%, the maximum jet velocity allowing for stabilisation through flame propagation would only be 11.3\,m/s (an order-of-magnitude lower than the actual value). Comparing the integral of fuel consumption in the core region to the overall fuel consumption also suggests that 86.7\% of fuel consumption occurs in the core.

In summary, the flame cannot be stabilised by turbulent premixed flame propagation at these conditions, despite the vast majority of the fuel being consumed by deflagration in the core flame zone.



%% file: stabilisation/autoignition.tex
To examine the evolution of ignition events, a line of sight diagnostic has been constructed, taking a one-dimensional integral of a reaction term, i.e.
\begin{equation}\label{eq:LOS}
    \dot{\omega}_{\textsc{los}} = \frac{1}{L_{i}}\int_{0}^{L_{i}}\dot{\omega}_{k}\,\text{d}x_{i},
\end{equation}
where $i$ could be either $x$ or $y$, and $\dot{\omega}_{k}$ could be a species consumption or heat release rate. This removes the possibility of misleading out-of-plane effects, and so allows for easy identification of ignition spots and tracking them as they move through multiple planes. 

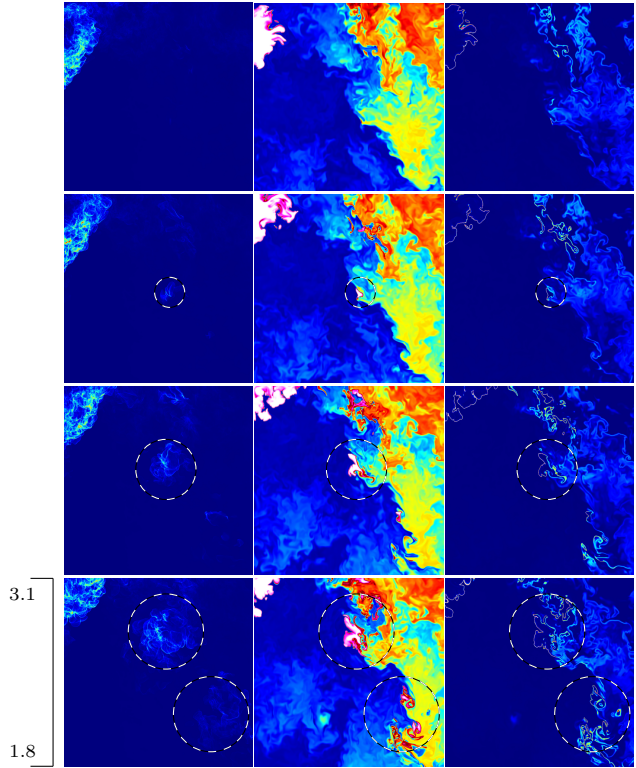
\begin{figure}[ht!]
    \centering
    \begin{tikzpicture}
    \node[inner sep=0pt] (image) at (0,0) {
    \includegraphics[width=25mm]{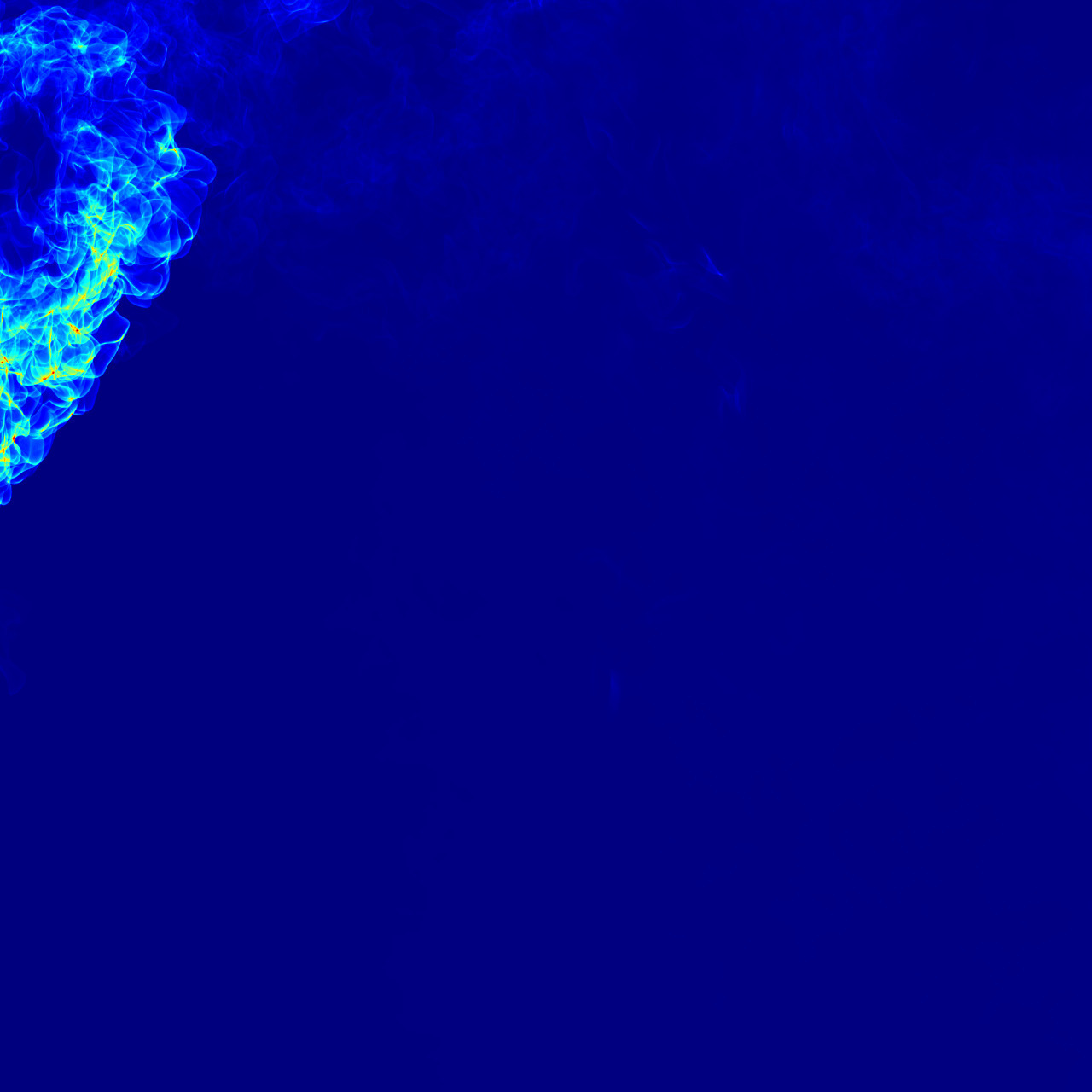}};
    \draw[white] (-14mm,-12.5mm) -- (-14mm,12.5mm);
    \draw[white] (-17mm,-12.5mm) -- (-14mm,-12.5mm);
    \node[anchor=south east,text=white] at (-15mm, -12.5mm) {\scriptsize 1.8};
    \draw[white] (-17mm,12.5mm) -- (-14mm, 12.5mm);
    \node[anchor=north east,text=white] at (-15mm, 12.5mm) {\scriptsize 3.1};
    \end{tikzpicture}
    \hspace{-2mm}
    \begin{tikzpicture}
    \node[inner sep=0pt] (image) at (0,0) {
    \includegraphics[width=25mm]{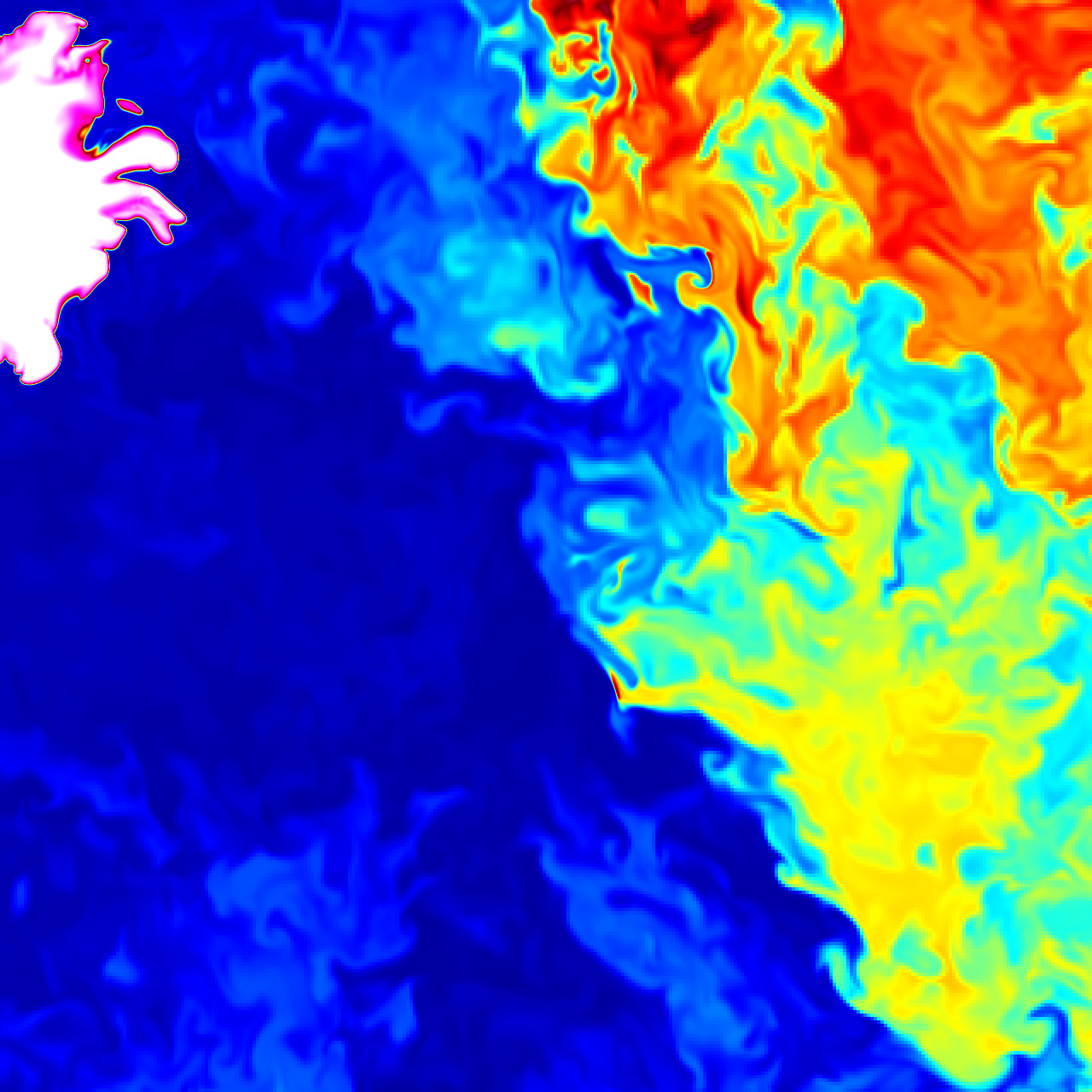}};
    \end{tikzpicture}
    \hspace{-2mm}    
    \begin{tikzpicture}
    \node[inner sep=0pt] (image) at (0,0) {
    \includegraphics[width=25mm]{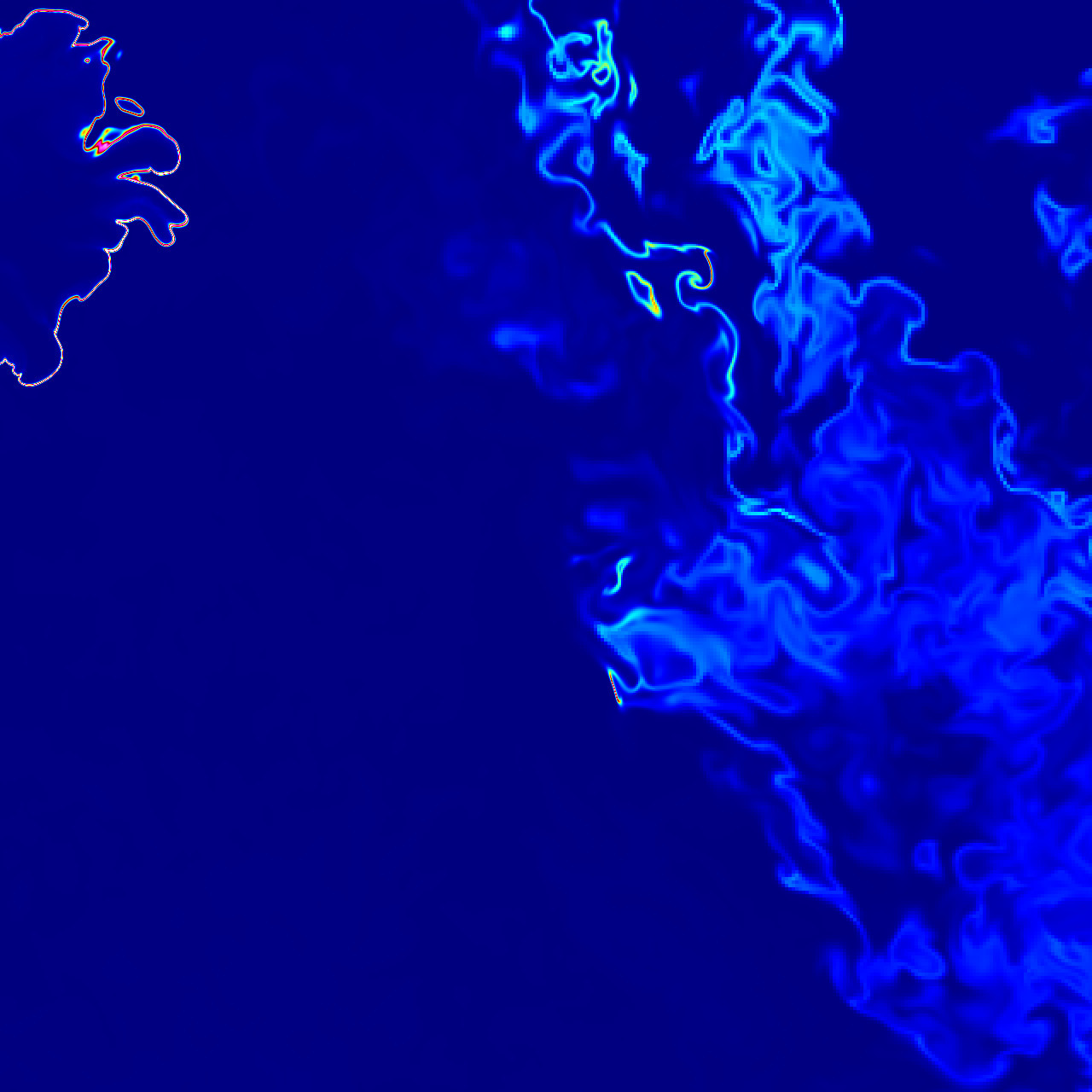}};
    \end{tikzpicture}
    
    \begin{tikzpicture}
    \node[inner sep=0pt] (image) at (0,0) {
    \includegraphics[width=25mm]{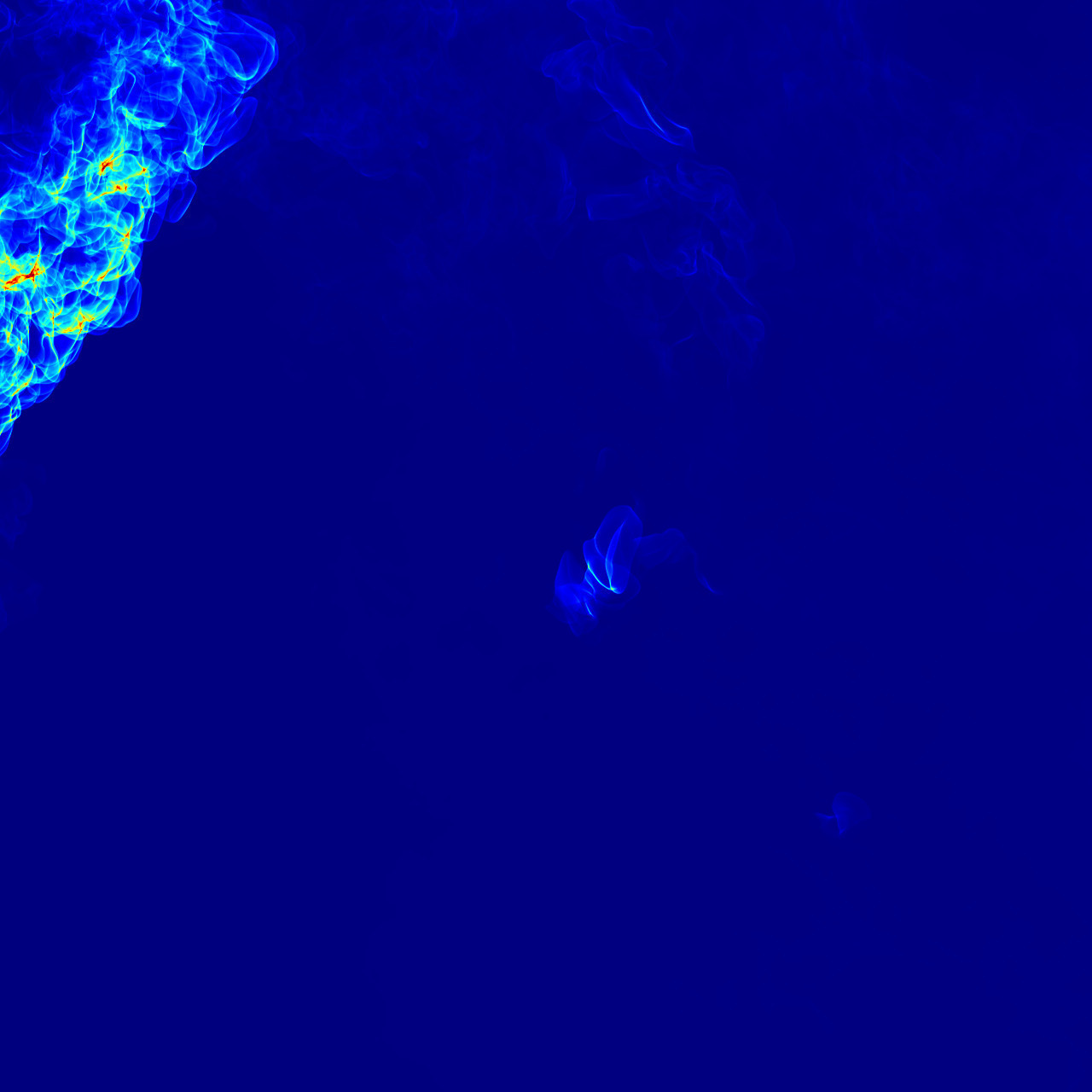}};
    \draw[white] (-14mm,-12.5mm) -- (-14mm,12.5mm);
    \draw[white] (-17mm,-12.5mm) -- (-14mm,-12.5mm);
    \node[anchor=south east,white] at (-15mm, -12.5mm) {\scriptsize 1.8};
    \draw[white] (-17mm,12.5mm) -- (-14mm, 12.5mm);
    \node[anchor=north east,white] at (-15mm, 12.5mm) {\scriptsize 3.1};
    \draw[draw=black] (image.south) ++(1.5mm,12mm) circle (2mm);
    \draw[draw=white, dashed] (image.south) ++(1.5mm,12mm) circle (2mm);
    \end{tikzpicture}
    \hspace{-2mm}
    \begin{tikzpicture}
    \node[inner sep=0pt] (image) at (0,0) {
    \includegraphics[width=25mm]{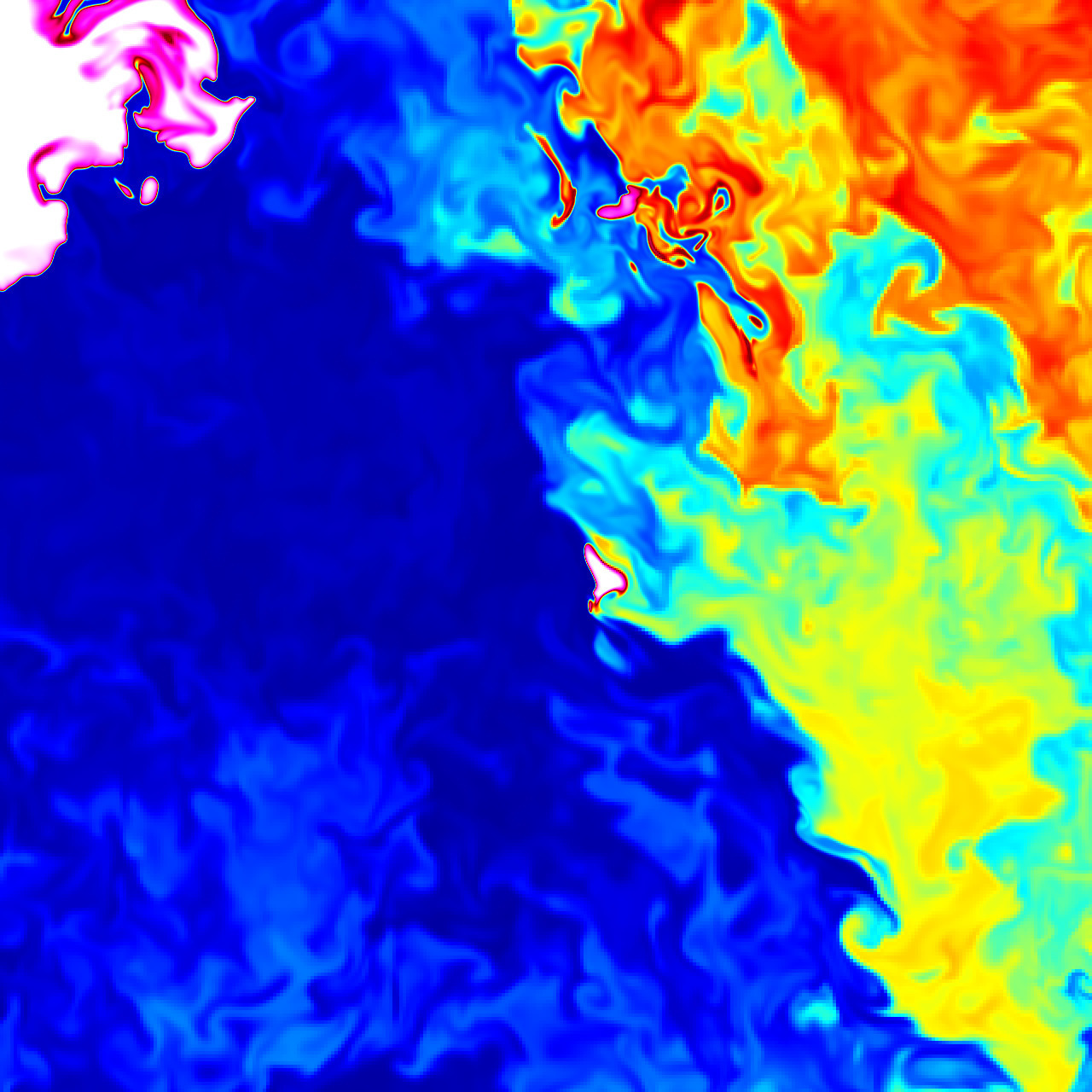}};
    \draw[draw=black] (image.south) ++(1.5mm,12mm) circle (2mm);
    \draw[draw=white,dashed] (image.south) ++(1.5mm,12mm) circle (2mm);
    \end{tikzpicture}
    \hspace{-2mm}
    \begin{tikzpicture}
    \node[inner sep=0pt] (image) at (0,0) {
    \includegraphics[width=25mm]{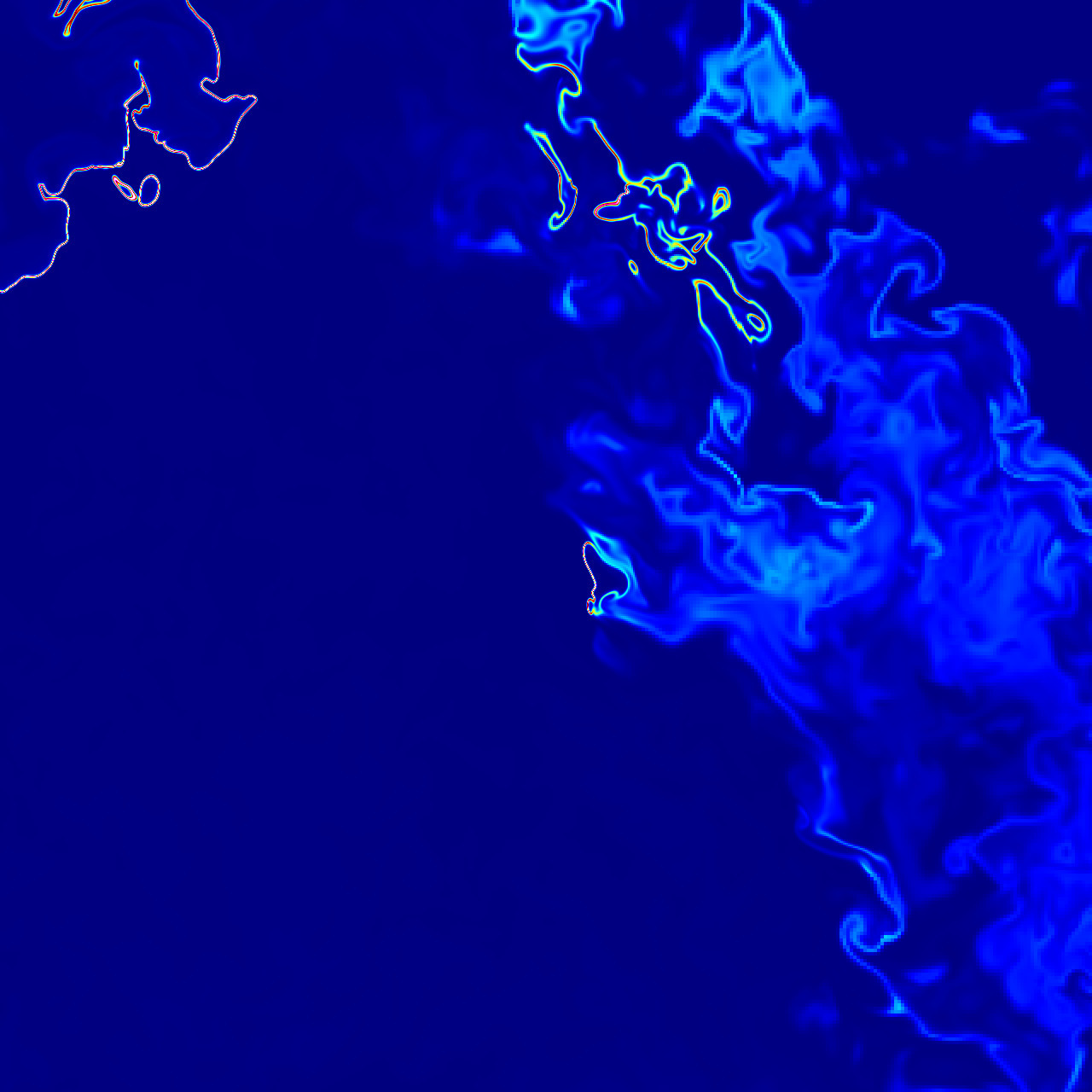}};
    \draw[draw=black] (image.south) ++(1.5mm,12mm) circle (2mm);
    \draw[draw=white,dashed] (image.south) ++(1.5mm,12mm) circle (2mm);
    \end{tikzpicture}
    
    \begin{tikzpicture}
    \node[inner sep=0pt] (image) at (0,0) {
    \includegraphics[width=25mm]{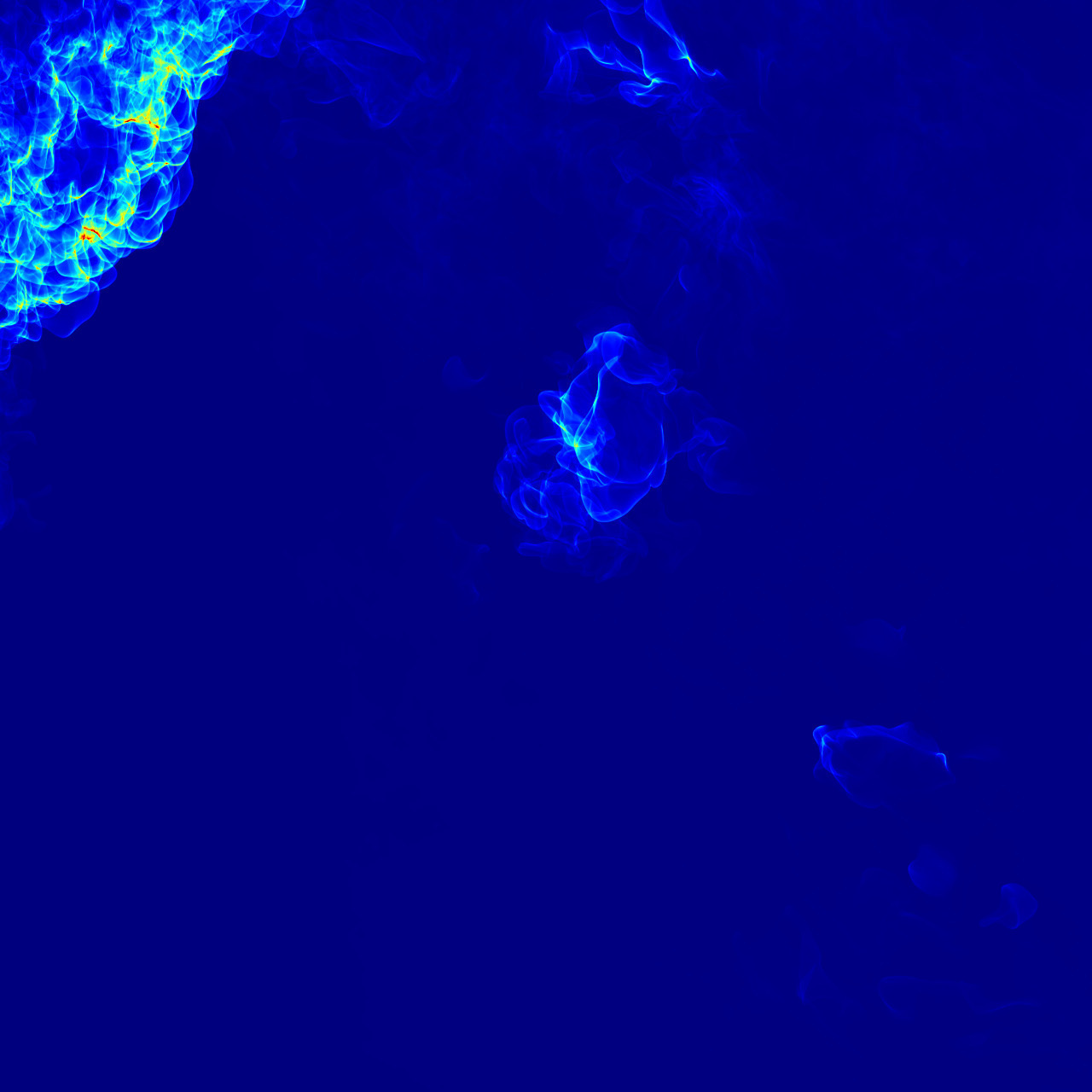}};
    \draw[white] (-14mm,-12.5mm) -- (-14mm,12.5mm);
    \draw[white] (-17mm,-12.5mm) -- (-14mm,-12.5mm);
    \node[anchor=south east,text=white] at (-15mm, -12.5mm) {\scriptsize 1.8};
    \draw[white] (-17mm,12.5mm) -- (-14mm, 12.5mm);
    \node[anchor=north east,text=white] at (-15mm, 12.5mm) {\scriptsize 3.1};
    \draw[draw=black] (image.south) ++(1mm,14mm) circle (4mm);
    \draw[draw=white,dashed] (image.south) ++(1mm,14mm) circle (4mm);
    \end{tikzpicture}
    \hspace{-2mm}
    \begin{tikzpicture}
    \node[inner sep=0pt] (image) at (0,0) {
    \includegraphics[width=25mm]{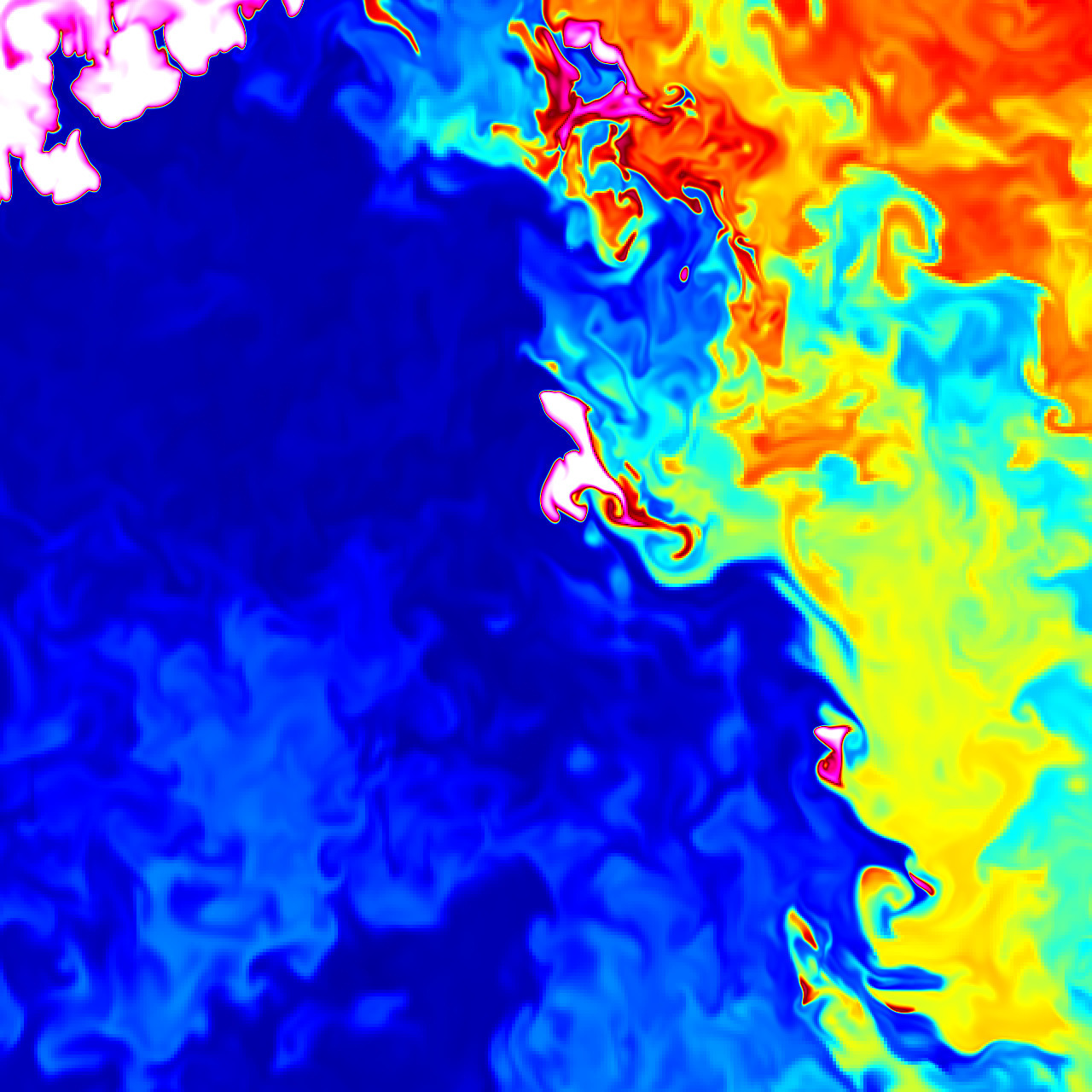}};
    \draw[draw=black] (image.south) ++(1mm,14mm) circle (4mm);
    \draw[draw=white,dashed] (image.south) ++(1mm,14mm) circle (4mm);
    \end{tikzpicture}
    \hspace{-2mm}
    \begin{tikzpicture}
    \node[inner sep=0pt] (image) at (0,0) {
    \includegraphics[width=25mm]{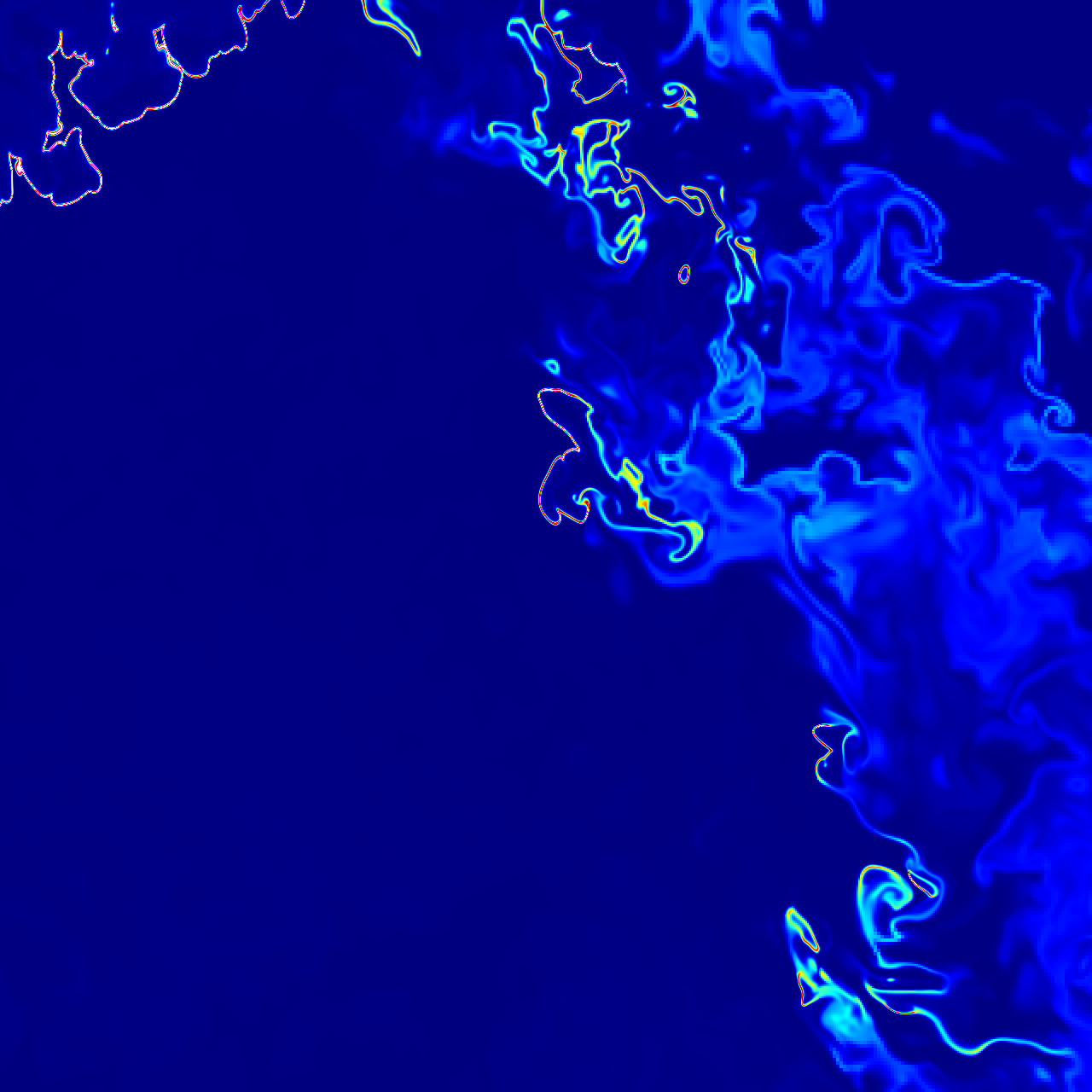}};
    \draw[draw=black] (image.south) ++(1mm,14mm) circle (4mm);
    \draw[draw=white,dashed] (image.south) ++(1mm,14mm) circle (4mm);
    \end{tikzpicture}
    
     \begin{tikzpicture}
    \node[inner sep=0pt] (image) at (0,0) {
    \includegraphics[width=25mm]{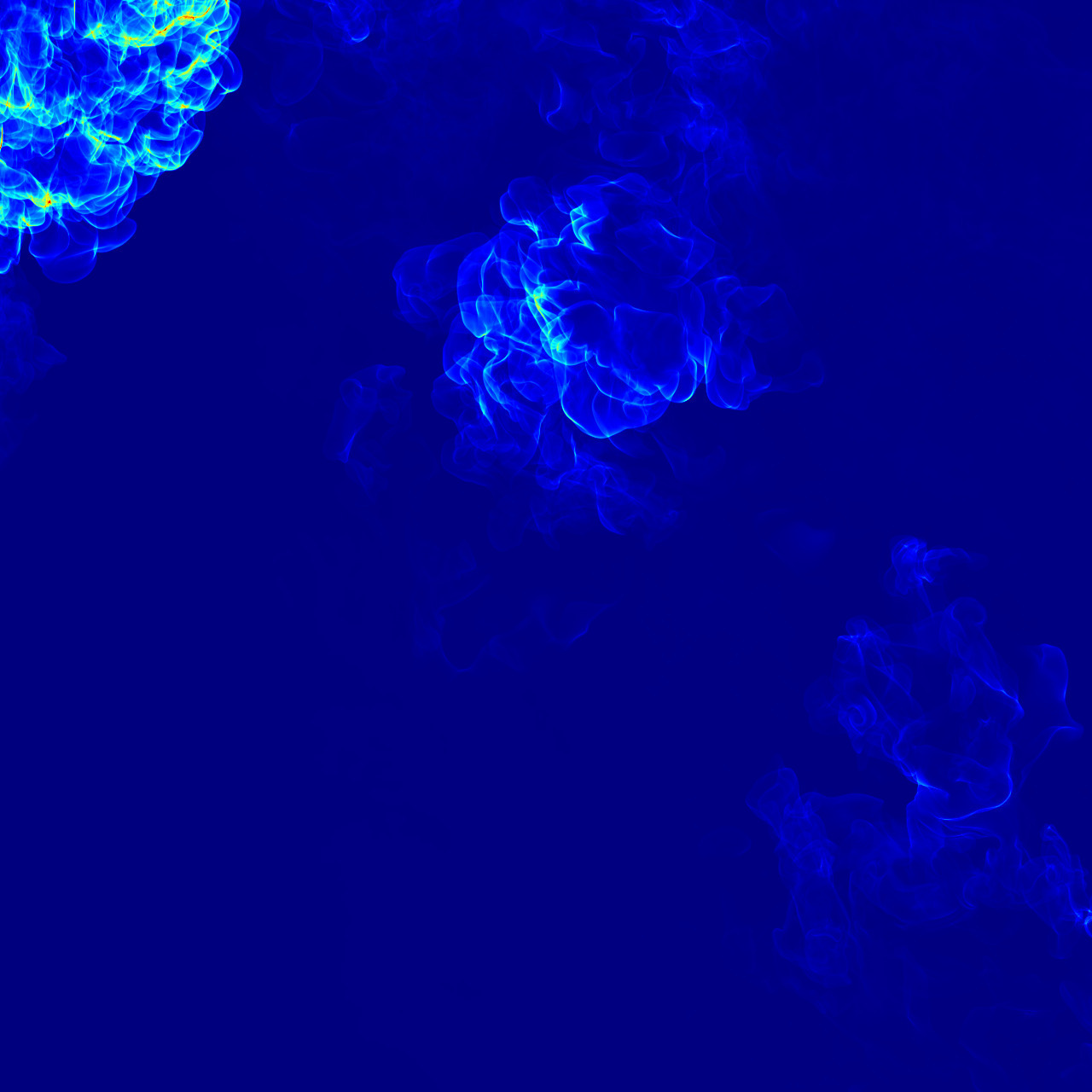}};
    \draw (-14mm,-12.5mm) -- (-14mm,12.5mm);
    \draw(-17mm,-12.5mm) -- (-14mm,-12.5mm);
    \node[anchor=south east] at (-15mm, -12.5mm) {\scriptsize 1.8};
    \draw (-17mm,12.5mm) -- (-14mm, 12.5mm);
    \node[anchor=north east] at (-15mm, 12.5mm) {\scriptsize 3.1};
    \draw[draw=black] (image.south) ++(1mm,18mm) circle (5mm);
    \draw[draw=white,dashed] (image.south) ++(1mm,18mm) circle (5mm);
     \draw[draw=black] (image.south) ++(7mm,7mm) circle (5mm);
     \draw[draw=white,dashed] (image.south) ++(7mm,7mm) circle (5mm);
    \end{tikzpicture}
    \hspace{-2mm}
    \begin{tikzpicture}
    \node[inner sep=0pt] (image) at (0,0) {
    \includegraphics[width=25mm]{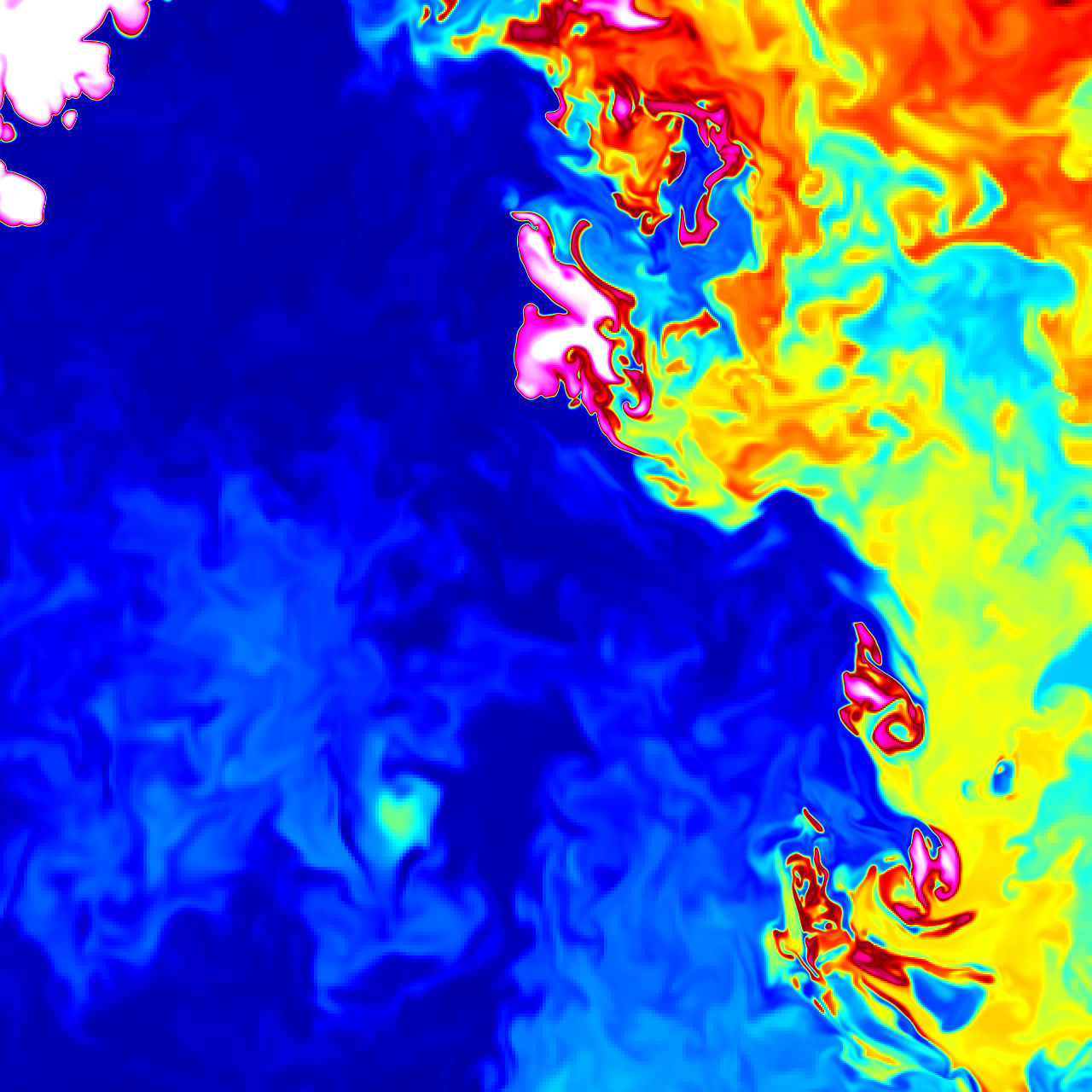}};
    \draw[draw=black] (image.south) ++(1mm,18mm) circle (5mm);
    \draw[draw=white,dashed] (image.south) ++(1mm,18mm) circle (5mm);
     \draw[draw=black] (image.south) ++(7mm,7mm) circle (5mm);
     \draw[draw=white,dashed] (image.south) ++(7mm,7mm) circle (5mm);
    \end{tikzpicture}
    \hspace{-2mm}
    \begin{tikzpicture}
    \node[inner sep=0pt] (image) at (0,0) {
    \includegraphics[width=25mm]{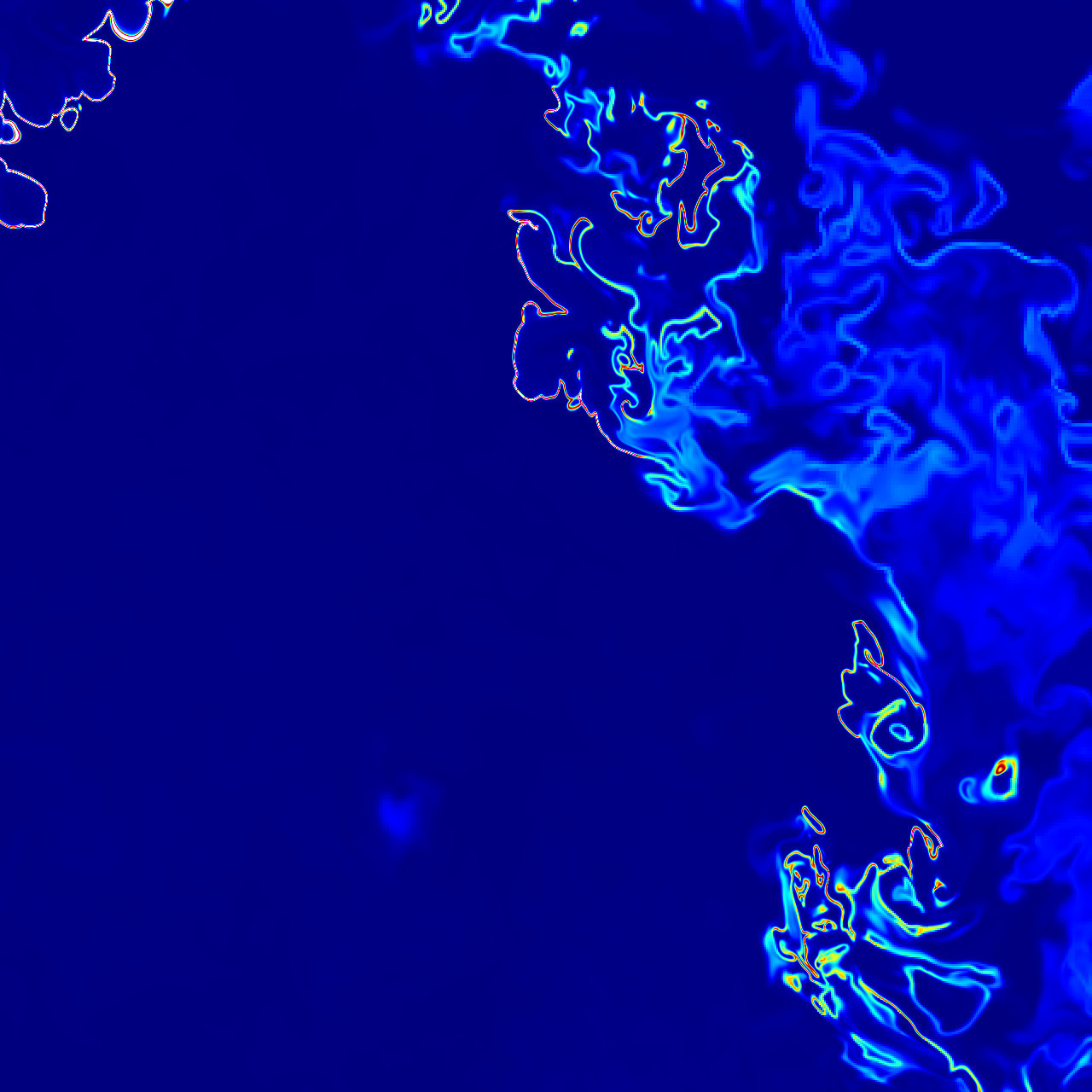}};
    \draw[draw=black] (image.south) ++(1mm,18mm) circle (5mm);
    \draw[draw=white,dashed] (image.south) ++(1mm,18mm) circle (5mm);
     \draw[draw=black] (image.south) ++(7mm,7mm) circle (5mm);
     \draw[draw=white,dashed] (image.south) ++(7mm,7mm) circle (5mm);
    \end{tikzpicture}
    \caption{Evolution of an igniting kernel from formation to flame sheet. The circles point to the formation of the initial kernel and the secondary kernels that form in the KH roller.  Time difference between slices $\Delta t = 10.3$\,$\mu$s. }
    \label{fig:LOSkernel}
\end{figure}

\begin{figure}[ht!]
    \centering

     \begin{tikzpicture}
    \node[inner sep=0pt] (image) at (0,0) {
    \includegraphics[width=25mm]{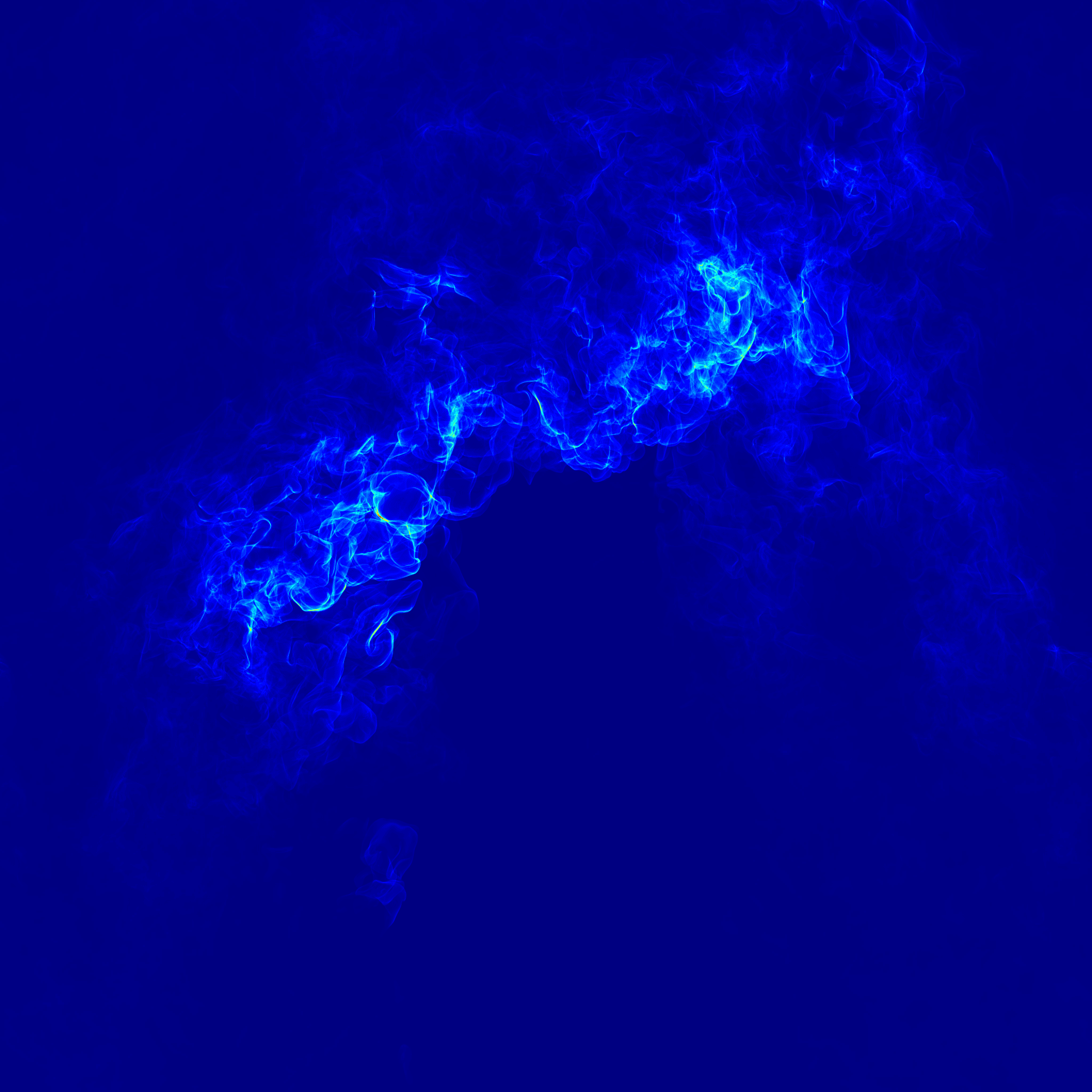}};
    \draw[white] (-14mm,-12.5mm) -- (-14mm,12.5mm);
    \draw[white] (-17mm,-12.5mm) -- (-14mm,-12.5mm);
    \node[anchor=south east,white] at (-15mm, -12.5mm) {\scriptsize 2.4};
    \draw[white] (-17mm,12.5mm) -- (-14mm, 12.5mm);
    \node[anchor=north east,white] at (-15mm, 12.5mm) {\scriptsize 4.4};
    \draw[rotate=60] (image.south) ++(4.5mm,6.5mm) ellipse (7.5mm and 3mm);
    \draw[rotate=60,draw=white,dashed] (image.south) ++(4.5mm,6.5mm) ellipse (7.5mm and 3mm);
    \end{tikzpicture}
    \hspace{-2.3mm}
    \begin{tikzpicture}
    \node[inner sep=0pt] (image) at (0,0) {
    \includegraphics[width=25mm]{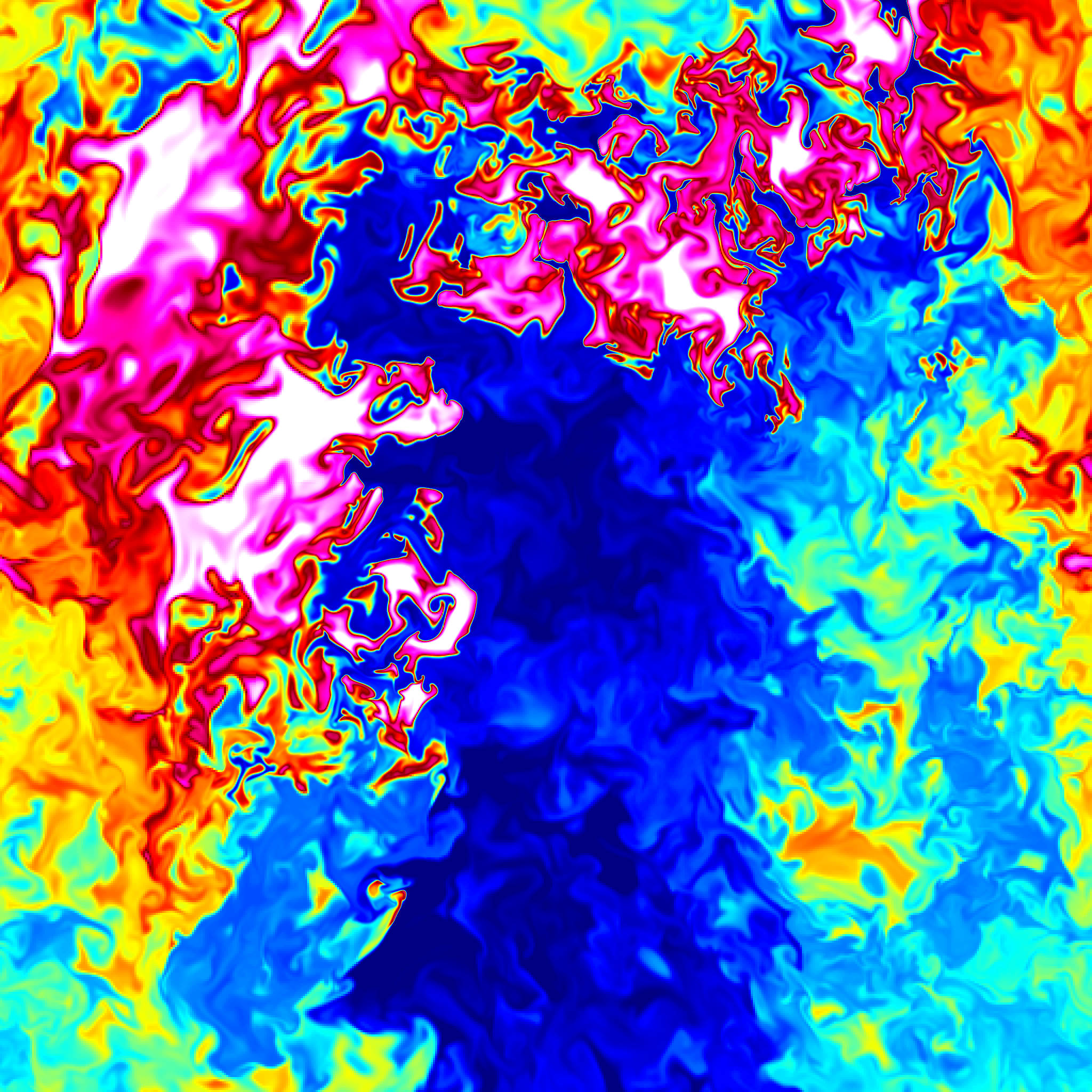}};
    \draw[rotate=60] (image.south) ++(4.5mm,6.5mm) ellipse (7.5mm and 3mm);
    \draw[rotate=60,draw=white,dashed] (image.south) ++(4.5mm,6.5mm) ellipse (7.5mm and 3mm);
    \end{tikzpicture}
    \hspace{-2.3mm}
    \begin{tikzpicture}
    \node[inner sep=0pt] (image) at (0,0) {
    \includegraphics[width=25mm]{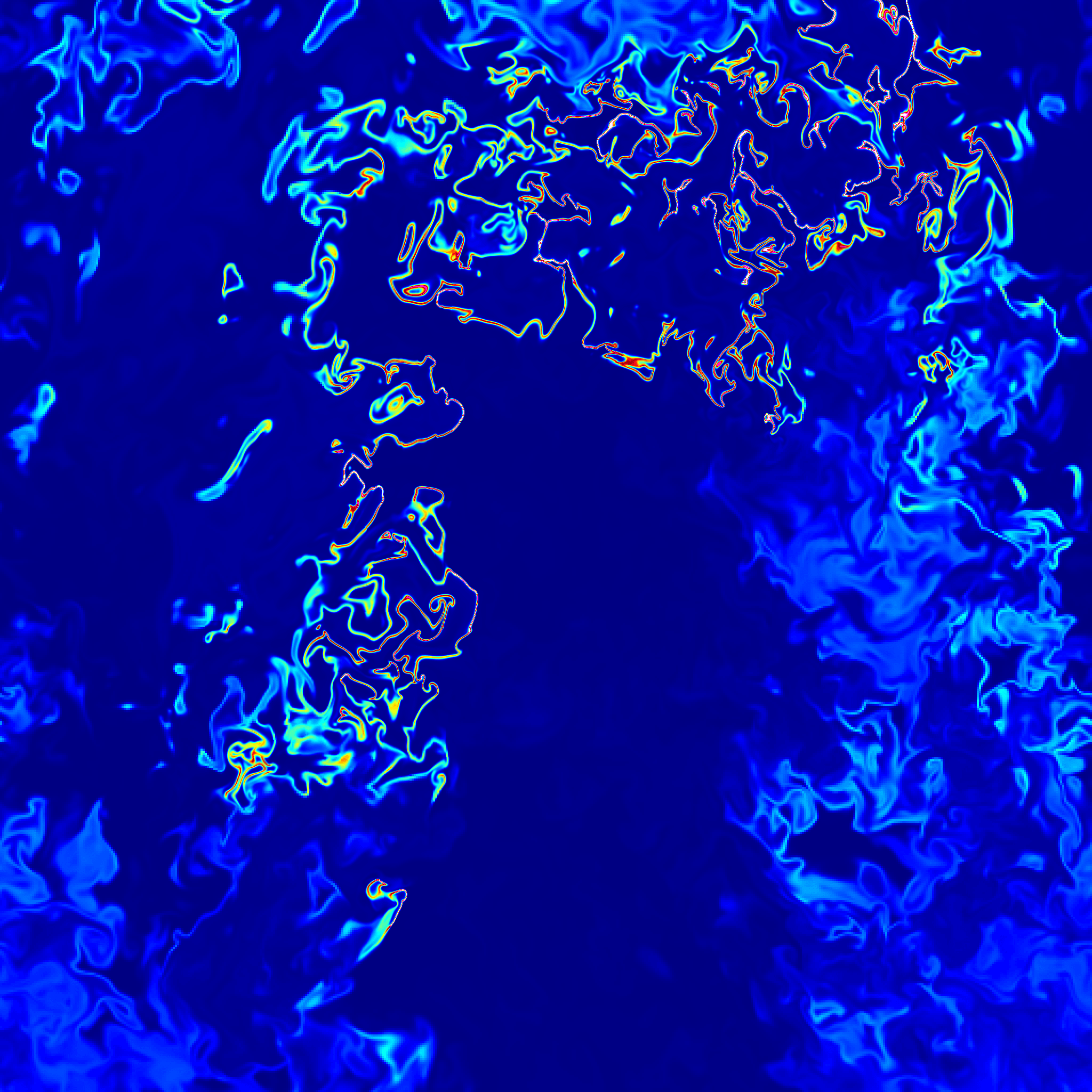}};
    \draw[rotate=60] (image.south) ++(4.5mm,6.5mm) ellipse (7.5mm and 3mm);
    \draw[rotate=60,draw=white,dashed] (image.south) ++(4.5mm,6.5mm) ellipse (7.5mm and 3mm);
    \end{tikzpicture}

    \begin{tikzpicture}
    \node[inner sep=0pt] (image) at (0,0) {
    \includegraphics[width=25mm]{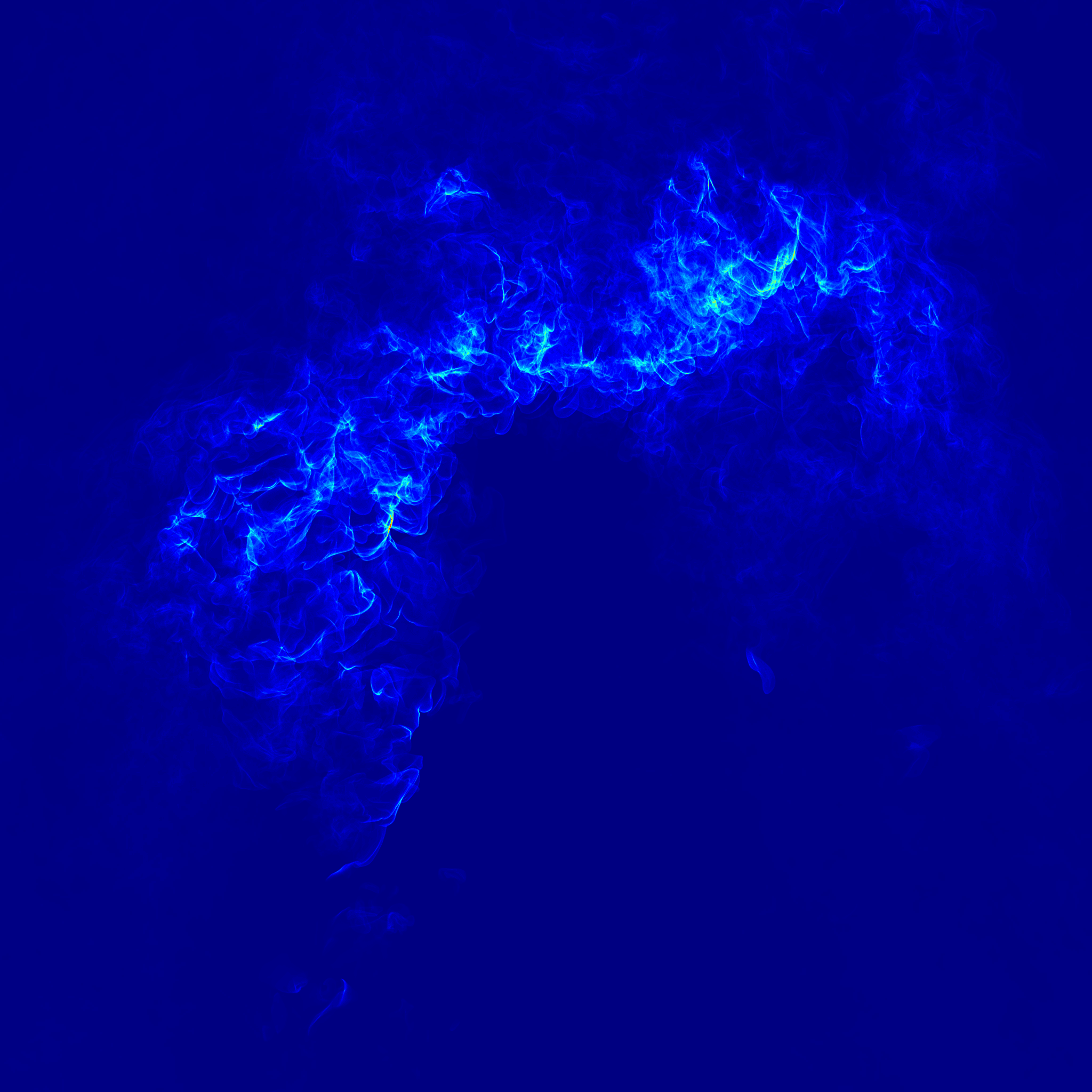}};
    \draw[white] (-14mm,-12.5mm) -- (-14mm,12.5mm);
    \draw[white] (-17mm,-12.5mm) -- (-14mm,-12.5mm);
    \node[anchor=south east,white] at (-15mm, -12.5mm) {\scriptsize 2.4};
    \draw[white] (-17mm,12.5mm) -- (-14mm, 12.5mm);
    \node[anchor=north east,white] at (-15mm, 12.5mm) {\scriptsize 4.4};
    \draw[rotate=60] (image.south) ++(4.5mm,6.5mm) ellipse (7.5mm and 3mm);
    \draw[rotate=60,draw=white,dashed] (image.south) ++(4.5mm,6.5mm) ellipse (7.5mm and 3mm);
    \end{tikzpicture}
    \hspace{-2.3mm}
    \begin{tikzpicture}
    \node[inner sep=0pt] (image) at (0,0) {
    \includegraphics[width=25mm]{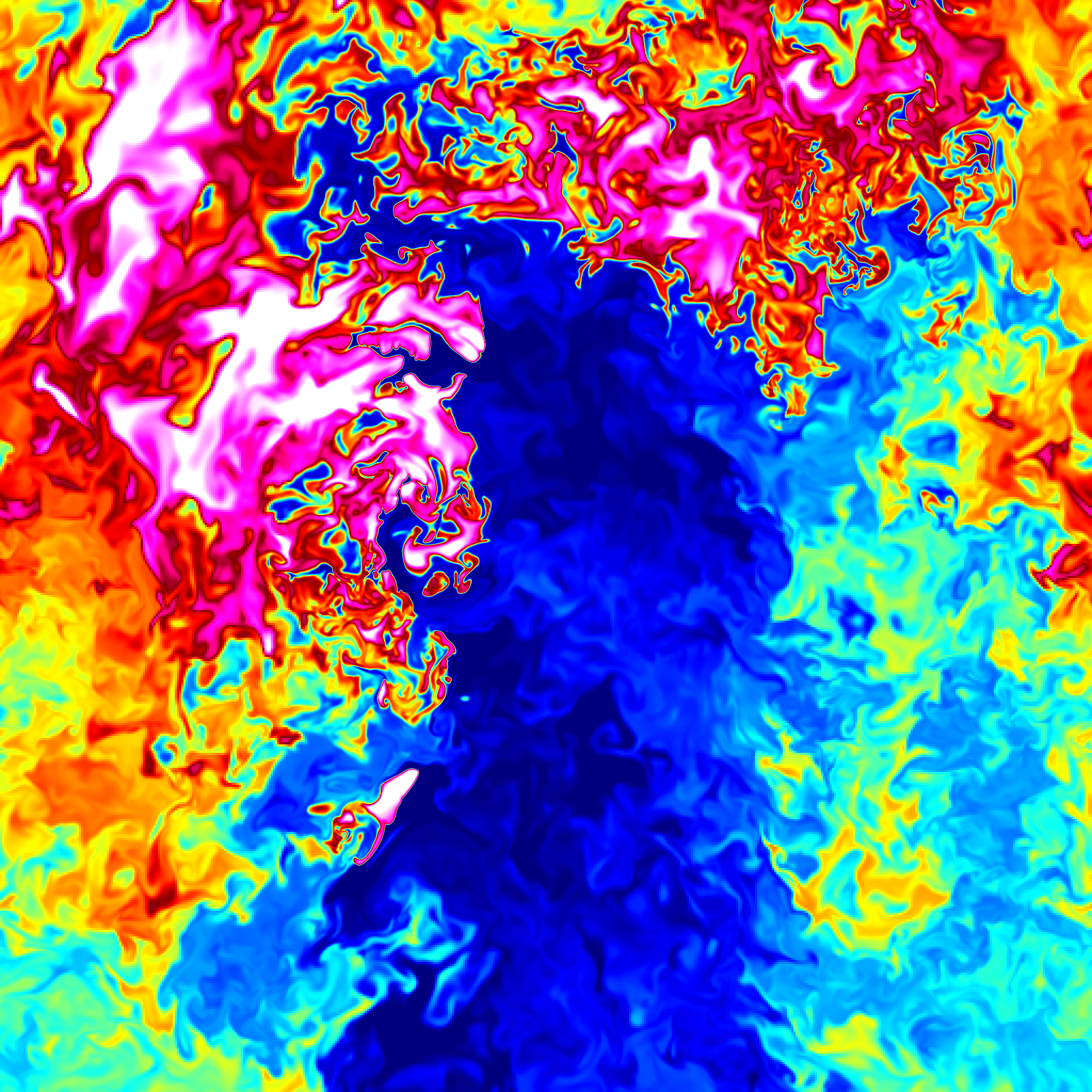}};
    \draw[rotate=60] (image.south) ++(4.5mm,6.5mm) ellipse (7.5mm and 3mm);
    \draw[rotate=60,draw=white,dashed] (image.south) ++(4.5mm,6.5mm) ellipse (7.5mm and 3mm);
    \end{tikzpicture}
    \hspace{-2.3mm}
    \begin{tikzpicture}
    \node[inner sep=0pt] (image) at (0,0) {
    \includegraphics[width=25mm]{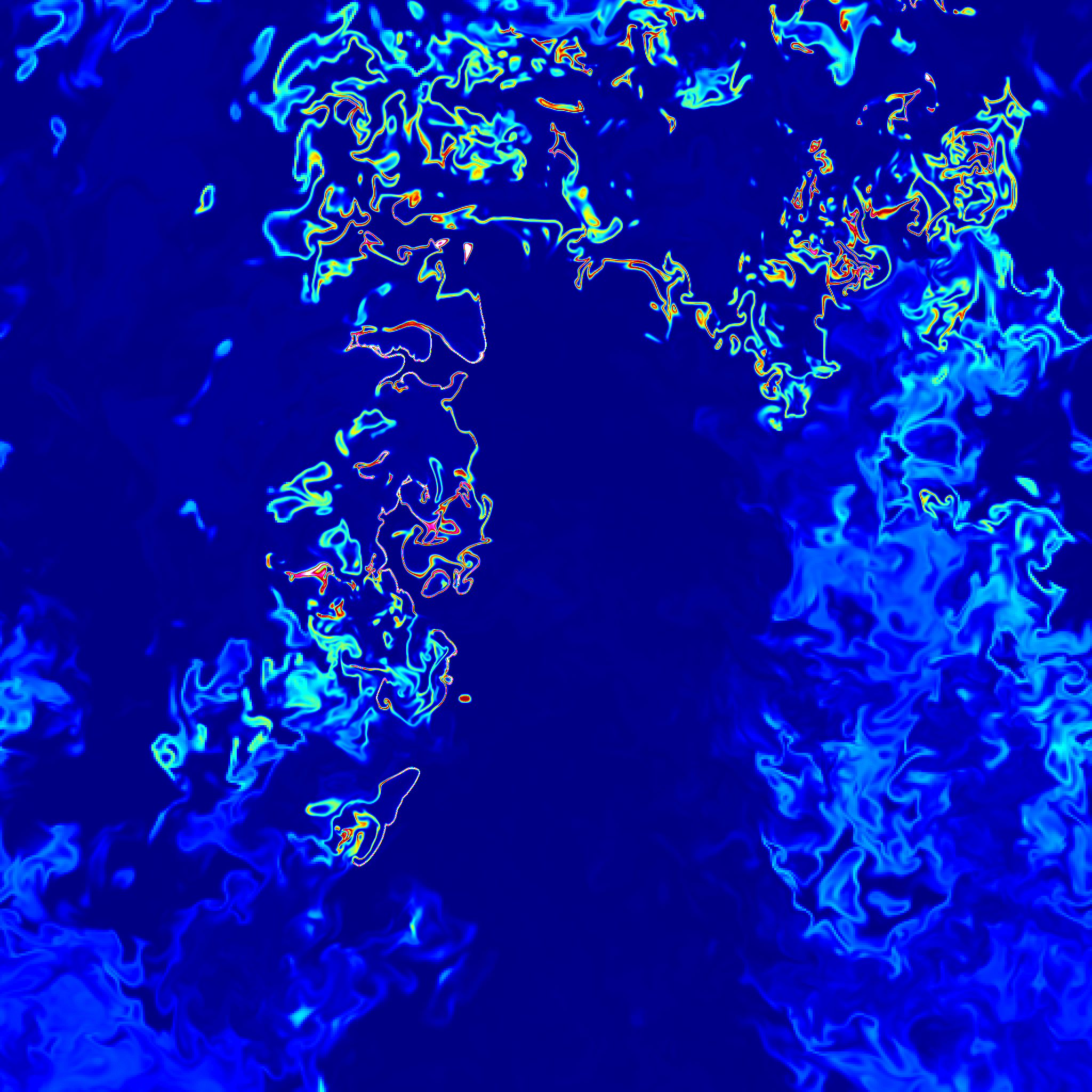}};
    \draw[rotate=60] (image.south) ++(4.5mm,6.5mm) ellipse (7.5mm and 3mm);
    \draw[rotate=60,draw=white,dashed] (image.south) ++(4.5mm,6.5mm) ellipse (7.5mm and 3mm);
    \end{tikzpicture}
    
    \begin{tikzpicture}
    \node[inner sep=0pt] (image) at (0,0) {
    \includegraphics[width=25mm]{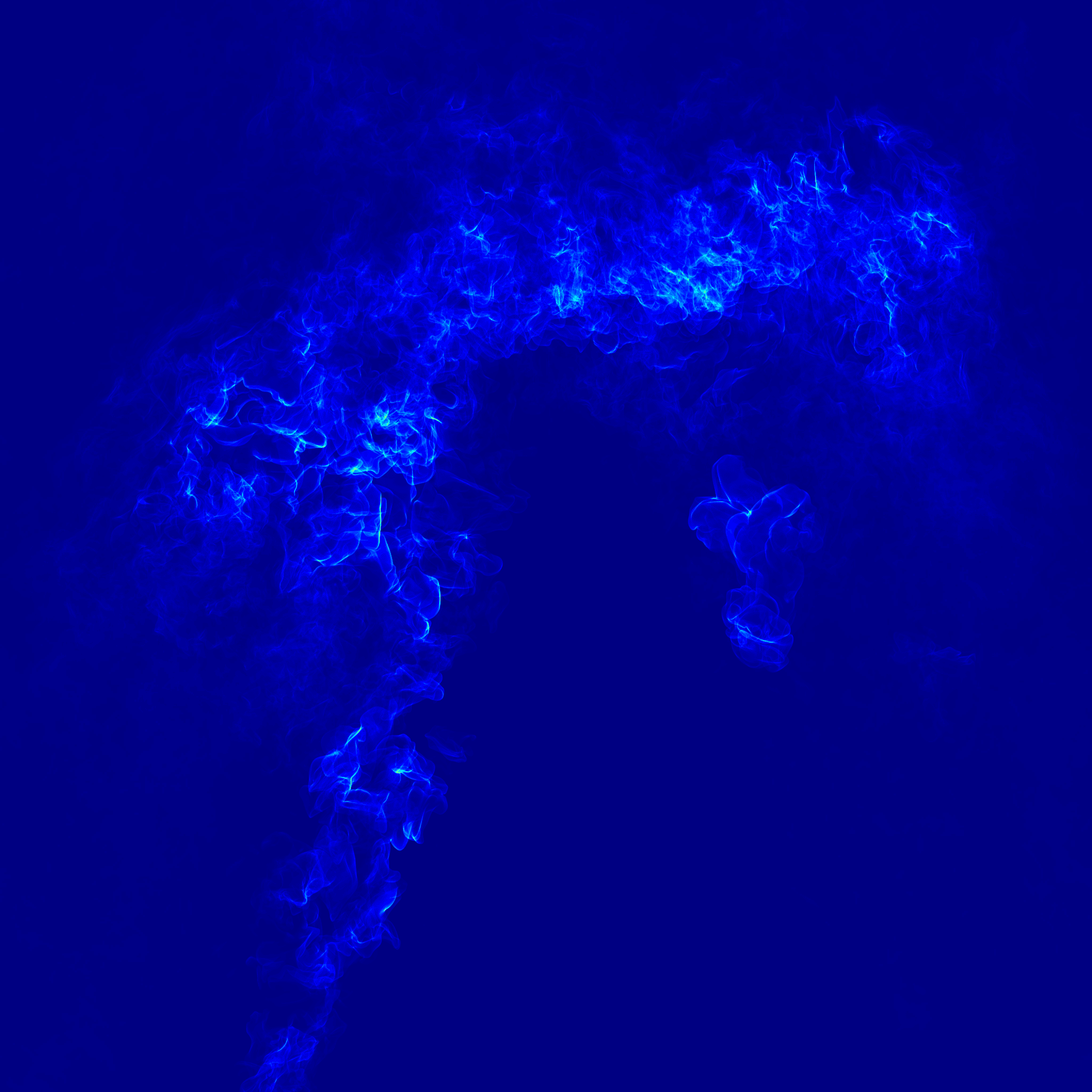}};
    \draw[white] (-14mm,-12.5mm) -- (-14mm,12.5mm);
    \draw[white] (-17mm,-12.5mm) -- (-14mm,-12.5mm);
    \node[anchor=south east,white] at (-15mm, -12.5mm) {\scriptsize 2.4};
    \draw[white] (-17mm,12.5mm) -- (-14mm, 12.5mm);
    \node[anchor=north east,white] at (-15mm, 12.5mm) {\scriptsize 4.4};
    \draw[rotate=60] (image.south) ++(4.5mm,6.5mm) ellipse (7.5mm and 3mm);
    \draw[rotate=60,draw=white,dashed] (image.south) ++(4.5mm,6.5mm) ellipse (7.5mm and 3mm);
    \end{tikzpicture}
    \hspace{-2.3mm}
    \begin{tikzpicture}
    \node[inner sep=0pt] (image) at (0,0) {
    \includegraphics[width=25mm]{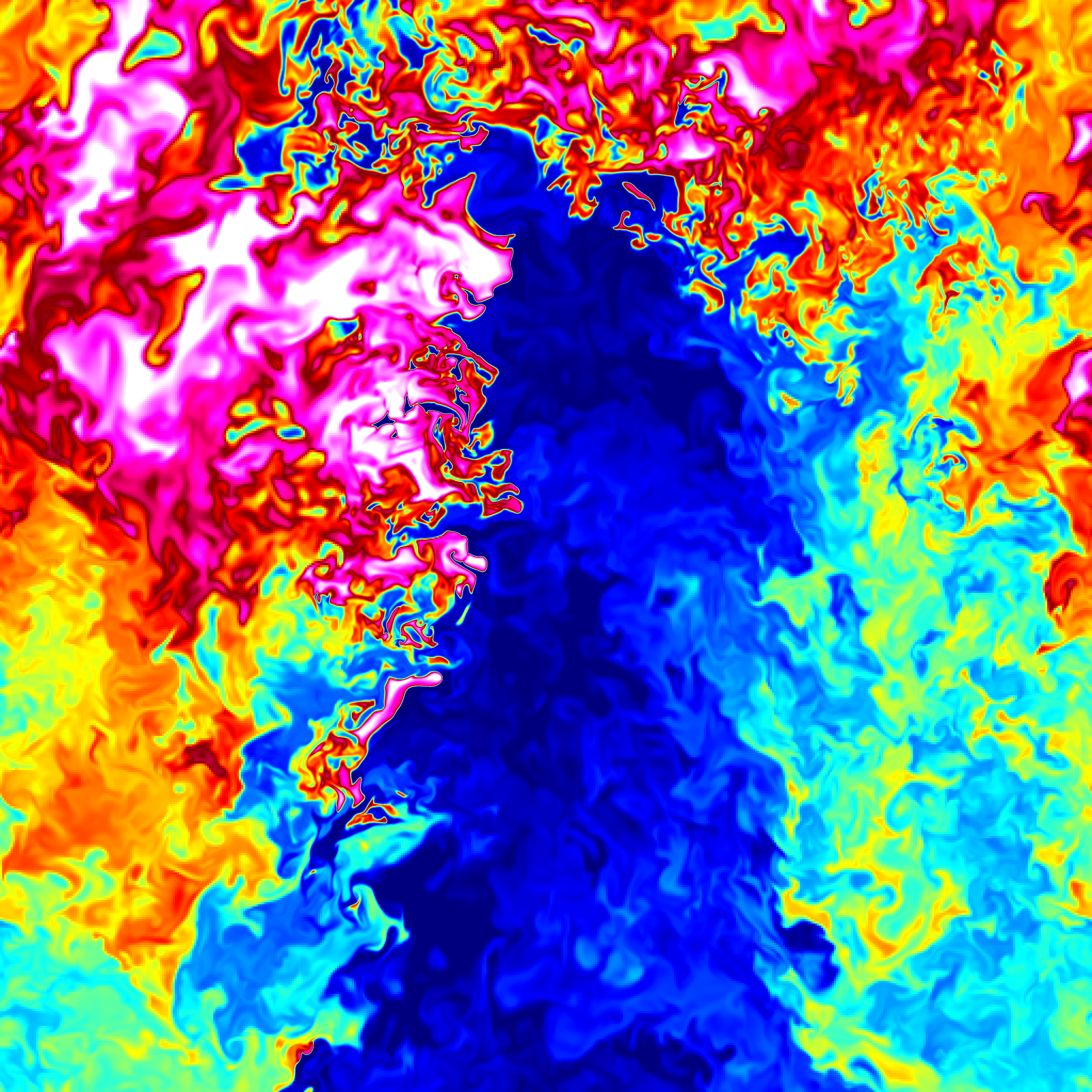}};
    \draw[rotate=60] (image.south) ++(4.5mm,6.5mm) ellipse (7.5mm and 3mm);
    \draw[rotate=60,draw=white,dashed] (image.south) ++(4.5mm,6.5mm) ellipse (7.5mm and 3mm);
    \end{tikzpicture}
    \hspace{-2.3mm}
    \begin{tikzpicture}
    \node[inner sep=0pt] (image) at (0,0) {
    \includegraphics[width=25mm]{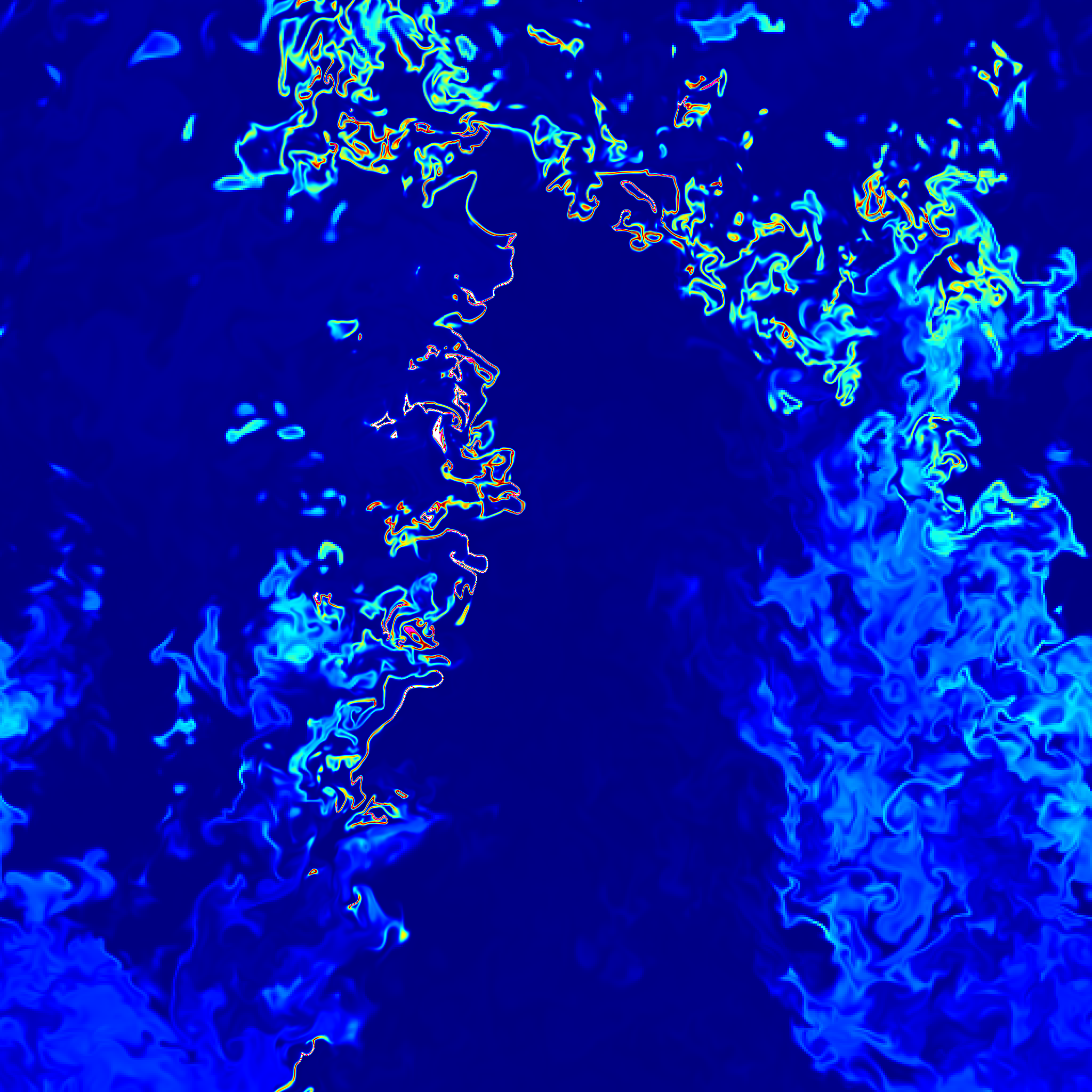}};
    \draw[rotate=60] (image.south) ++(4.5mm,6.5mm) ellipse (7.5mm and 3mm);
    \draw[rotate=60,draw=white,dashed] (image.south) ++(4.5mm,6.5mm) ellipse (7.5mm and 3mm);
    \end{tikzpicture}
    
    \begin{tikzpicture}
    \node[inner sep=0pt] (image) at (0,0) {
    \includegraphics[width=25mm]{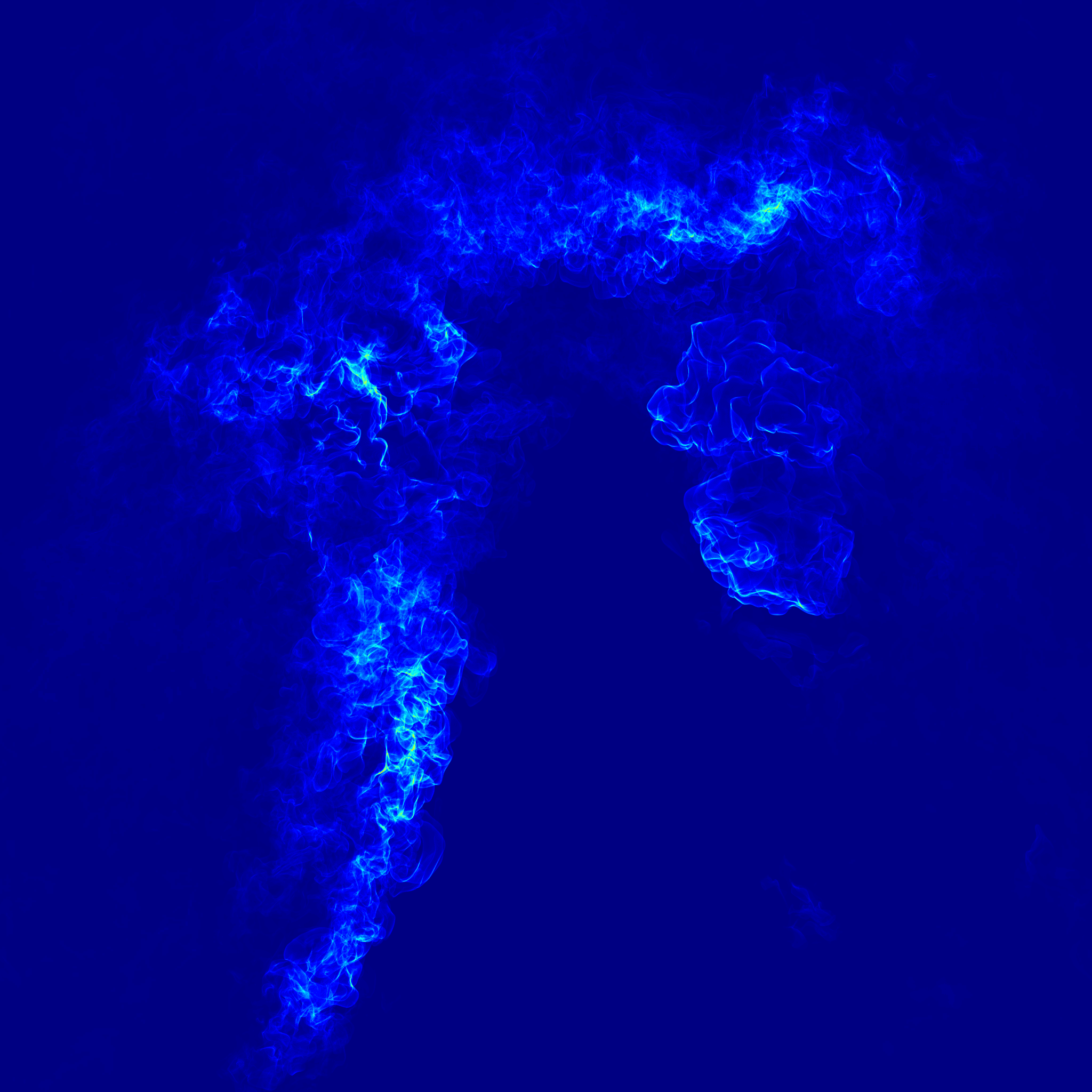}};
    \draw[white] (-14mm,-12.5mm) -- (-14mm,12.5mm);
    \draw[white] (-17mm,-12.5mm) -- (-14mm,-12.5mm);
    \node[anchor=south east,white] at (-15mm, -12.5mm) {\scriptsize 2.4};
    \draw[white] (-17mm,12.5mm) -- (-14mm, 12.5mm);
    \node[anchor=north east,white] at (-15mm, 12.5mm) {\scriptsize 4.4};
    \draw[rotate=60] (image.south) ++(4.5mm,6.5mm) ellipse (7.5mm and 3mm);
    \draw[rotate=60,draw=white,dashed] (image.south) ++(4.5mm,6.5mm) ellipse (7.5mm and 3mm);
    \end{tikzpicture}
    \hspace{-2.3mm}
    \begin{tikzpicture}
    \node[inner sep=0pt] (image) at (0,0) {
    \includegraphics[width=25mm]{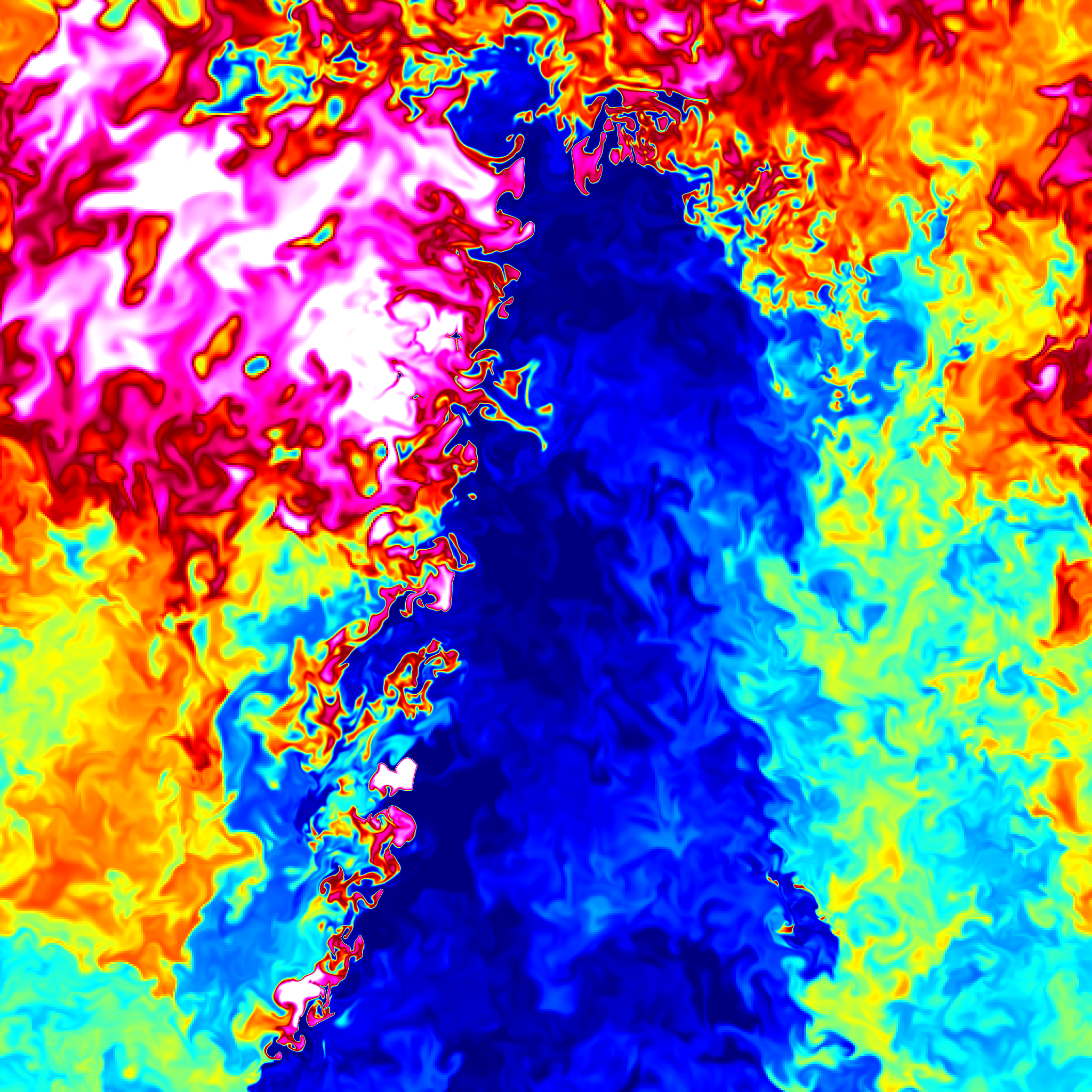}};
    \draw[rotate=60] (image.south) ++(4.5mm,6.5mm) ellipse (7.5mm and 3mm);
    \draw[rotate=60,draw=white,dashed] (image.south) ++(4.5mm,6.5mm) ellipse (7.5mm and 3mm);
    \end{tikzpicture}
    \hspace{-2.3mm}
    \begin{tikzpicture}
    \node[inner sep=0pt] (image) at (0,0) {
    \includegraphics[width=25mm]{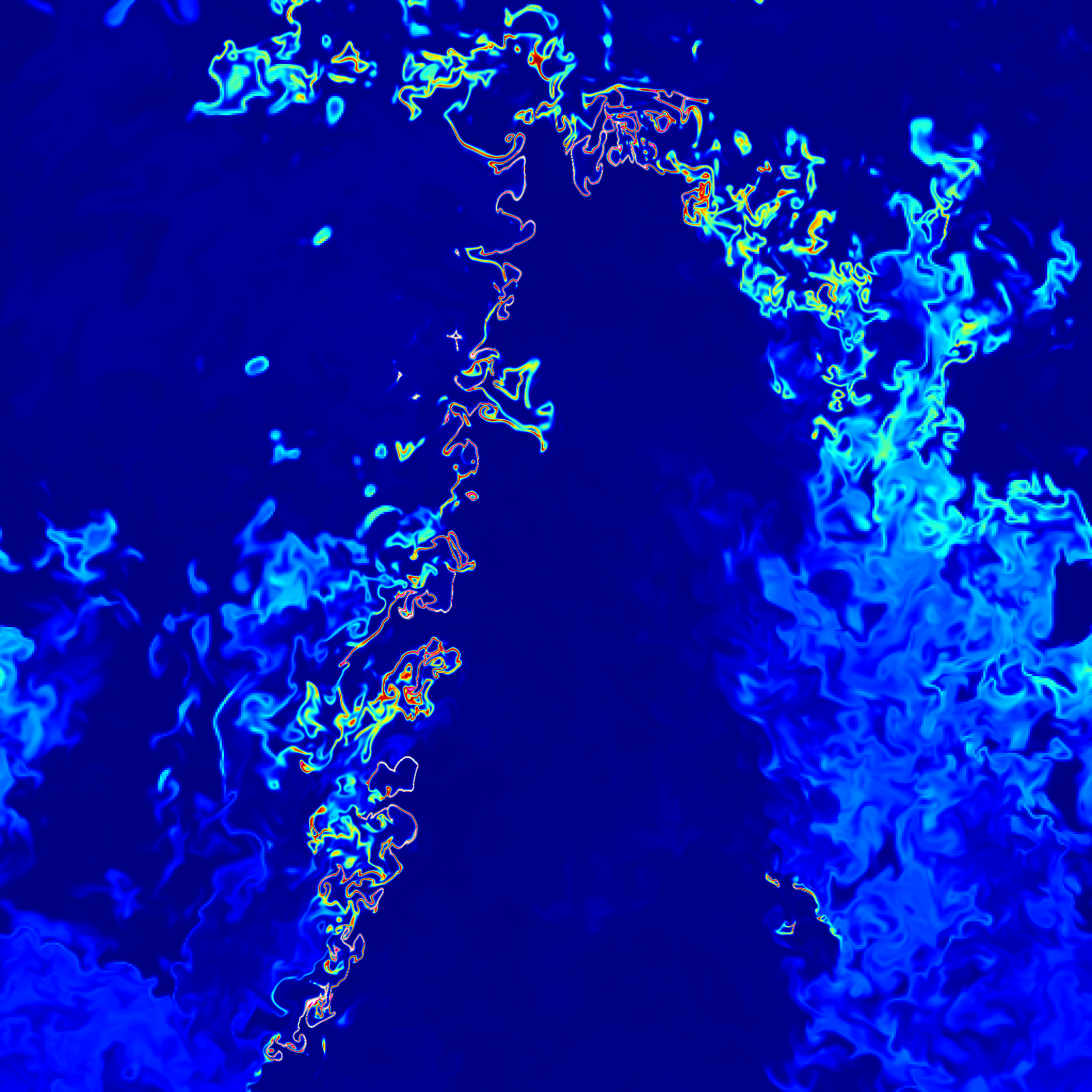}};
    \draw[rotate=60] (image.south) ++(4.5mm,6.5mm) ellipse (7.5mm and 3mm);
    \draw[rotate=60,draw=white,dashed] (image.south) ++(4.5mm,6.5mm) ellipse (7.5mm and 3mm);
    \end{tikzpicture}

    \begin{tikzpicture}
    \node[inner sep=0pt] (image) at (0,0) {
    \includegraphics[width=25mm]{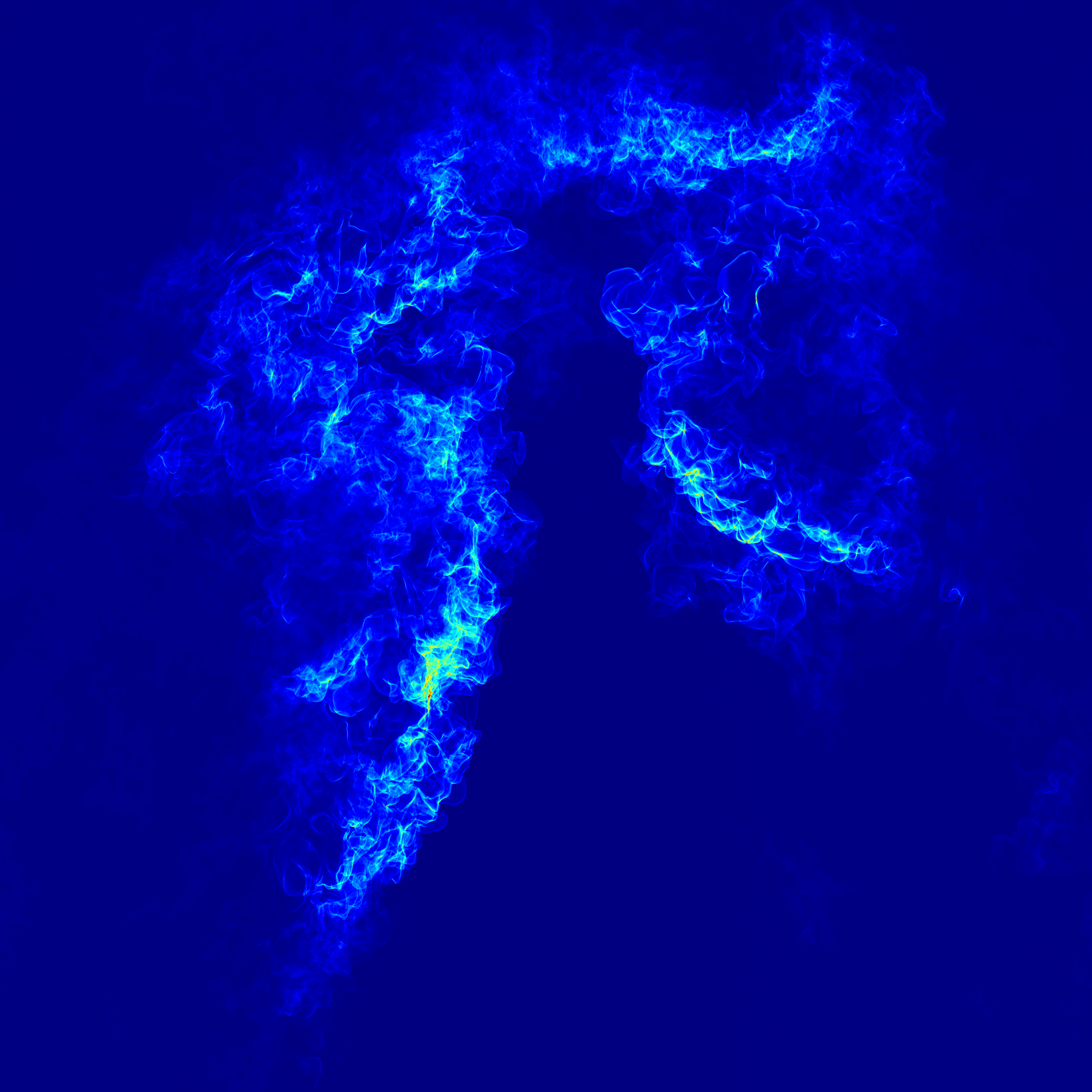}};
    \draw[white] (-14mm,-12.5mm) -- (-14mm,12.5mm);
    \draw[white] (-17mm,-12.5mm) -- (-14mm,-12.5mm);
    \node[anchor=south east,white] at (-15mm, -12.5mm) {\scriptsize 2.4};
    \draw[white] (-17mm,12.5mm) -- (-14mm, 12.5mm);
    \node[anchor=north east,white] at (-15mm, 12.5mm) {\scriptsize 4.4};
    \draw[rotate=60] (image.south) ++(4.5mm,6.5mm) ellipse (7.5mm and 3mm);
    \draw[rotate=60,draw=white,dashed] (image.south) ++(4.5mm,6.5mm) ellipse (7.5mm and 3mm);
    \end{tikzpicture}
    \hspace{-2.3mm}
    \begin{tikzpicture}
    \node[inner sep=0pt] (image) at (0,0) {
    \includegraphics[width=25mm]{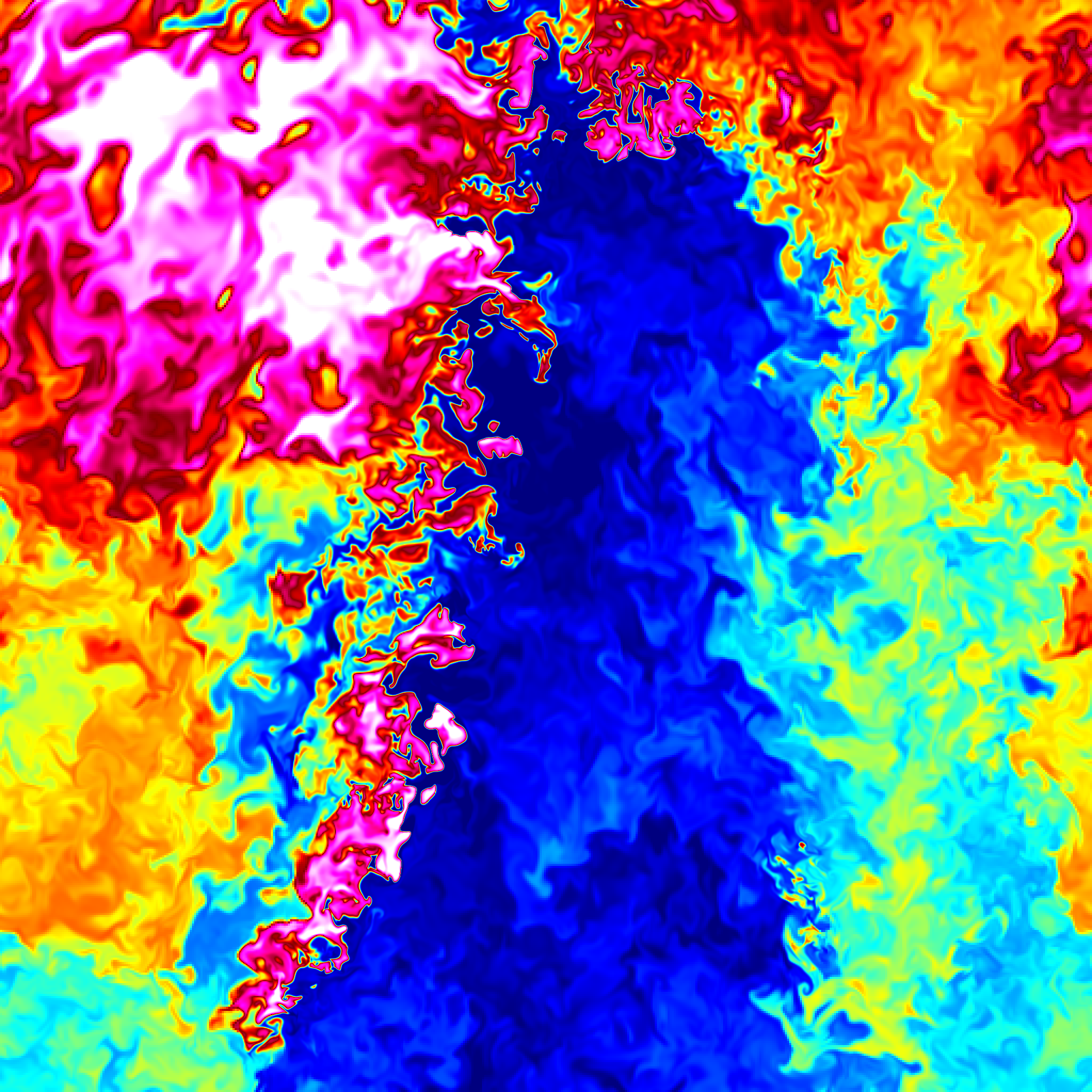}};
    \draw[rotate=60] (image.south) ++(4.5mm,6.5mm) ellipse (7.5mm and 3mm);
    \draw[rotate=60,draw=white,dashed] (image.south) ++(4.5mm,6.5mm) ellipse (7.5mm and 3mm);
    \end{tikzpicture}
    \hspace{-2.3mm}
    \begin{tikzpicture}
    \node[inner sep=0pt] (image) at (0,0) {
    \includegraphics[width=25mm]{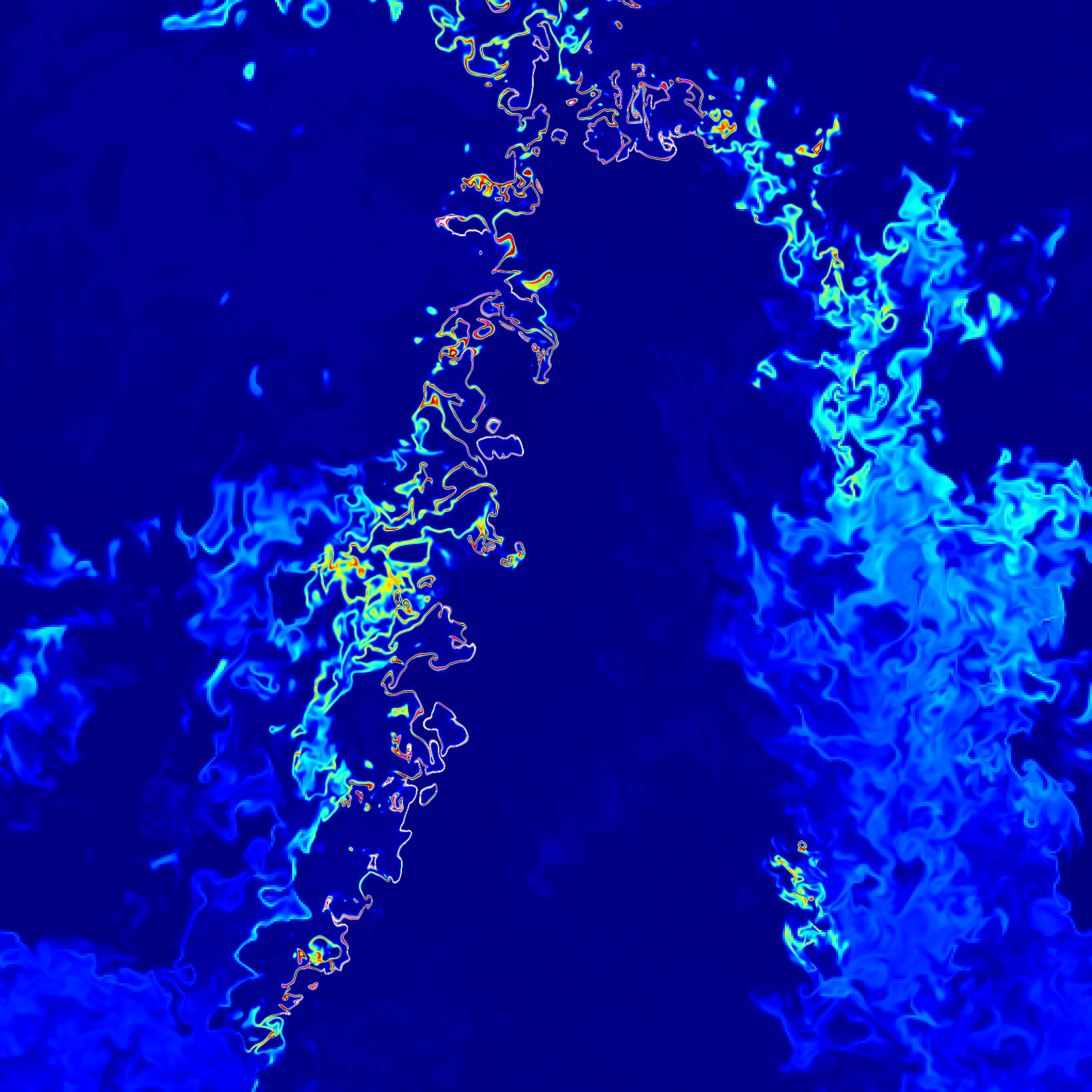}};
    \draw[rotate=60] (image.south) ++(4.5mm,6.5mm) ellipse (7.5mm and 3mm);
    \draw[rotate=60,draw=white,dashed] (image.south) ++(4.5mm,6.5mm) ellipse (7.5mm and 3mm);
    \end{tikzpicture}

    \begin{tikzpicture}
    \node[inner sep=0pt] (image) at (0,0) {
    \includegraphics[width=25mm]{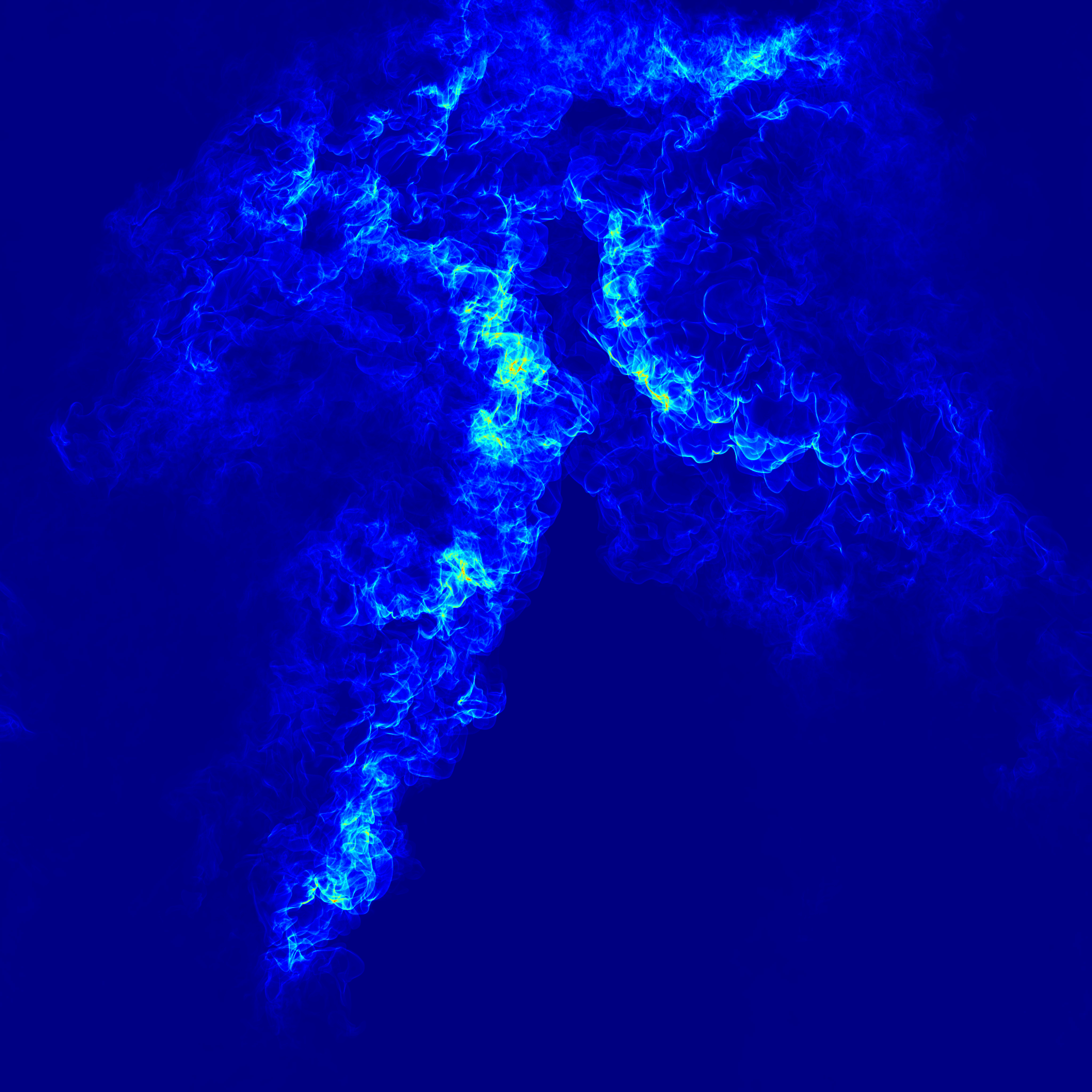}};
    \draw (-14mm,-12.5mm) -- (-14mm,12.5mm);
    \draw (-17mm,-12.5mm) -- (-14mm,-12.5mm);
    \node[anchor=south east] at (-15mm, -12.5mm) {\scriptsize 2.4};
    \draw (-17mm,12.5mm) -- (-14mm, 12.5mm);
    \node[anchor=north east] at (-15mm, 12.5mm) {\scriptsize 4.4};
    \draw[rotate=60] (image.south) ++(4.5mm,6.5mm) ellipse (7.5mm and 3mm);
    \draw[rotate=60,draw=white,dashed] (image.south) ++(4.5mm,6.5mm) ellipse (7.5mm and 3mm);
    \end{tikzpicture}
    \hspace{-2.3mm}
    \begin{tikzpicture}
    \node[inner sep=0pt] (image) at (0,0) {
    \includegraphics[width=25mm]{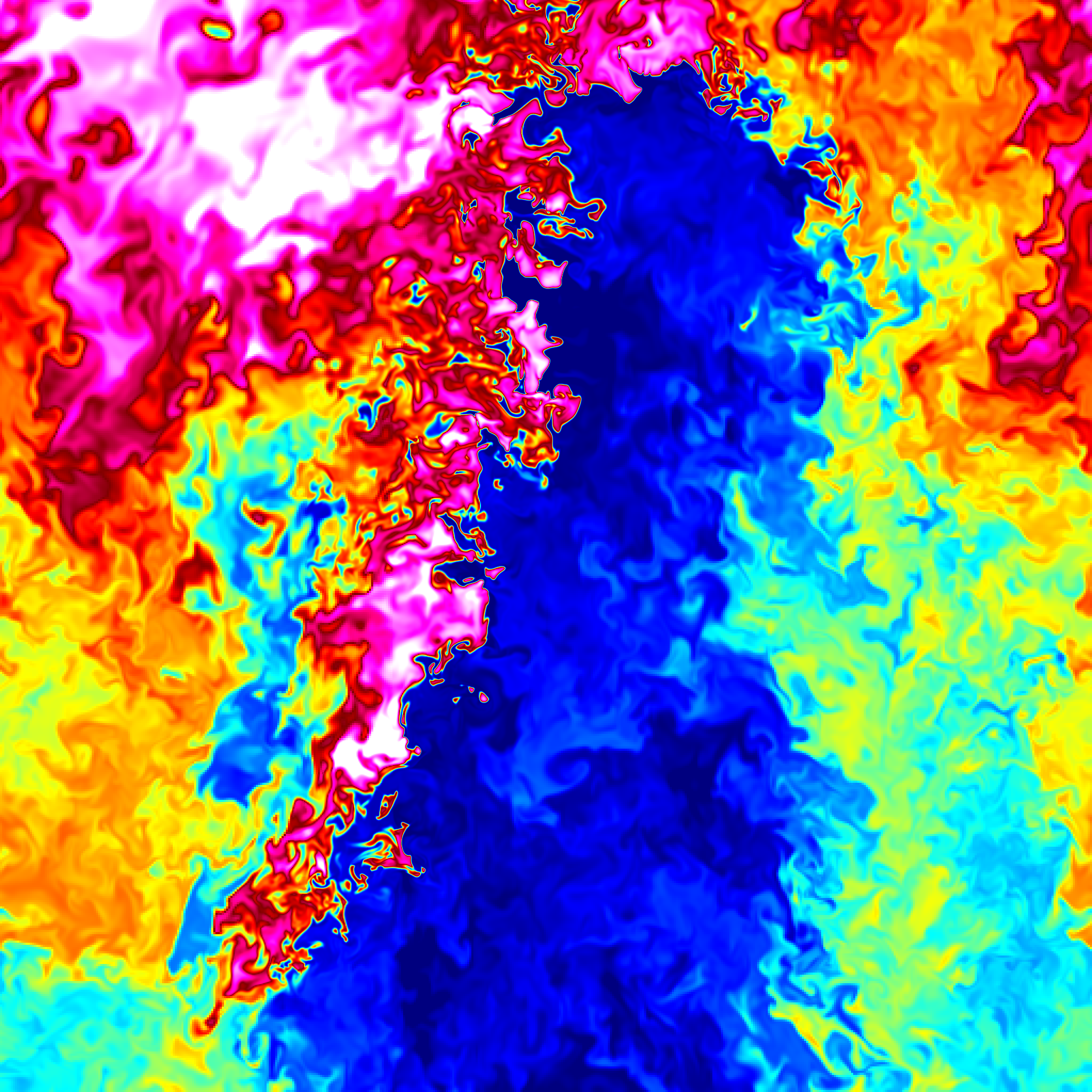}};
    \draw[rotate=60] (image.south) ++(4.5mm,6.5mm) ellipse (7.5mm and 3mm);
    \draw[rotate=60,draw=white,dashed] (image.south) ++(4.5mm,6.5mm) ellipse (7.5mm and 3mm);
    \end{tikzpicture}
    \hspace{-2.3mm}
    \begin{tikzpicture}
    \node[inner sep=0pt] (image) at (0,0) {
    \includegraphics[width=25mm]{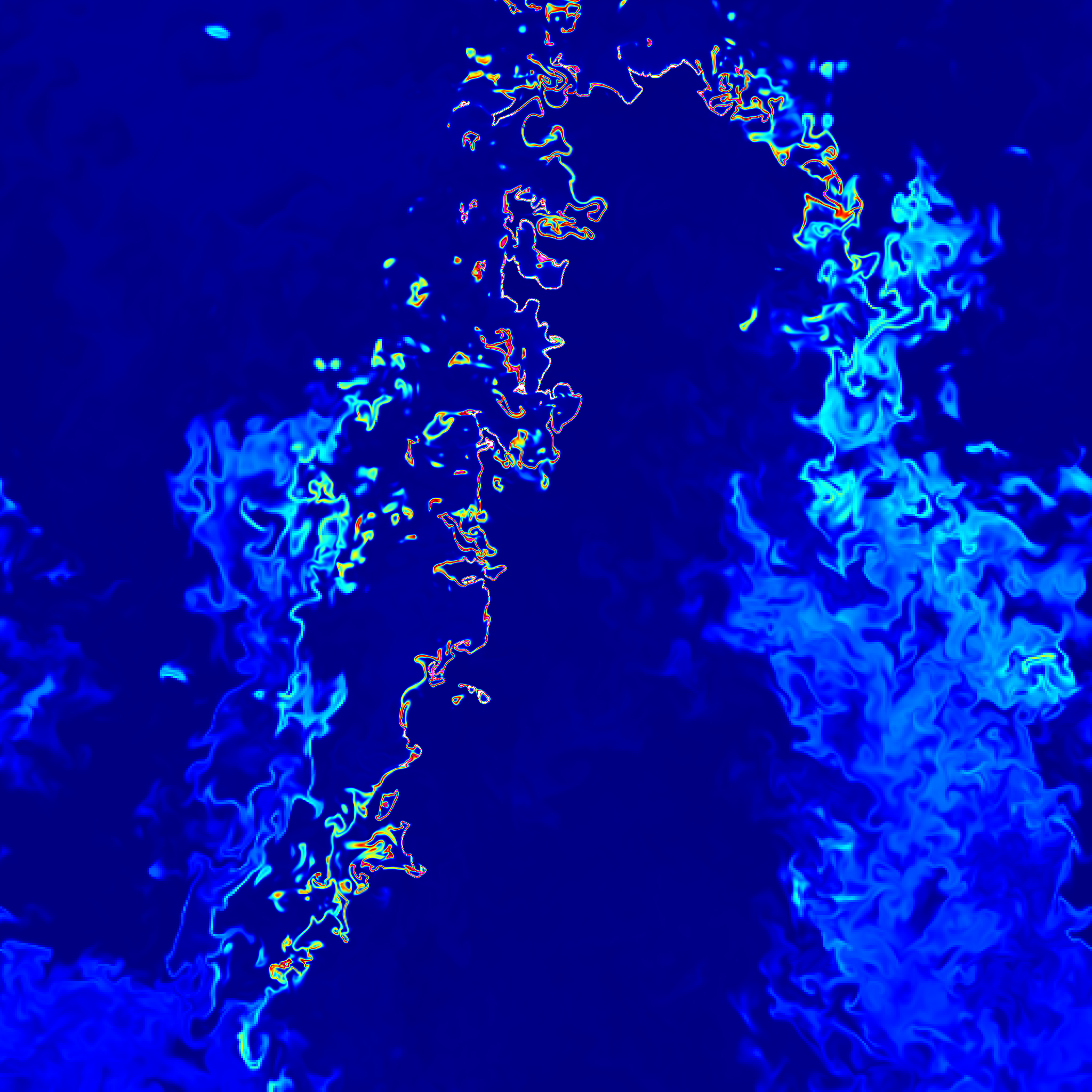}};
    \draw[rotate=60] (image.south) ++(4.5mm,6.5mm) ellipse (7.5mm and 3mm);
    \draw[rotate=60,draw=white,dashed] (image.south) ++(4.5mm,6.5mm) ellipse (7.5mm and 3mm);
    \end{tikzpicture}
    
    \caption{Formation of multiple flame kernels simultaneously leading to a sheet that propagates inwards and blows downstream much more slowly compared to isolated kernels. Time difference between slices $\Delta t = 15.6$\,$\mu$s.}
    \label{fig:LOSsheet}
\end{figure}

Figure \ref{fig:LOSkernel} shows the formation and evolution of an isolated kernel. The left-hand column shows the line of sight diagnostic (using heat release rate as the reaction rate variable), where a small flame kernel appears spontaneously and grows with time (from top to bottom). The formation and growth of the flame kernel can also be seen in the temperature slices in the middle column.  The accumulation of HO$_{2}$ seen in the right-hand column demonstrates the presence of low-temperature chemistry and eventual thermal runaway associated with ignition.  Additional ignition events can be observed at later times in the lower-right corner of the images in figure \ref{fig:LOSkernel}. The kernel formation appears to coincide with Kelvin-Helmholtz rollers; large-scale shear-driven mixing brings fuel into contact with recirculated heat, leading to ignition. Not all large scale KH mixing events lead to ignition, but when an ignition event occurs, it appears to coincide with a KH roller. Ignition events happen more frequently with increasing height; sporadic kernels are seen to appear below the flame stabilisation point, whereas more frequently formed kernels form adjacent to the peripheral flame. Ignition events can also form in clusters or sheets; figure \ref{fig:LOSsheet} shows the simultaneous formation of several flame kernels, which propagate towards the core of the jet, evolving into a flame sheet. The left column is again the line of sight where several small kernels appear before connecting and forming a relatively-stable flame sheet on the side of the jet. This can be seen further in the temperature and $Y_{\HOtwo}$ slices, and this sheet persists for far longer than than the sporadic kernels that appear. All of these phenomena can be seen more clearly in the accompanying animations provided as supplementary materials.

Animations of the flame (provided as supplementary material) show that the core flame is stabilised by these repeated, sometimes isolated, ignition events. The peripheral flame itself is approximately situated in a (highly-turbulent) stagnation region. However, the convection time from the fuel injector to the flame base (around 0.2 ms based on centreline velocity) is orders-of-magnitude smaller than the autoignition delay time of hydrogen at these conditions, precluding flame stabilisation by autoignition of the jet fluid alone. Rather, the appearance of isolated ignition kernels can be explained by mixing between the jet fluid and recirculated lean low-temperature combustion products. This mixing and the appearance of ignition kernels occurs in the region where the spread of unburned fuel within the jet intersects with the recirculation zone.

%% file: stabilisation/entrainment.tex
\begin{figure}[ht!]
    \centering
    \includegraphics[width=88mm]{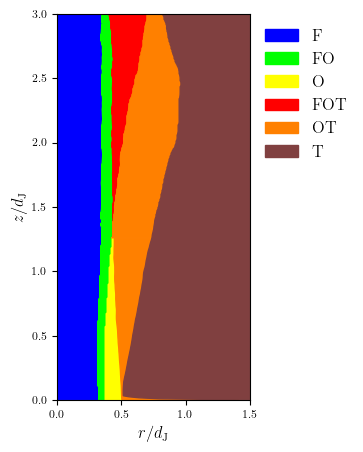}
    \caption{Zonal arrangement of the F/FO/FOT/O/OT/T regions. Colour matches that used in figures \ref{fig:flux_zone}.}
    \label{fig:combustor_rz_zones}
\end{figure}

Since flame ignition and stabilisation appears to be driven by the mixing of fuel from the jet with recirculated heat, this section considers the mechanisms by which mixing occurs. Since the combustor is lean everywhere, there is an abundance of oxygen, and so any ignition events and hence flame stabilisation point is determined by the mixing of fuel and heat. Based on azimuthally and temporally averaged data, different mixing zones can be classified as follows:
\begin{itemize}
    \item F - Unburned fuel-air mixture with $Y_{\Htwo} > 0.0133$. \\
    Defined by $0<r<R_{1}(z)$.
    \item O - Essentially pure air which acts as a non-reactive buffer between the regions with fuel and hot recirculating fluid. \\
    Defined by $R_{2}(z) < r < R_{3}(z)$ and $z < z_{c}$.
    \item T - Hot recirculating products from the flame above, mostly comprised of steam and air.\\ Defined by $r > R_{4}(z)$.
    \item FO - Mixing region between the inner fuel/air region (F) and pure air region (O).\\
    Defined by $R_{1} < r < R_{2}$ (where $z < z_{c}$), 
    and by $R_{1}(z) < r < R_{3}(z)$ (where $z > z_{c}$). 
    \item OT - Mixing region between the pure air region (O) and hot recirculation region (T). \\
    Defined by $R_{3}(z) < r < R_{4}(z)$ (where $z < z_{c}$), 
    and by $R_{2}(z) < r < R_{4}(z)$ (where $z > z_{c}$).
    \item FOT - Region where there is a flammable mixture of species spread from the core of the jet and high temperatures entrained from the recirculation region. Ignition is possible in this region.\\ Defined by $R_{3}(z) < r < R_{2}(z)$ and  $z > z_{c}$.
\end{itemize}
The height $z_{c}$ is defined to be the point where $R_{2}(z) = R_{3}(z)$, and the radius definitions are given by:
\begin{align}
    R_{1}(z) &= r(Y_{\Htwo} = 0.0133),\\
    R_{2}(z) &= r(Y_{\Htwo} = 0.003),\\
    R_{3}(z) &= r(T = 850K),\\
    R_{4}(z) &= r(T = 1150K).
\end{align}
The mass fractions are associated with the mean equivalence ratio and flammability limit of hydrogen, which also dictates the lower temperature limit. The upper limit of the temperature is set to reflect the temperature of the recirculation zone. Figure \ref{fig:combustor_rz_zones} shows the partitioning of the domain after applying the conditioning. 

For each zone $\Omega_{i}$, the integrated mass $\mathcal{Q}$, species $\mathcal{S}$ and heat $\mathcal{T}$ fluxes are given by
\begin{align}
    \mathcal{Q}_{\Omega_{i}}(z) &= \int_{R_{\textsc{l},i}}^{R_{\textsc{r},i}} \overline{\rho u_{z}} f(r)\,\text{d}r,\\
    \mathcal{S}_{\Omega_{i}}(z) &= \int_{R_{\textsc{l},i}}^{R_{\textsc{r},i}} \overline{\rho Y_{\Htwo}u_{z}} f(r)\,\text{d}r,\\
    \mathcal{T}_{\Omega_{i}}(z) &= \int_{R_{\textsc{l},i}}^{R_{\textsc{r},i}} \overline{T u_{z}} f(r)\,\text{d}r,
\end{align}
where $R_{\textsc{r},i}$ and $R_{\textsc{l},i}$ are the radii associated with the left and right boundaries of zone $\Omega_i$, respectively, and the function $f(r)$ is given by
\begin{equation}
    f(r) = \begin{cases}2\pi r &\mbox{ if } \displaystyle r < \frac{L_{x}}{2},\\ \displaystyle 2\pi r-8r\cos^{-1}\bigg(\frac{L_{x}}{2r}\bigg) &\mbox{ otherwise.}\end{cases}
\end{equation}

\begin{figure}[ht!]
    \centering
    \includegraphics[width=75mm]{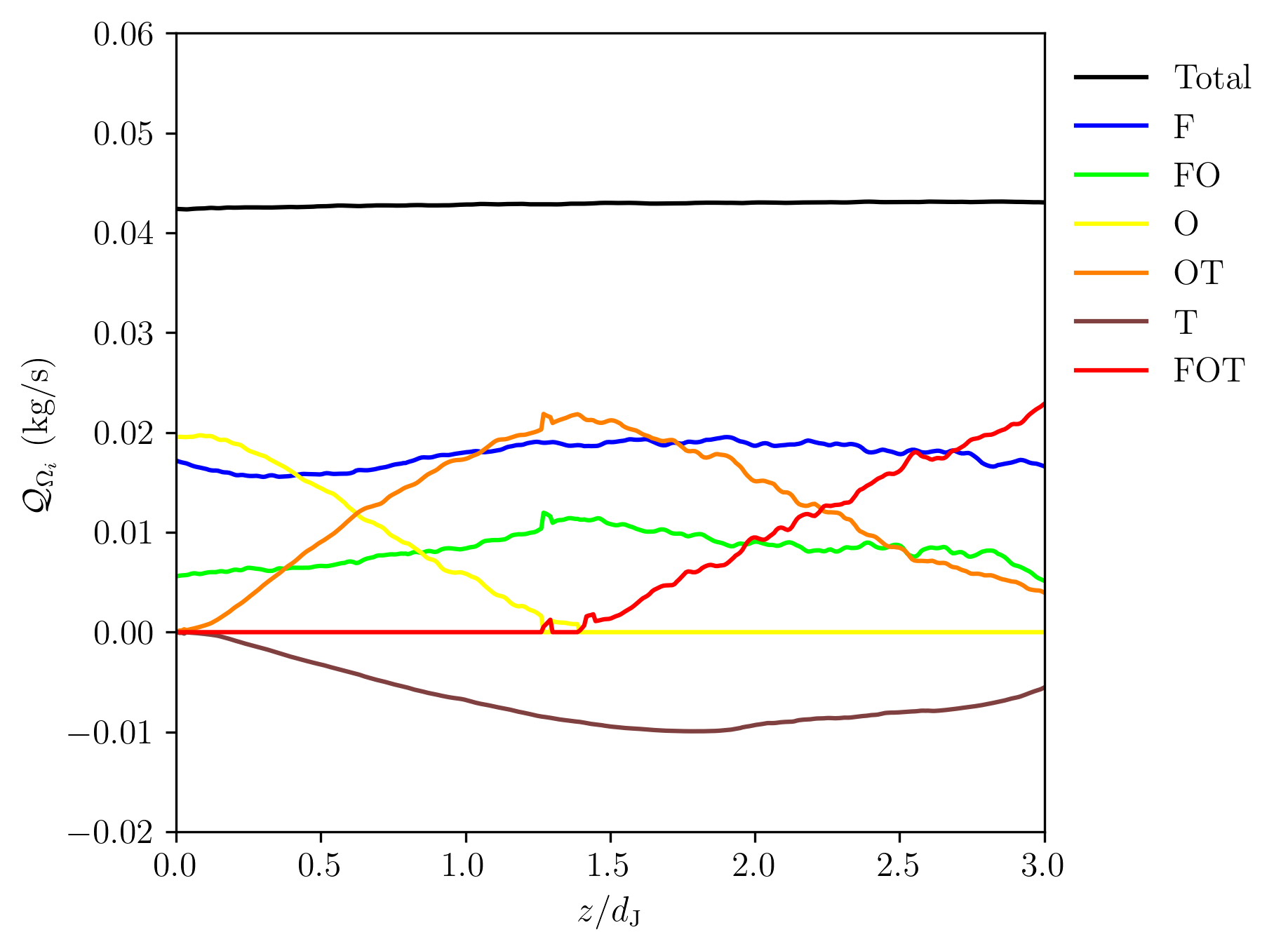}
    
    \includegraphics[width=75mm]{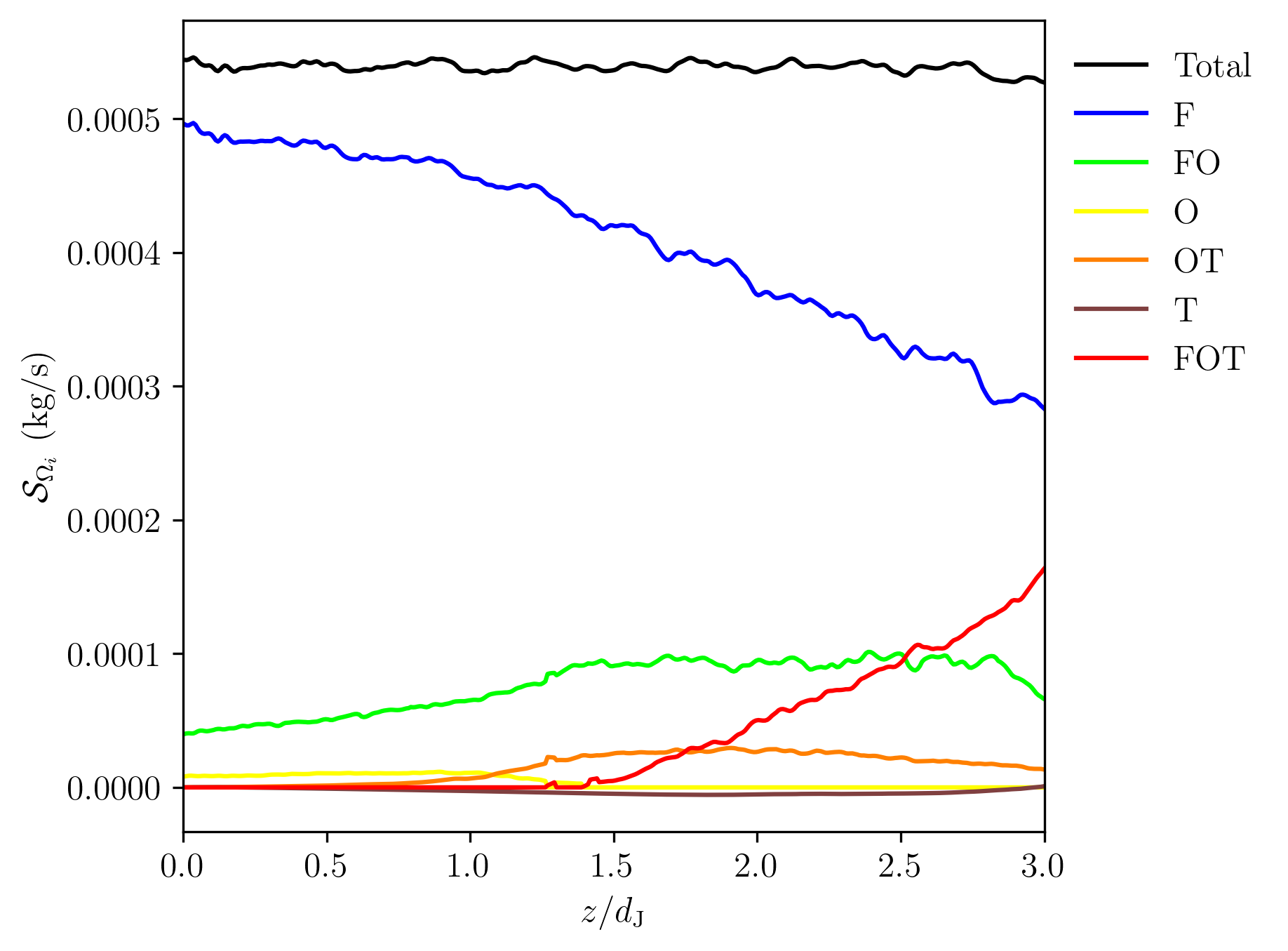}

    \includegraphics[width=75mm]{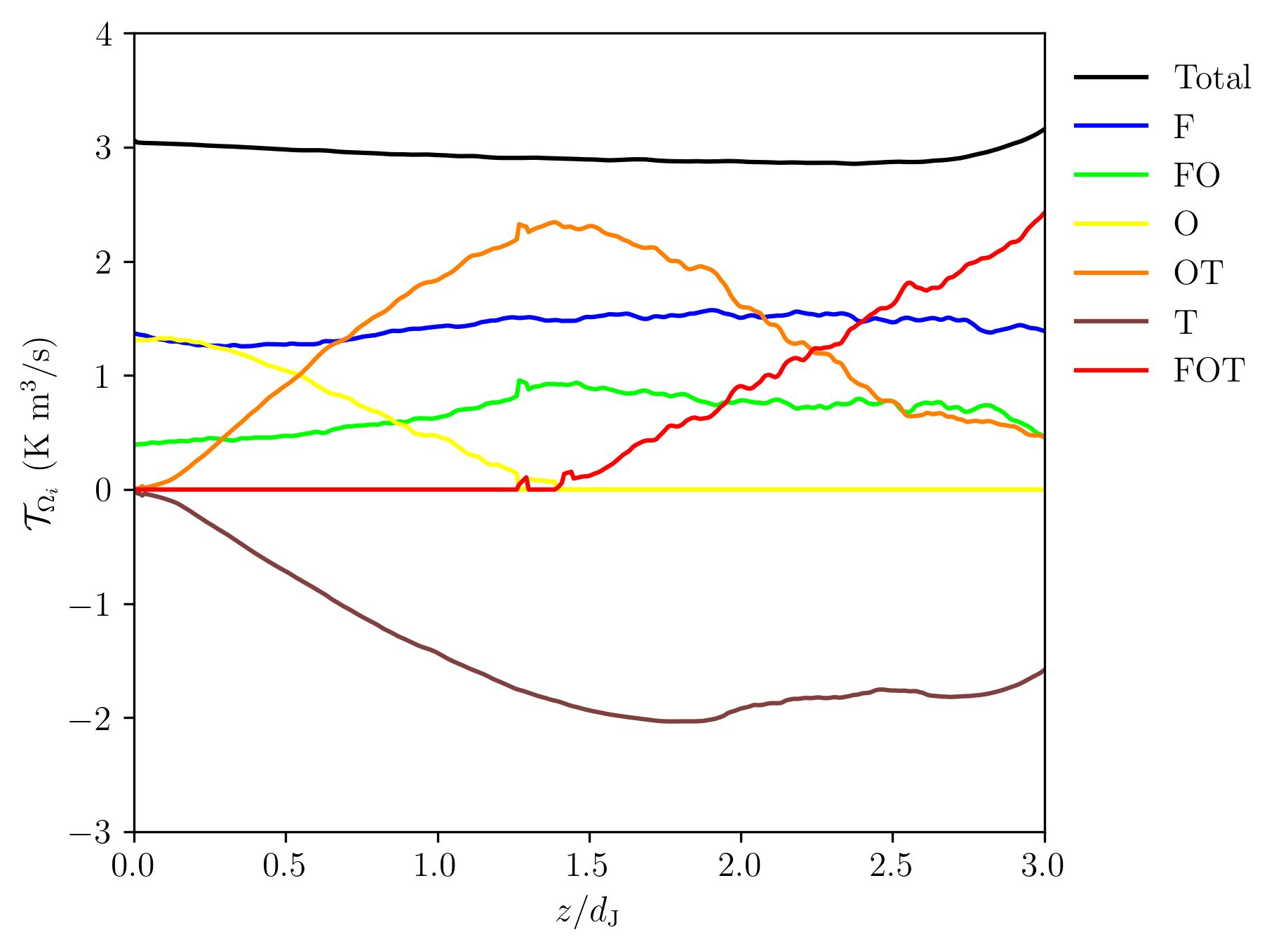}

    \caption{Integrated mass, species and thermal fluxes as a function of streamwise distance.}
    \label{fig:flux_zone}
\end{figure}

By partitioning the domain in this way, mass, species and heat exchange between each zone can be evaluated; integrated fluxes are presented in figure \ref{fig:flux_zone}. In the fuel region (F), the mass and heat flux stay constant, while the hydrogen flux decreases as it moves to the FO region. In the FO region, the mass and heat flux again are relatively constant, while the hydrogen flux increases up to $z_{c}$ and remains constant as hydrogen moves to the FOT region. The recirculation region (T) loses mass and heat to the OT region, which can be seen through the negative (positive) values of mass and heat flux in the T (OT) region; there is no transport of fuel in either of these regions. Beyond $z_{c}$ the OT region transfers mass and heat to the FOT region.

In each panel in figure \ref{fig:flux_zone}, the totals are essentially constant (there is practically no fuel consumption or heat release since the analysis region is upstream of the flame base, and supports the use of temperature as an approximation for heat flux); therefore, the type and direction of species and heat transfer can be examined by integrating the cylindrical form of the continuity, species and temperature equation (for simplicity, assuming constant specific heat capacities, no reactions, and neglecting molecular diffusion), giving
\begin{align}
    \ddz\mathcal{Q}_{\Omega_{i}} &= 
    \Bigl[\overline{\rho \textbf{u}\cdot\textbf{n}}\Bigr]_{R_{\textsc{l},i}}^{R_{\textsc{r},i}},\\
    \ddz \mathcal{S}_{\Omega_{i}} &= 
    \Bigl[\overline{\rho Y_{\Htwo}\textbf{u}\cdot\textbf{n}}\Bigr]_{R_{\textsc{l},i}}^{R_{\textsc{r},i}},\\
    \ddz \mathcal{T}_{\Omega_{i}} &= 
    \Bigl[\overline{T \textbf{u} \cdot \textbf{n}}\Bigr]_{R_{\textsc{l},i}}^{R_{\textsc{r},i}},
\end{align}
where $[q(r)]_{R_1}^{R_2}=q(R_2)-q(R_1)$ is used to denote the difference in horizontal fluxes in and out of the region.

The equations can be decomposed further by specifying the normal as
\begin{equation}
    \textbf{n} 
    = \begin{bmatrix}
        n_{r} \\ n_{z}
    \end{bmatrix}
    = \frac{\pm\nabla c}{|\nabla c|}
    = \frac{\pm1}{\sqrt{(\partial_r c)^{2} + (\partial_z c)^{2}}}\begin{bmatrix}
        \partial_r c \\ \partial_z c
    \end{bmatrix},
\end{equation}
where $c$ is the field chosen to define the boundary, either $Y_{\Htwo}$ or $T$ in this particular case; the sign of the normal is chosen such that the direction is always to the right (i.e. $n_{r} \geq 0$).  The resulting decomposition can be written as
\begin{align}
    \ddz \mathcal{Q}_{\Omega_{i}} = & 
    \Bigl[(\overline{\rho} \tilde{u}_{z})n_{z} +
    (\overline{\rho}\tilde{u}_{r})n_{r}
    \Bigr]_{R_{\textsc{l},i}}^{R_{\textsc{r},i}},\\
    \ddz \mathcal{S}_{\Omega_{i}} = & 
    \Bigl[ \left( \overline{\rho}\tilde{Y}_{\Htwo}\tilde{u}_{z}
    + \overline{\rho}\widetilde{Y_{\Htwo}^{\prime\prime}u_{z}^{\prime\prime}} \right)n_{z}\nonumber\\&
    + \left( \overline{\rho}\tilde{Y}_{\Htwo}\tilde{u}_{r} 
    + \overline{\rho}\widetilde{Y_{\Htwo}^{\prime\prime}u_{r}^{\prime\prime}}\right)n_{r}
    \Bigr]_{R_{\textsc{l},i}}^{R_{\textsc{r},i}},\\
    \ddz \mathcal{T}_{\Omega_{i}} = &
    \Bigr[\left(\overline{T}\overline{u}_{z} 
    + \overline{T^{\prime}u_{z}^{\prime}}\right)n_{z} \nonumber\\&
    + \left(\overline{T}\overline{u}_{r} 
    + \overline{T^{\prime}u_{r}^{\prime}}\right)n_{r}
    \Bigr]_{R_{\textsc{l},i}}^{R_{\textsc{r},i}}.
\end{align}

\begin{figure}[ht!]
    \centering
    \includegraphics[width=88mm]{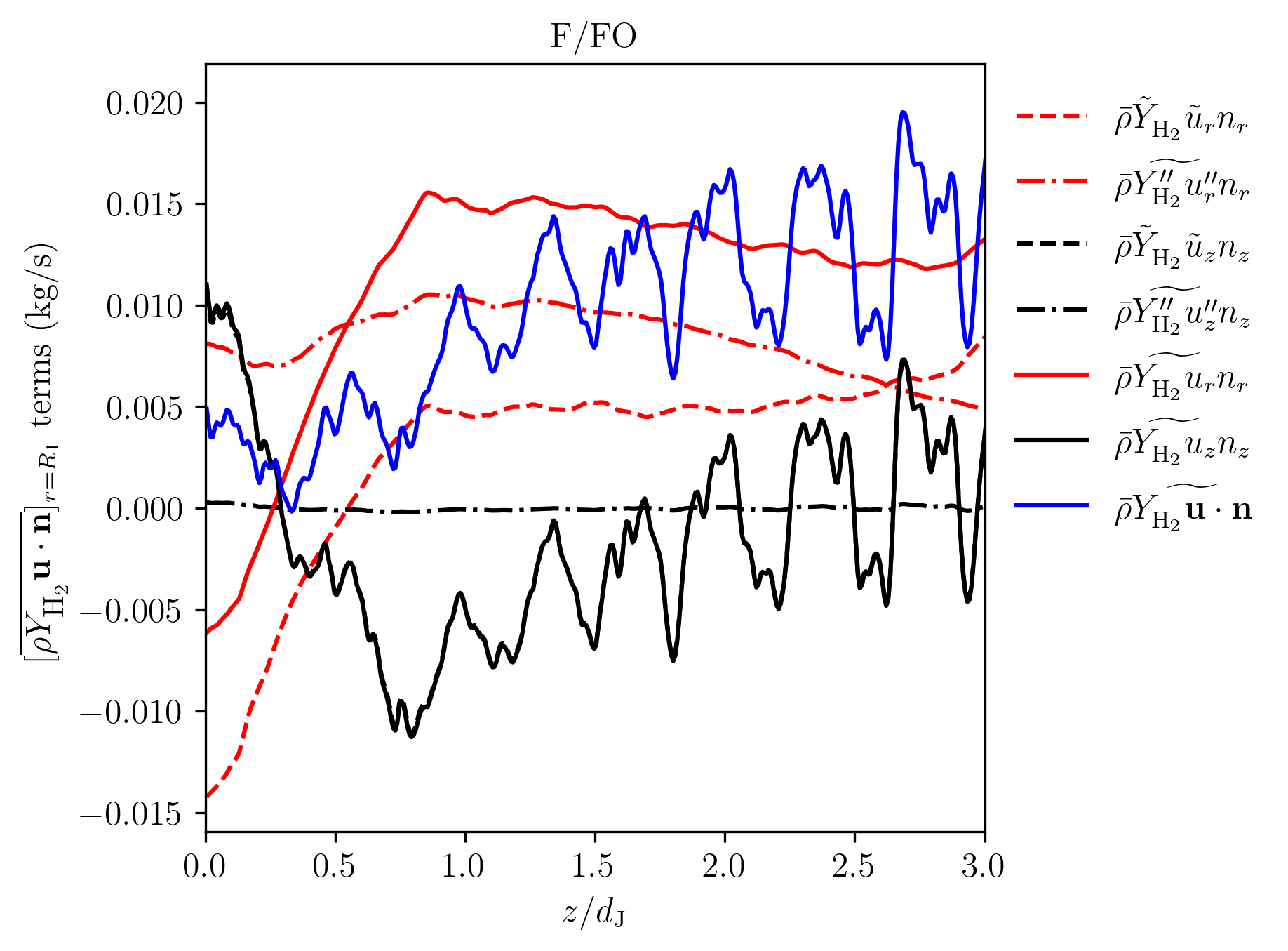}

    \includegraphics[width=88mm]{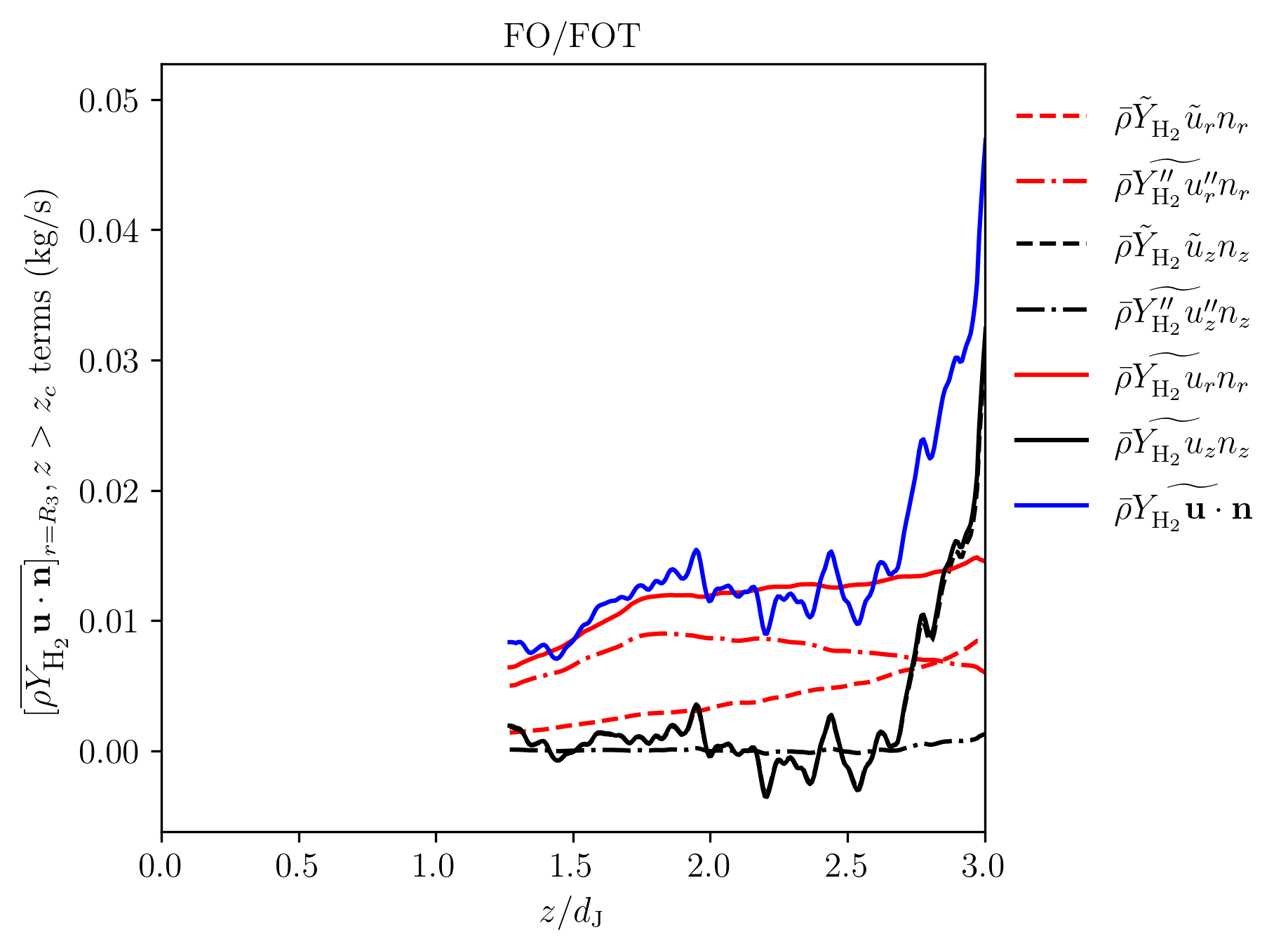}
    \caption{Hydrogen flux across F/FO and FO/FOT boundary}
    \label{fig:boundaryH2flux}
\end{figure}

\begin{figure}[ht!]
    \centering
    \includegraphics[width=88mm]{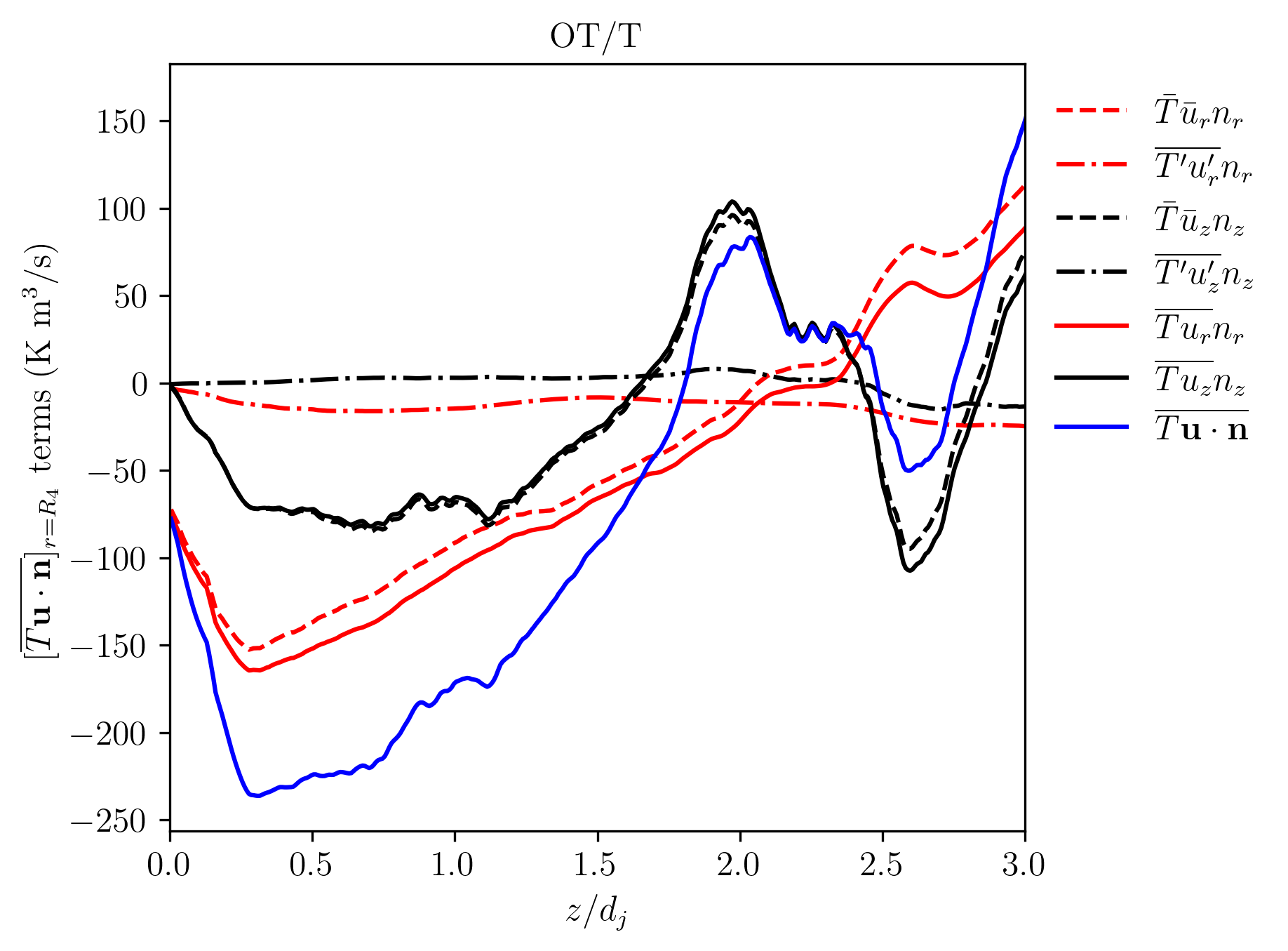}
    
    \includegraphics[width=88mm]{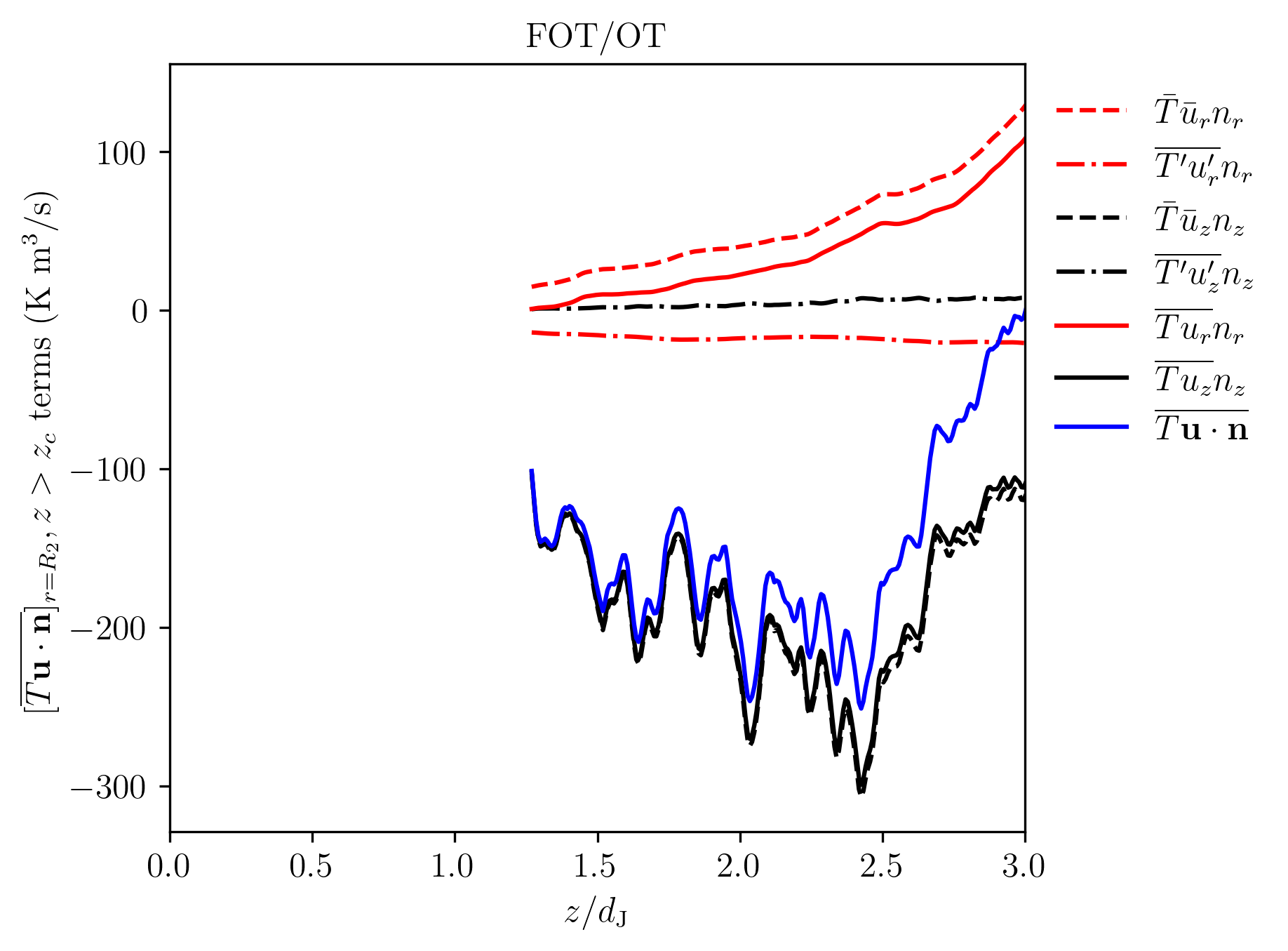}
    \caption{Heat flux across OT/T and FOT/OT boundary}
    \label{fig:boundaryTflux}
\end{figure}

The exchanges of particular interest are the species fluxes flowing from the F to FO to FOT zones, as well as the heat flux flowing from T to OT to FOT zones. Figure \ref{fig:boundaryH2flux} shows the evaluation of the boundary hydrogen fluxes between the F and FO, and FO and FOT zones. In both cases, the main driver of species transport is the turbulent radial term ($\overline{\rho}\widetilde{Y_{\Htwo}^{\prime\prime}u_{r}^{\prime\prime}}n_{r}$), followed by mean radial transport driven ($\overline{\rho}\tilde{Y}_{\Htwo}\tilde{u}_{r}n_{r}$), with very little streamwise transport. From figure \ref{fig:boundaryTflux}, the transport of heat from the T to OT region (below $z/d_{\textsc{j}}=2$) is mostly driven by mean transport (both $\overline{T}\overline{u}_{r}n_{r}$ and $\overline{T}\overline{u}_{z}n_{z}$), with little contribution from the turbulent terms. Finally, the heat transport between the OT and FOT region is driven by mean streamwise transport ($\overline{T}\overline{u}_{z}n_{z}$).


%% file: conclusion.tex
A high-pressure hydrogen micromix combustor has been investigated using three-dimensional DNS with detailed chemistry to examine the flame structure and stabilisation mechanism. The combustor geometry follows the archetypal NASA micromix design from Schefer \cite{Schefer2003EvaluationBurner}, and the high-pressure and temperature operating conditions are relevant to aerospace applications. The flame stabilisation process at these conditions is summarised in section \ref{sec:conclusion:structure}, and shown schematically in figure \ref{fig:combustor_schematic}.  A more general discussion of the main features of micromix flame stabilisation is given in section \ref{sec:conclusion:conjecture}, along with implications for design across different operating conditions.

\subsection{Flame structure and stabilisation mechanism}\label{sec:conclusion:structure}
The present simulation employs opposed jet-in-crossflow injection of hydrogen fuel with momentum ratio $J=17.2$. At these conditions, the hydrogen jets meet at the mid-plane ($y=0$) and roll up into stream-wise vortex tubes. At the inlet plane, the resulting profiles of fuel mass fraction are shaped like two kidneys near the centreline (figure \ref{fig:meanSlicesSpecies}), with a clear air gap between the fuel and the walls of the air port. Planar averaging gives one-dimensional streamwise profiles (figure \ref{fig:1dprofiles}) that indicate a liftoff height for the flame base between approximately 3 and 4.5 jet diameters (21 to 31.5\,mm). Azimuthally-averaged profiles in $r-z$ space (figure \ref{fig:rz_profiles}) and streamlines (figure \ref{fig:streamlines}) indicate the existence of regions of recirculation in the corners of the domain (also see figure \ref{fig:meanSlicesZvel}b), with the recirculation bubble reaching up to 4 jet diameters downstream. Importantly, the recirculation region extends sufficiently far to result in upstream transport of combustion products (i.e.\ heat).

Instantaneous temperature and intermediate species fields (figures \ref{fig:tempInstant} and \ref{fig:HO2instant}) suggest there are multiple burning regimes in the domain (figure \ref{fig:borghi}). Construction of fuel- and products-based progress variables (figure \ref{fig:cInstant}) concur with these findings, suggesting vastly different turbulence-flame interactions occur within each flame. The progress variables also agree that a non-reacting air buffer is present at the inlet and a short distance into the domain and that combustion products are present in the recirculation zone. There is a thin core flame situated directly above the main jet region, accounting for over 85\% of the total fuel consumption, and a much thicker peripheral flame shrouding the core. By partitioning the domain into 5 categories (jet fluid, core flame, peripheral flame, recirculating fluid and products, figure \ref{fig:combustor_rz_zones}), conditional analyses were performed to examine the combustion regimes. It was found that the core flame burns in an inhomogeneously-premixed mode, with equivalence ratios varying from 0.4--1.2, whereas the peripheral flame burns in the narrower and leaner range of 0.18--0.3 (figure \ref{fig:phiConditioned}). PDFs of normalised flame index demonstrate that no diffusion flamelets are present (figure \ref{fig:flameIndex}).  The core flame was classified in the thin reaction zone, and the peripheral flame in the distributed burning regime.  The stark difference in turbulence-flame interaction arises due to the high pressure of the system, which has resulted in a large Damk\"ohler number in the core flame, and a large Karlovitz number in the peripheral flame.

Given the essentially-premixed nature of the flame, the role of turbulent burning velocity on the flame stabilisation is assessed. By estimating an upper bound for jet velocity that would allow flame stabilisation through flame propagation, it was concluded that this flame cannot be stabilised through turbulent premixed flame propagation. Evaluation of mean local flame speeds and turbulent intensities supports this argument, and suggests that the jet is an order-of-magnitude too fast for flame propagation to be able to stabilise the flame at these conditions. The flame speed computed is comparable to the one-dimensional flame speed, implying that, although the core flame is not stabilising, it is deflagrative (rather than ignitive) in nature.

The autoignition delay time of the inlet condition is orders-of-magnitude too large to allow for the incoming mixture to autoignite with the prescribed domain size and inlet speed. However, sporadic ignition kernels are observed ahead of the flame base and appear to be correlated with shear-driven Kelvin-Helmoltz vortices. By using a line-of-sight diagnostic (figure \ref{fig:LOSkernel}), infrequent and isolated events are observed, which grow into rapidly-burning flame kernels. Further downstream, as the fuel spreads into the recirculation region, the likelihood of ignition increased, and it was possible to observe near-simultaneous ignition events more spatially distributed, which appeared more like an ignition sheet (figure \ref{fig:LOSsheet}). These sheets are situated between the peripheral flame and core flame and result from the entrainment of the distributed reactions into the jet.

Since there is an abundance of oxygen everywhere (due to the lean global equivalence ratio), the formation of ignition spots relies on mixing fuel from the main jet with heat from the recirculation zone. Partitioning the domain below the flame stabilisation point (using values of hydrogen mass fraction and temperature corresponding to the flammability limits) identified five regions: fuel (F), oxidiser (O), temperature (T), fuel+oxidiser (FO), oxidiser+temperature (OT), and fuel+oxidiser+temperature (FOT). This partitioning was used to analyse boundary fluxes between each of the zones, which showed that the spreading of hydrogen to the point of ignition primarily results from turbulent mixing in the radial direction ($\bar{\rho}\widetilde{Y_{\Htwo}^{\prime\prime}u_r^{\prime\prime}}n_r$; figure \ref{fig:boundaryH2flux}). On the other hand, the transport of heat was shown to be driven by the mean flow ($\bar{T}\bar{u}$; figure \ref{fig:boundaryTflux}) from the recirculation to the ignition zone. Hypothetically, a one-dimensional model could be formulated to predict the structure of this zonal arrangement using appropriate models for turbulent mixing of fuel inside the jet and entrainment of hot fluid, and will be the subject of future work.

An idealised schematic (assuming symmetry across the left boundary) of the combustor is shown in figure \ref{fig:combustor_schematic}.

\subfile{tikz/structure}

\subsection{Principles for micromix combustor design}\label{sec:conclusion:conjecture}

\subsubsection{Present configuration}
In the specific configuration presented here, the momentum flux ratio $J$ is a critically-important factor affecting lift-off height and general flame dynamics. Ignoring direct interaction with other fuel injectors, increasing the value of $J$ will naturally increase the size of the air annulus and effect the point at which fuel encounters necessary heat to ignite. It will also, as discussed in section \ref{sec:background}, increase the degree of mixing occurring before entering the combustor and within the jet. As such, times associated with the ignition process would be expected to decrease as hot fluid will encounter richer mixtures more readily upon entrainment into the jet. The precise relationship between $J$ and the lift-off height is beyond the scope of this paper and would be more suited to either large eddy simulation, experimental measurement or direct numerical simulation at lower pressures where resolution requirements are less stringent. However, flame stability will also be dictated by the heat loss at the sides of the combustor, which are not considered here through the periodic boundary conditions, where extinguishing of outer flames may lead to full failure of the flame. Furthermore, the usage of a low Mach number solver has neglected the effect that thermoacoustic excitation will have on the flame. Future work could focus on any of these numerous factors that can affect flame stability in micromix combustors.

\subsubsection{General case}
While the present simulation represents a single combustor design and operating conditions, its analysis points to several principles of flame stabilisation in micromix combustors that are likely applicable more generally. Given the design goal of low-NO$_{x}$ emission, a general objective is to achieve satisfactory flame stabilisation at low overall equivalence ratios, and to provide short fluid residence time in any regions with locally elevated temperatures. Flame stabilisation behaviour is satisfactory if a design provides a low risk of flashback/blowoff or flame attachment, acceptable ignition behaviour, and low susceptibility to thermoacoustic excitation.

Hydrogen micromix combustion can be implemented in many different ways: the fuel can be introduced into the airflow by any number of jets, which may be aligned with or angled into the airflow; the air jets need not be circular in cross-section, and might be arranged as a regular array (e.g.\ square or close-packed), or some other pattern intended to aid manufacture or improve performance; and the fuel/air ratio of individual air jets may in principle be controlled in order to manage the overall emissions, heat release, and stability.  

Using established experimental and computational methods, it is feasible to tailor various different types of micromix combustor designs to achieve desired mixing features. Even simple models could be constructed configuration for both the recirculation bubble size (e.g.\cite{Massey2019AFlows,Kallifronas2022InfluencesBody}), or potentially the lift-off height by adapting models used for premixed bluff-body stabilised flames (e.g.\cite{Most2002LiftedFlames}). For all of the different design possibilities, the common objective is to achieve recirculation zones and an inlet fuel-air premixture that provides the following features:

\begin{enumerate}
    \item At the inlet to the combustor, the fuel-air premixture should be adequately surrounded by a non-flammable mixture (e.g.~air) to prevent flame attachment or flashback to the combustor lip.
    \item The lean flammability isosurface of the fuel plume should intersect with the recirculation zone; otherwise, the flame could easily blow off.
    \item The equivalence ratio of the recirculated fluid should be sufficiently high that there is enough heat to ignite the fuel as it is entrained into the jet.
    \item For low-NO$_{x}$ operation, the fuel-air premixture should be flammable but overall fuel-lean, preferably without any rich or stoichiometric mixture, by the point that it reaches the core flame.
\end{enumerate}

%% file: tikz/structure.tex
\begin{figure}[t!]
    \centering
    \resizebox{88mm}{!}{
    \begin{tikzpicture}[scale=4]
\draw[dotted] (0,0) -- (0,5);

\draw (1,0) -- (1,0.4);
\fill[pattern=north east lines] (1,0) rectangle (1.05,0.35);
\draw (1,0.6) -- (1,1.5) -- (2,1.5);
\fill[pattern=north east lines] (1,0.65) rectangle (1.05,1.45);
\fill[pattern=north east lines] (1,1.45) rectangle (2,1.5);

\draw[dotted](2,1) -- (2,5);

\draw (1,0.6) -- (2,0.6);
\fill[pattern=north east lines] (1,0.6) rectangle (2,0.65);
\draw (1,0.4) -- (2,0.4);
\fill[pattern=north east lines] (1,0.4) rectangle (2,0.35);
\draw[dashed] (1,0.6) to[out=180, in=270] (0.4,1.5);
\draw[dashed] (1,0.4) to[out=180, in=270] (0,1.5);

\draw[red] (0,4.5) to[out=0,in=180] (2,4.8);
\draw[red] (0,4.4) to[out=0,in=180] (2,3);

\fill[red!30]  (0,4.5) to[out=0,in=180] (2,4.8) to[out=270,in=90] (2,3) to[out=180,in=0] (0,4.4);

\draw[dashed] (1,1.5) to[out=100,in=180,looseness=0.8] (2,4.8);
\draw[dashed,color=blue] (0.4,1.5) -- (1.1,3.6);

\draw[-{Stealth[length=3mm, width=2mm]},dashdotted,ultra thick,color=red] (1.3,4.3) to[out=45,in=45,looseness=0.8] (1.5,1.8) to [out=225,in=280,looseness=0.8] (1,2.7);

\draw[thick,postaction={decorate},
      decoration={
        markings,
        mark=between positions 0.1 and 0.9 step 0.25 with  {\arrow[scale=1]{>}}
      }
    ] (0.5,1.9) arc (300:0:0.1cm);
\draw[thick,postaction={decorate},
      decoration={
        markings,
        mark=between positions 0.1 and 0.9 step 0.25 with  {\arrow[scale=1]{>}}
      }
    ] (0.7,2.4) arc (300:0:0.1cm);

\node[anchor=west] at (2.1,0.5) {\LARGE Fuel injector}; 
\draw[thick,->] (2.1,0.5) -- (1.5,0.5);
\node[anchor=east] at (-0.1,1.4) {\LARGE Fuel-air premixture}; 
\draw[thick,->] (-0.1,1.4) -- (0.2,1.4);
\node[anchor=east] at (-0.1,1) {\LARGE Air buffer}; 
\draw[thick,->] (-0.1,1) -- (0.7,1.4);
\node[anchor=west] at (2.1,1.8) {\LARGE Recirculated products}; 
\draw[thick,->] (2.1,1.8) -- (1.4,2.2);
\node[anchor=east] at (-0.1,1.7) {\LARGE Lean flammability limit};
\draw[thick,->] (-0.1,1.7) -- (0.4,1.7);
\node[anchor=east] at (-0.1,2.3) {\LARGE Turbulent mixing of fuel}; 
\draw[thick,->] (-0.1,2.3) -- (0.3,2);
\draw[thick,->] (-0.1,2.3) -- (0.5,2.4);
\node[anchor=west] at (2.1,2.7) {\LARGE Entrainment of hot products}; 
\draw[thick,->] (2.1,2.7) -- (1.1,2.7);
\node[anchor=west] at (2.1,3.15) {\LARGE Ignition kernel}; 
\draw[thick,->] (2.1,3.15) -- (1.1,3.15);
\node[anchor=west] at (2.1,3.5) {\LARGE Ignition sheets}; 
\draw[thick,->] (2.1,3.5) -- (1.3,3.5);
\node[anchor=east] at (-0.1,4.45) {\LARGE Core flame}; 
\node[anchor=west] at (2.1,4) {\LARGE Peripheral flame}; 

\node[color=red] at (0.95,3.15) {\huge x};
\node[color=red] at (1.2,3.5) {\huge x};
\node[color=red] at (1.15,3.575) {\huge x};
\node[color=red] at (1.1,3.65) {\huge x};
\node[color=red] at (1.05,3.725) {\huge x};
\end{tikzpicture}
}
\caption{Idealised half-plane schematic of the present a micromix combustor.}
\label{fig:combustor_schematic}
\end{figure}
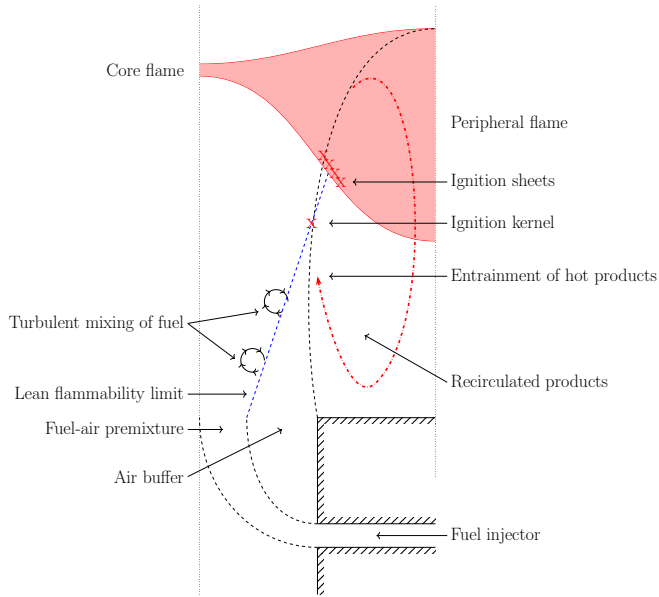

%% file: acknowledgements.tex
This project is funded by an EPSRC Industrial CASE Studentship and Reaction Engines Ltd.; the authors are grateful for discussions with Dr. Vladeta Zmijanovic. The authors are grateful to EPSRC (grant number: EP/R029369/1) and ARCHER2 for computational support as a part of their funding to the UK Consortium on Turbulent Reacting Flows. This research made use of the Rocket High Performance Computing Service at Newcastle University. Additionally, this research used resources of the Argonne Leadership Computing Facility, which is a DOE Office of Science User Facility supported under Contract DE-AC02-06CH11357.

%% file: appendices/resolution.tex
\begin{figure*}
\begin{subfigure}{50mm}
    
    \centering
    \includegraphics[width=50mm]{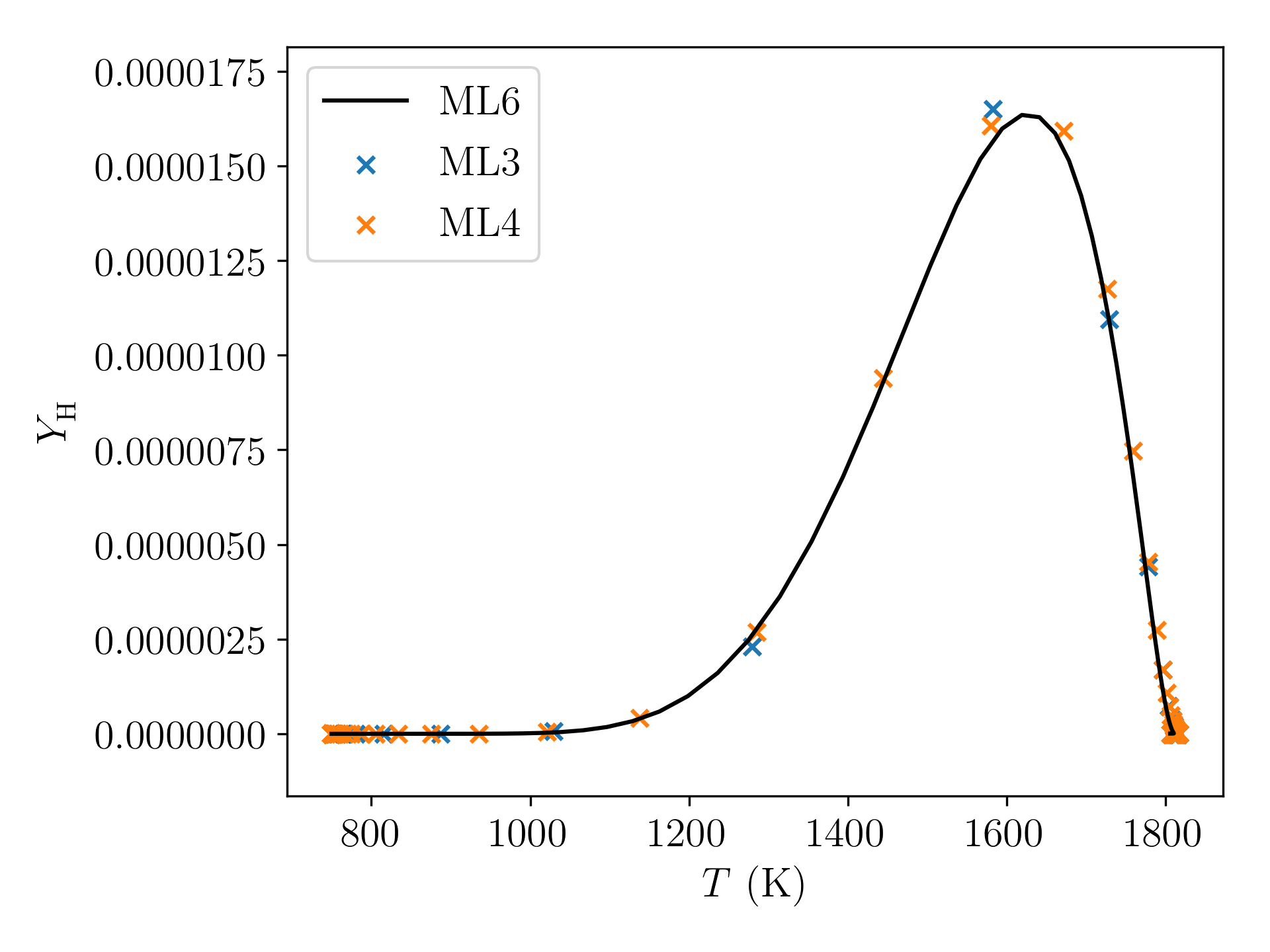}
    \includegraphics[width=50mm]{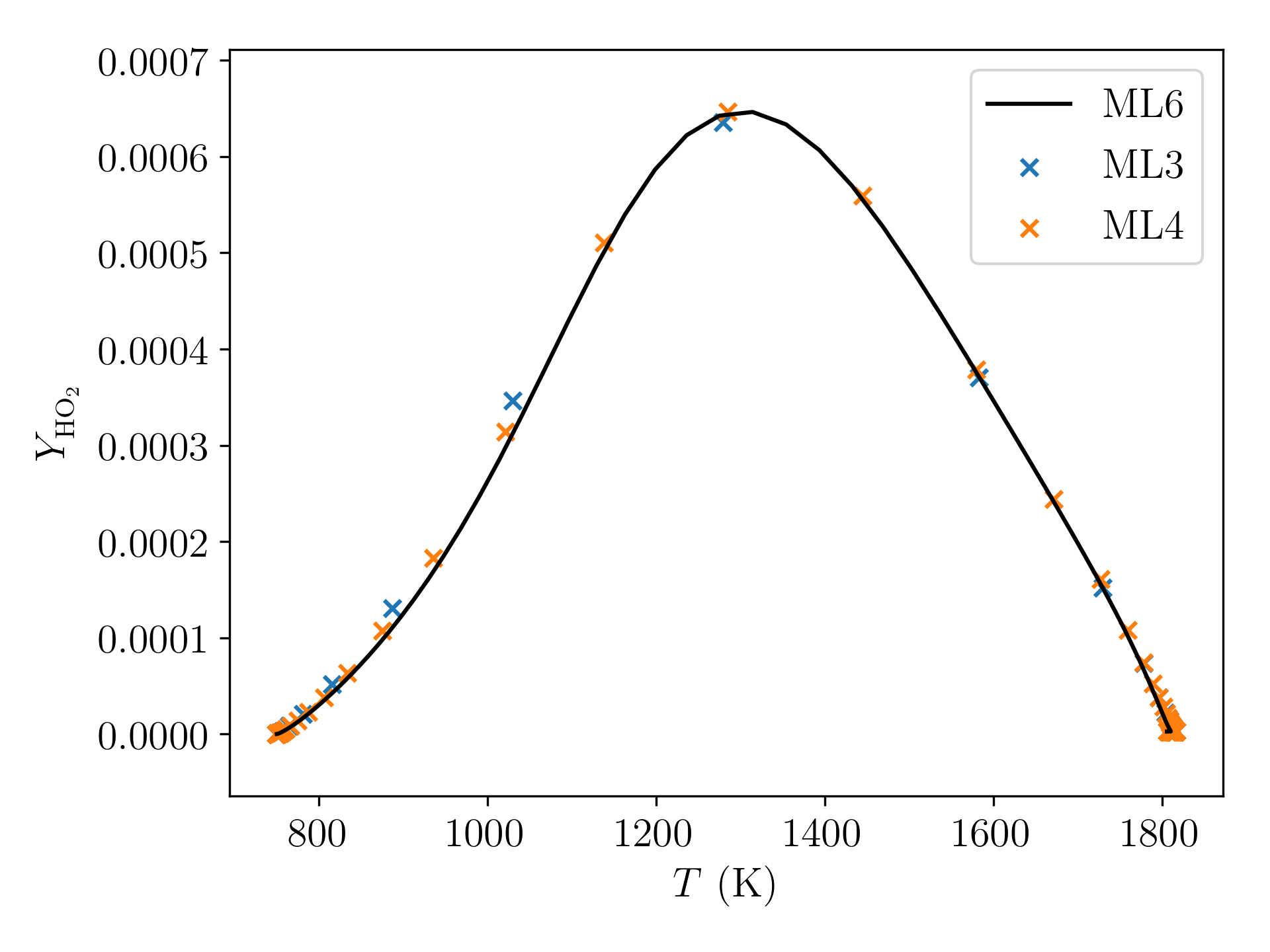}
    \includegraphics[width=50mm]{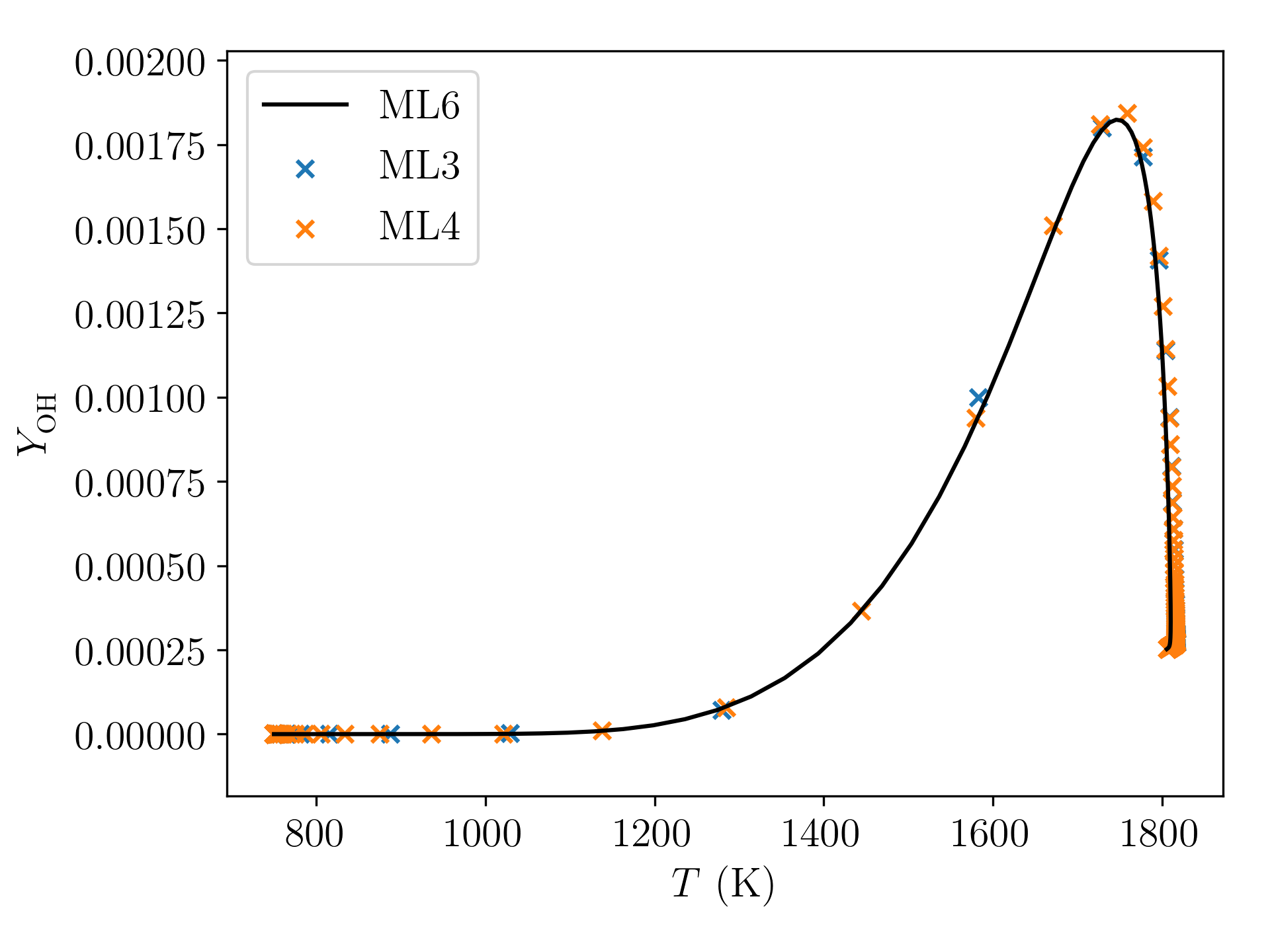}
    \includegraphics[width=50mm]{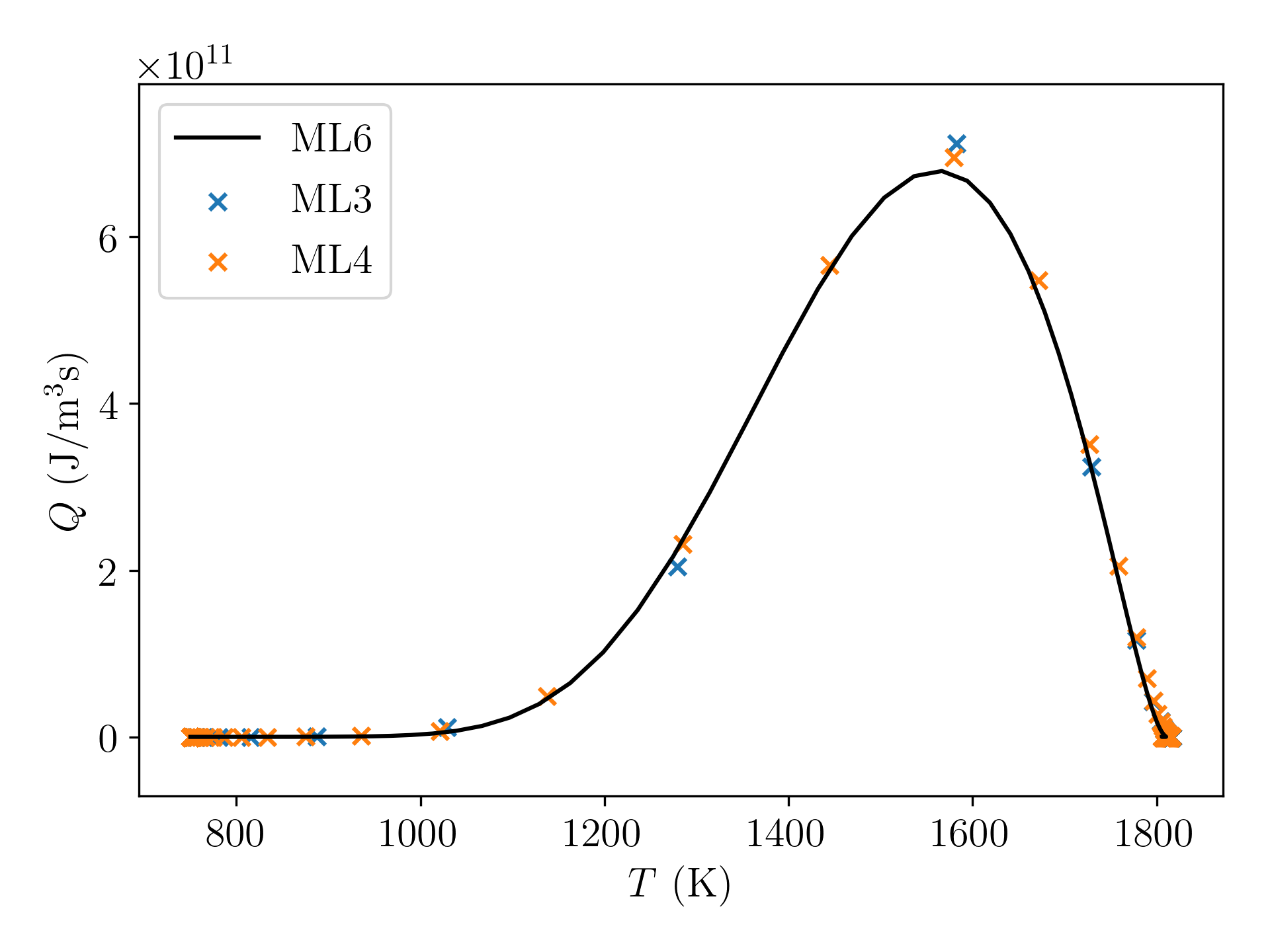}
    
    \caption{1D profiles for $\phi=0.4$}
    \label{fig:04res}
    \end{subfigure}
\begin{subfigure}{50mm}
    \centering
    \includegraphics[width=50mm]{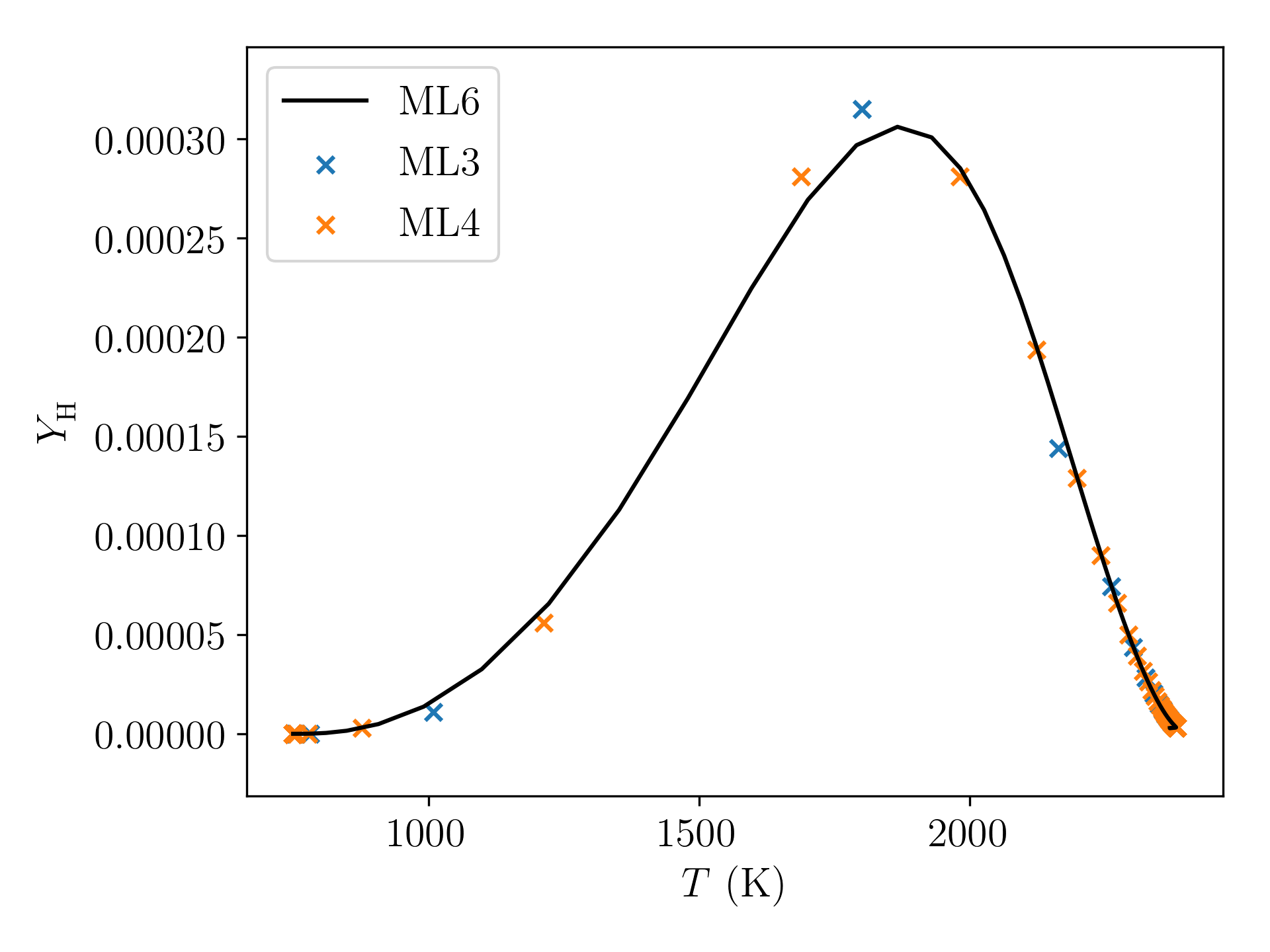}
    \includegraphics[width=50mm]{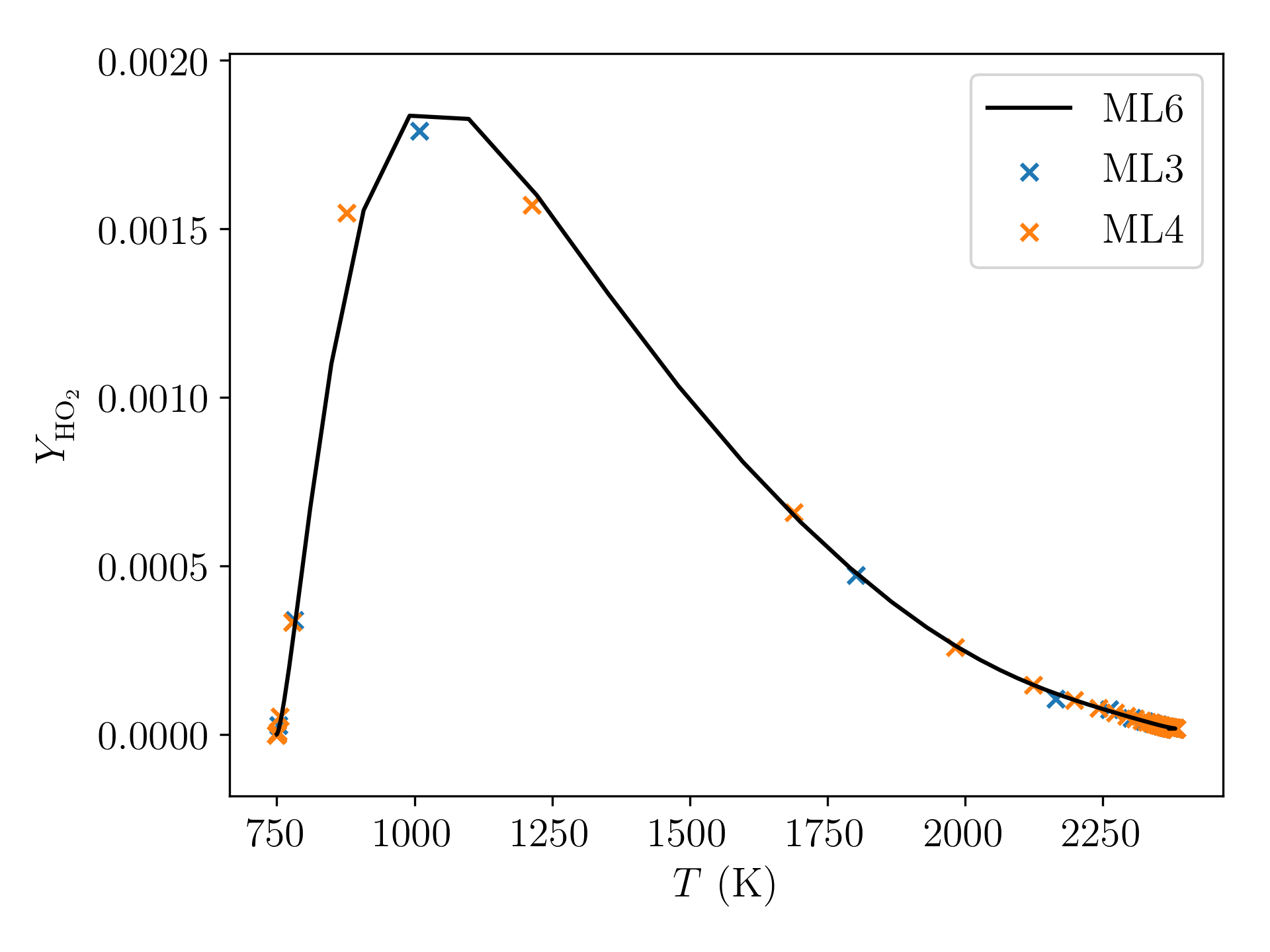}
    \includegraphics[width=50mm]{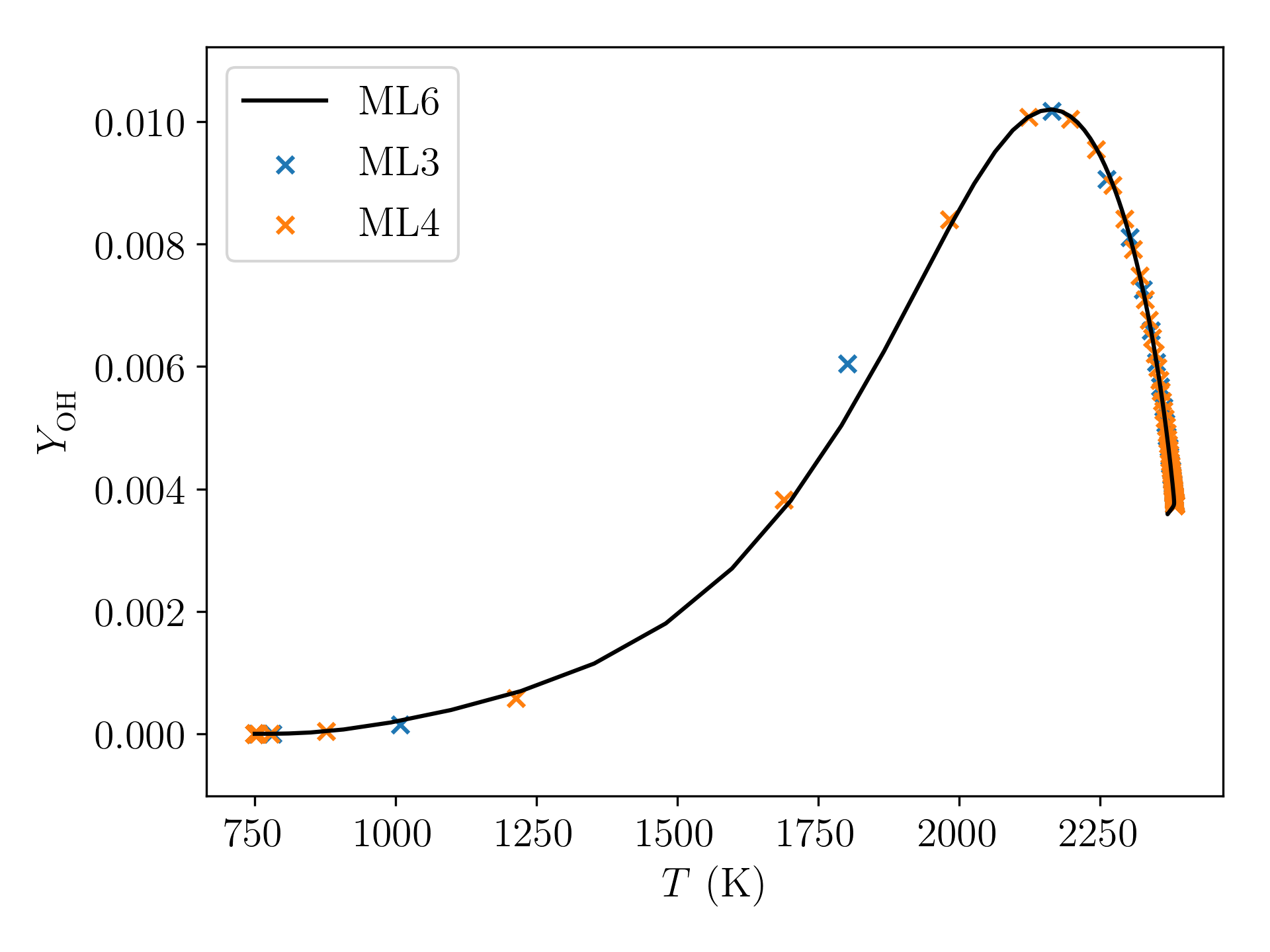}
    \includegraphics[width=50mm]{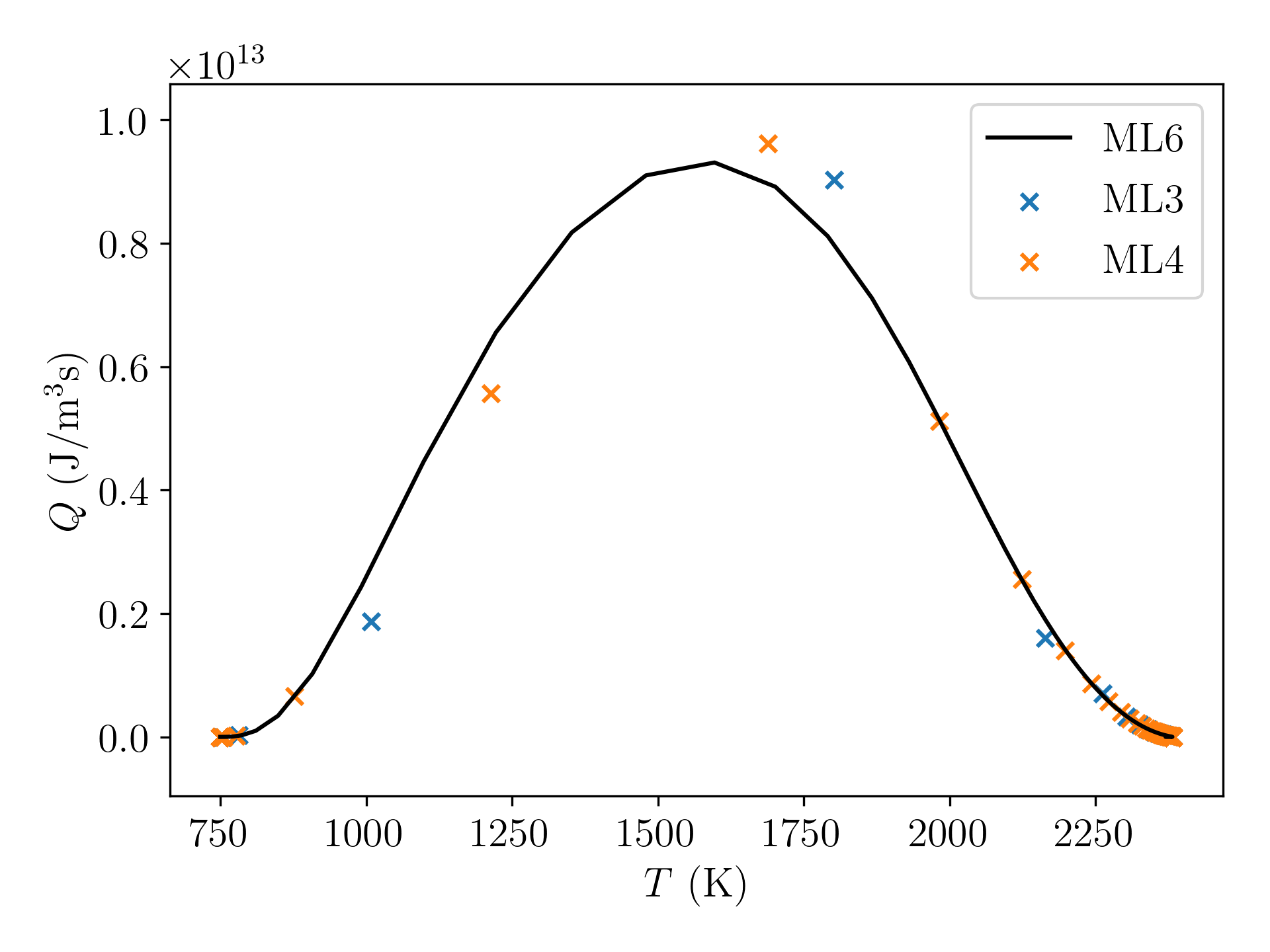}
    
    \caption{1D profiles for $\phi=0.7$}
    \label{fig:07res}
\end{subfigure}
\begin{subfigure}{50mm}
    \centering
    \includegraphics[width=50mm]{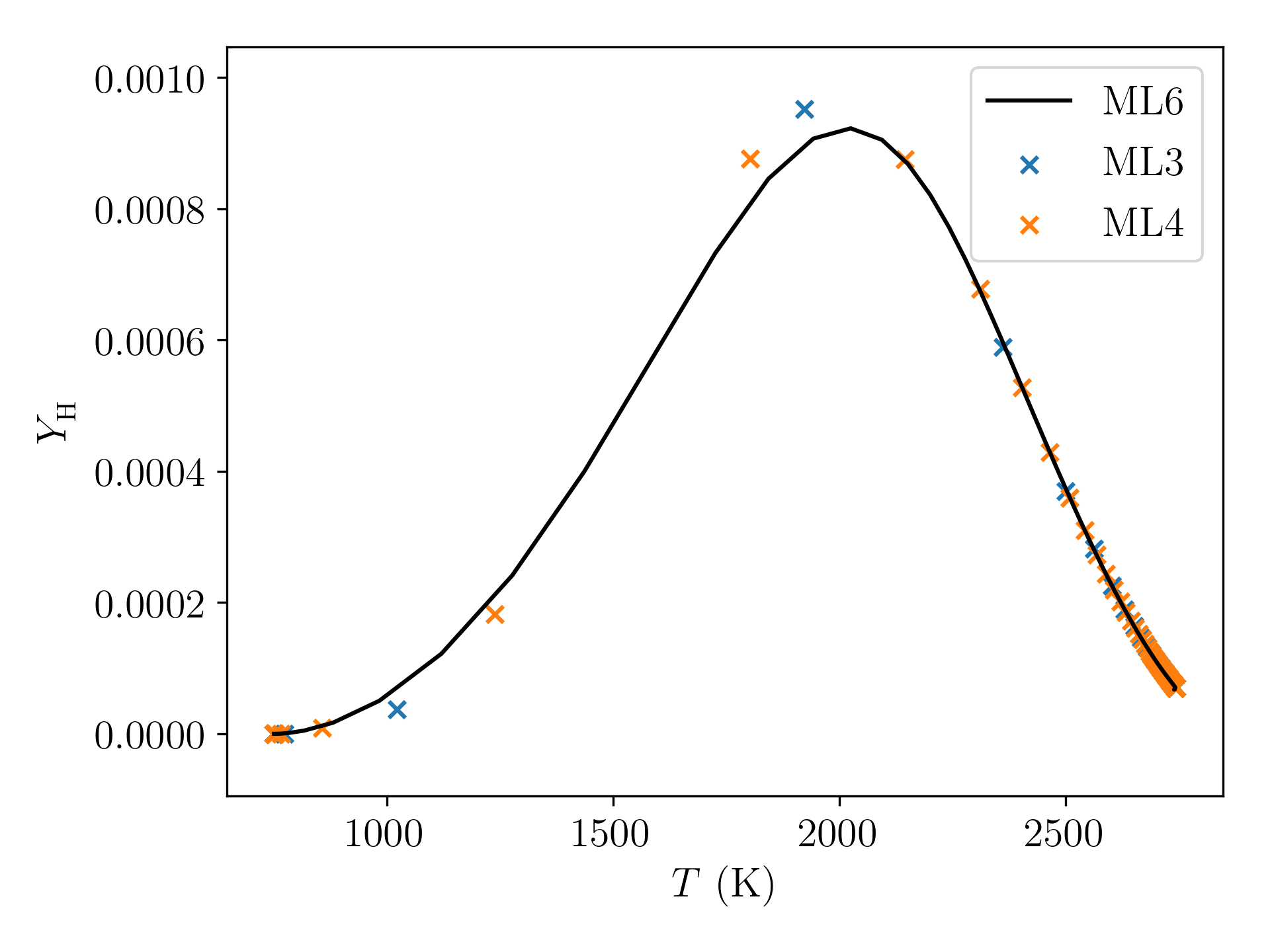}
    \includegraphics[width=50mm]{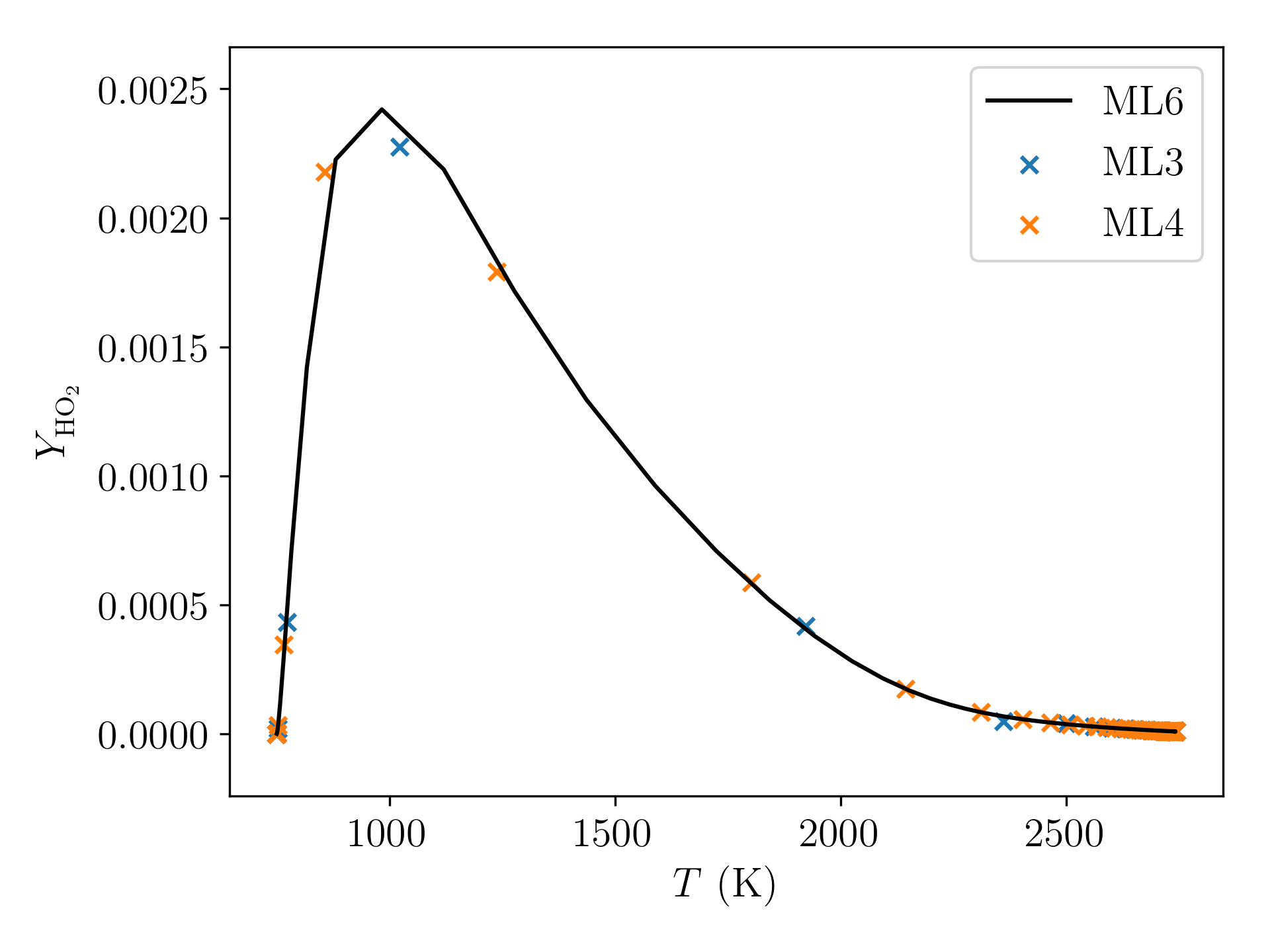}
    \includegraphics[width=50mm]{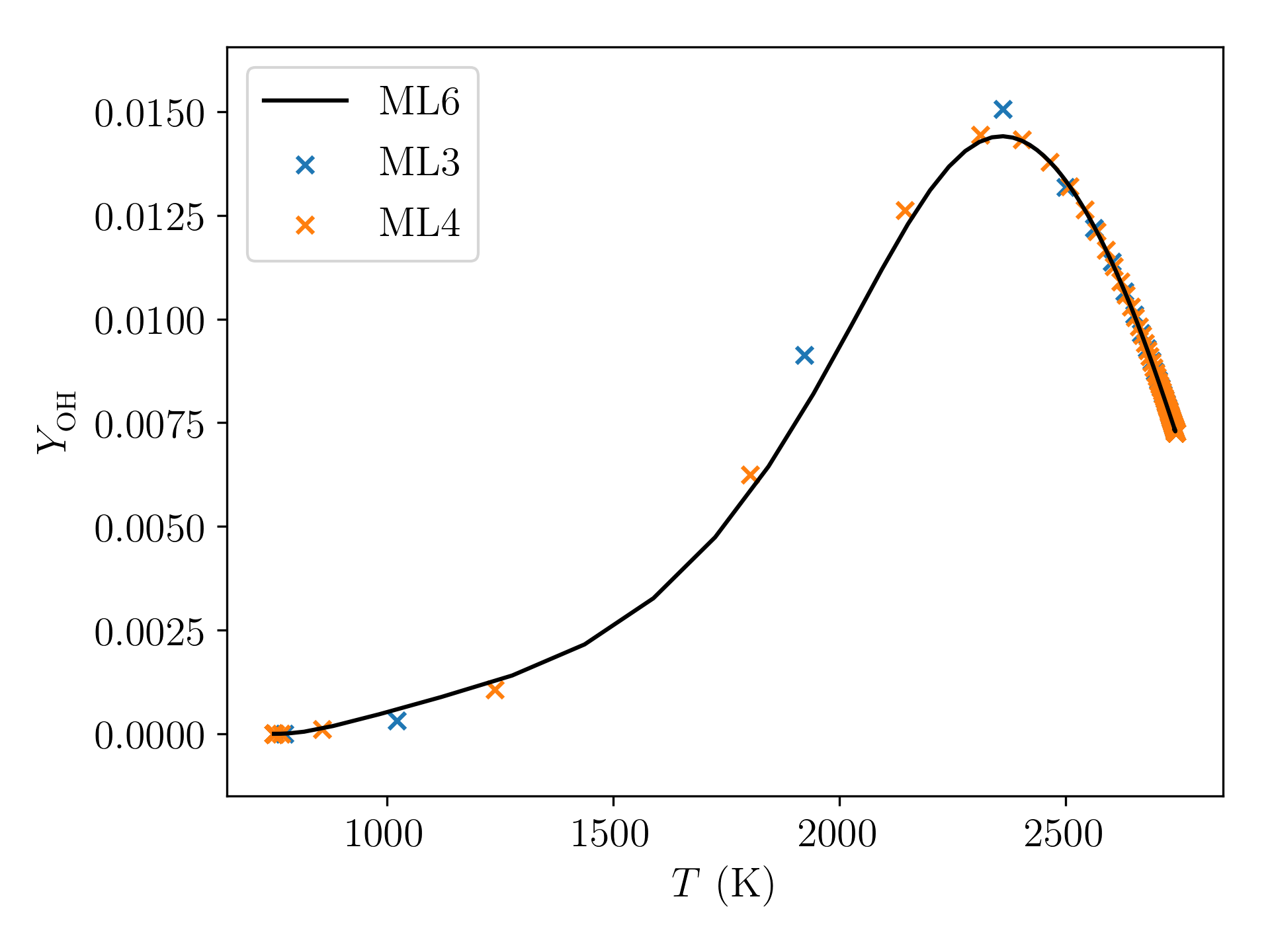}
    \includegraphics[width=50mm]{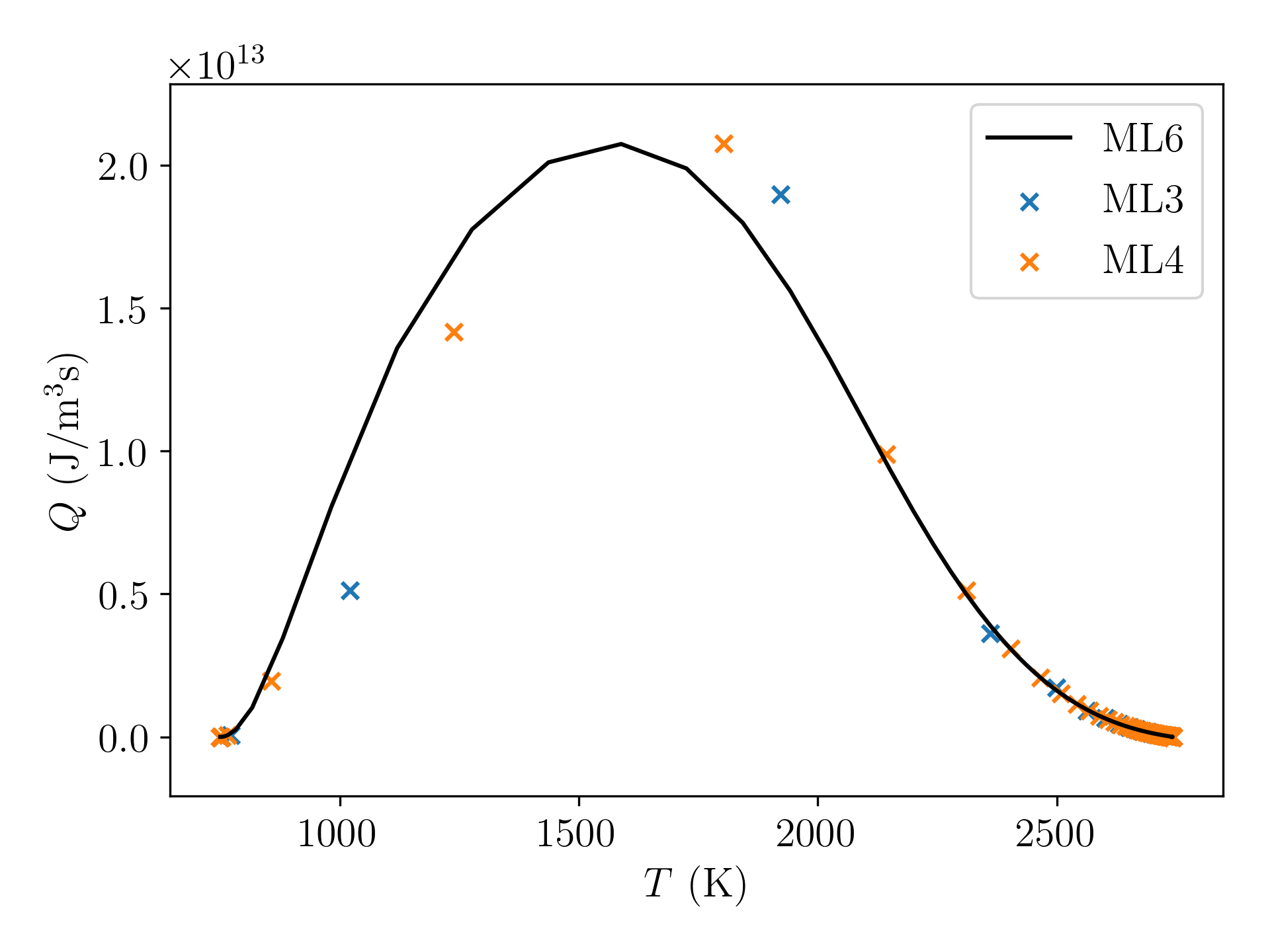}
    
    \caption{1D profiles for $\phi=1.0$}
    \label{fig:10res}
\end{subfigure}
\caption{Profiles of $Y_{\textsc{h}}$, $Y_{\textsc{ho}_{2}}$, $Y_{\textsc{oh}}$ and heat release rate $Q$ as functions of temperature for $\phi=0.4,0.7,1.0$. Reasonable agreement with high-resolution profile is seen throughout, bar small lags in the heat release profiles in the higher equivalence ratio cases.}
\end{figure*}

To establish resolution effects on the results presented in the paper, one-dimensional flames were simulated at three representative equivalence ratios each at three resolutions. The three resolutions presented correspond to ML3 ($\Delta x = 6.84$\,$\mu$m) that was used for data analysis, one extra level at ML4 and finally with two further levels ML6, which was taken as the reference for comparison. Figures \ref{fig:04res}, \ref{fig:07res} and \ref{fig:10res} show the profiles of the mass fractions of H, HO$_{2}$, OH and heat release rate $Q$ against temperature at three different equivalence ratios ($\phi=0.4$, 0.7 and 1.0). In the leanest case shown ($\phi=0.4$, figure \ref{fig:04res}) the thermal thickness is $23.2\mu$m which corresponds to $\Delta x/l_{L} = 3.39$. The one-dimensional profiles appear to be well-matched at both the ML3 and ML4 resolutions. In the middle case ($\phi=0.7$, figure \ref{fig:07res}), the thermal thickness is $10.8\mu$m which corresponds to $\Delta x/l_{L} = 1.58$. Again, the profiles in both ML3 and ML4 data appear to match well (despite the sparsity of data points), with a slight lag in the heat release noted. Finally, at the highest equivalence ratio ($\phi = 1$, figure \ref{fig:10res}), the thermal thickness of $10.7\mu$m which corresponds to $\Delta x/l_{L} = 1.56$. Similar observations are clear, with a nearly identical profile to the $\phi=0.7$ case, with good agreement, save for the slight lag in the heat release profiles. 

In each case, the flame speed was calculated as
\begin{equation}
    s_{\textsc{l},\textrm{lev}} = \frac{-1}{\rho_{u} Y_{\Htwo,u}}\int_{-\infty}^{\infty}\rho \dot{\omega}_{\Htwo} \,\textrm{d}x.
\end{equation}
The flame speeds at $\phi=0.4$ at ML3, 4 and 6 were 0.709, 0.729 and 0.724\,m/s, respectively (i.e.\ a difference of approximately 2\%).  The peak differences at $\phi=0.7$ and $\phi=1.0$ were 0.2\% and 1.4\%, respectively.
These small differences will lead to no different conclusions when examining flame speed statistics (i.e.\ section \ref{sec:stabilisation:propagation}).  Furthermore, the specific details of the internal flame structure are largely irrelevant for the global structure and stability, and the inherent conservative nature of the numerical algorithm guarantees that the temperature of the products (and therefore recirculation region) are accurate.

Additionally, a three-dimensional simulation of the combusting case was conducted at ML4 by restarting from a check point file in the statistically-stationary period.  The finest level matched the gridding criterion for ML3 and so focussed on the core flame; furthermore, the main jet was resolved at one level finer than in the production calculation. The resulting simulation had an effective resolution in excess of 400 billion computational cells, and was run on the Polaris facility at ALCF. It was not possible to run this simulation for long enough to be able to perform temporal averaging; nevertheless, no differences in flame behaviour were observed compared with the ML3 case analysed.